\documentclass[aps,prd,twocolumn,groupedaddress,showpacs,floatfix]{revtex4-1}
\usepackage[dvipdfm]{graphicx}
\usepackage{color}
\usepackage{amsmath,amssymb}

\begin{document}
\title{Gravitational waves from spinning black hole-neutron star
binaries: dependence on black hole spins and on neutron star equations
of state} \author{Koutarou Kyutoku, Hirotada Okawa, Masaru Shibata}
\affiliation{Yukawa Institute for Theoretical Physics, Kyoto University,
Kyoto 606-8502, Japan} \author{Keisuke Taniguchi} \affiliation{Graduate
School of Arts and Sciences, University of Tokyo, Komaba, Meguro, Tokyo
153-8902, Japan} \date{\today}

\begin{abstract}
 We study the merger of black hole-neutron star binaries with a variety
 of black hole spins aligned or antialigned with the orbital angular
 momentum, and with the mass ratio in the range $M_{\rm BH} / M_{\rm NS}
 = 2$--5, where $M_{\rm BH}$ and $M_{\rm NS}$ are the mass of the black
 hole and neutron star, respectively. We model neutron-star matter by
 systematically parametrized piecewise polytropic equations of
 state. The initial condition is computed in the puncture framework
 adopting an isolated horizon framework to estimate the black hole spin
 and assuming an irrotational velocity field for the fluid inside the
 neutron star. Dynamical simulations are performed in full general
 relativity by an adaptive-mesh refinement code, {\tt SACRA}. The
 treatment of hydrodynamic equations and estimation of the disk mass are
 improved. We find that the neutron star is tidally disrupted
 irrespective of the mass ratio when the black hole has a moderately
 large prograde spin, whereas only binaries with low mass ratios,
 $M_{\rm BH} / M_{\rm NS} \lesssim 3$, or small compactnesses of the
 neutron stars bring the tidal disruption when the black hole spin is
 zero or retrograde. The mass of the remnant disk is accordingly large
 as $\gtrsim 0.1 M_\odot$, which is required by central engines of short
 gamma-ray bursts, if the black hole spin is prograde. Information of
 the tidal disruption is reflected in a clear relation between the
 compactness of the neutron star and an appropriately defined ``cutoff
 frequency'' in the gravitational-wave spectrum, above which the
 spectrum damps exponentially. We find that the tidal disruption of the
 neutron star and excitation of the quasinormal mode of the remnant
 black hole occur in a compatible manner in high mass-ratio binaries
 with the prograde black hole spin. The correlation between the
 compactness and the cutoff frequency still holds for such cases. It is
 also suggested by extrapolation that the merger of an extremely
 spinning black hole and an irrotational neutron star binary does not
 lead to the formation of an overspinning black hole.
\end{abstract}
\pacs{04.25.D-, 04.30.-w, 04.40.Dg}

\maketitle

\section{introduction} \label{sec:intro}

Coalescing binaries composed of a black hole (BH) and/or a neutron star
(NS) are among the most promising sources of gravitational waves for
ground-based laser-interferometric gravitational-wave detectors such as
LIGO \cite{ligo2009} and VIRGO \cite{virgo2011}. Detections of
gravitational waves will be accomplished in a decade to come by planned
next-generation detectors such as advanced LIGO, advanced VIRGO, and
LCGT \cite{lcgt2010}. Because gravitational waves are much more
transparent to the absorption and scattering by the material than
electromagnetic waves and even neutrinos are, gravitational-wave
astronomy is expected to become a powerful and unique way to observe
strongly gravitating phenomena in our Universe. Among such phenomena,
the merger of a BH-NS binary plays an important role to investigate
properties of the NS such as the radius and the equation of state (EOS)
of a high-density nuclear matter
\cite{lindblom1992,vallisneri2000,rmsucf2009,fgp2010,ptr2011,pror2011}. An
important constraint on the EOS is obtained from detection of a $1.97
\pm 0.04 M_\odot$ NS, which is the most massive NS currently known
\cite{dprrh2010}, by recent pulsar-timing observation. However, we still
do not know the realistic EOS of the NS because there is no robust
measurement of the NS radius \cite{slb2010}. To constrain the NS radius
and EOS by observing gravitational waves from the BH-NS binary, we have
to prepare accurate theoretical templates of gravitational waveforms
employing a wide variety of the NS EOSs and other physical
parameters. For this purpose, numerical relativity is the unique
approach.

The merger of a BH-NS binary is also an important target for the
astrophysical study because it is a potential candidate for the
progenitor of short-hard gamma-ray bursts (GRBs) in the so-called merger
scenario (see Refs.~\cite{nakar,leeramirezruiz} and references therein
for reviews). If a NS is tidally disrupted during the merger of a BH-NS
binary, a system composed of a spinning BH and a hot, massive accretion
disk of $\gtrsim 0.01 M_\odot$ may be formed. Such an outcome could be a
central engine of the GRB, because it could radiate a large amount of
energy $\gtrsim 10^{48}$ erg via neutrino emission or the
Blandford-Znajek mechanism \cite{blandfordznajek1977} in a short time
scale of $\lesssim 2$ s and hence could launch a GRB jet. The merger
scenario of the GRB is attractive when a short-hard GRB is associated
with a galaxy of low star-formation rate
\cite{bergeretal2005,fbctlfgcf2010} because the collapsar model of GRBs
\cite{woosley1993} is not preferable. Only numerical relativity can
answer quantitatively the question whether the formation of a massive
accretion disk is possible in the BH-NS binary merger.

Fully general relativistic study of BH-NS binaries has achieved progress
in recent years, both in computations of quasiequilibrium states
\cite{grandclement2006,*grandclement2006e,tbfs2007,tbfs2008,fkpt2008,kst2009}
and in dynamical simulations of the merger
\cite{shibatauryu2006,shibatauryu2007,shibatataniguchi2008,eflstb2008,dfkpst2008,elsb2009,skyt2009,dfkot2010,kst2010,cabllmn2010}.
Although these include several simulations of spinning BH-NS binaries
with a qualitative $\Gamma = 2$ ideal-gas EOS, to date, only limited
number of simulations have been performed taking into account both the
nuclear-theory based EOS and the BH spin~\cite{dfkot2010}.  In
particular, we still do not understand the dependence of the merger
process and resulting gravitational waveforms on the BH spin and the EOS
of the NS in detail. One goal of current numerical relativity is to
clarify the effect of the BH spin on the BH-NS binary merger and
subsequent NS tidal disruption adopting a wide variety of the NS EOSs.

In this paper, we report our latest results obtained by
numerical-relativity simulations with a variety of the NS EOSs and the
BH spins. We employ five piecewise polytropic EOSs (see
Sec.~\ref{subsec:ini_pwp}), all of which do not conflict with the
current observation of the $1.97 \pm 0.04 M_\odot$ NS. We systematically
choose physical parameters such as BH mass, NS mass, and BH spin in an
astrophysically realistic range. While we only consider relatively low
mass BHs in the previous work \cite{kst2010}, we adopt a wider range of
the BH mass because a large BH spin enhances the NS tidal disruption for
high mass-ratio binaries. We clarify the dependence on the BH spin and
on the NS EOS of the properties of the merger remnants and
characteristics of gravitational waves---in particular, the
gravitational-wave spectrum.

This paper is organized as follows. In Sec.~\ref{sec:ini}, we describe
methods for a solution of initial conditions, piecewise polytropic EOSs,
and models of BH-NS binaries employed in this paper. In
Sec.~\ref{sec:sim}, the formulation and methods of numerical simulations
are summarized. Section \ref{sec:res} presents the numerical results and
clarifies the effect of the BH spin and NS EOS on the tidal disruption,
merger remnants, and gravitational waves. Section \ref{sec:summary} is
devoted to a summary. Throughout this paper, we adopt the geometrical
units in which $G=c=1$, where $G$ and $c$ are the gravitational constant
and the speed of light, respectively. Our convention of notation for
physically important quantities is summarized in Table
\ref{table:notation}. The nondimensional spin parameter of the BH, total
mass of the system at infinite separation, mass ratio, and compactness
of the NS are defined as $a = S_{\rm BH} / M_{\rm BH}^2$, $m_0 = M_{\rm
BH} + M_{\rm NS}$, $Q = M_{\rm BH} / M_{\rm NS}$, and ${\cal C} = M_{\rm
NS} / R_{\rm NS}$, respectively. Latin and Greek indices denote spatial
and spacetime components, respectively.

\begin{table}[]
 \caption{Our convention of notation for physically important
 quantities.}
 \begin{tabular}{cc}
  \hline
  Symbol & \\
  \hline \hline
  $M_{\rm irr}$ & The irreducible mass of the BH \\
  $S_{\rm BH}$ & The magnitude of the BH spin angular momentum \\
  $M_{\rm BH}$ & The gravitational mass of the BH in isolation \\
  $M_{\rm NS}$ & The gravitational mass of the NS in isolation \\
  $R_{\rm NS}$ & The circumferential radius of the NS in isolation \\
  $M_0$ & The Arnowitt-Deser-Misner mass of the system \\
  $m_0$ & The total mass of the system at infinite separation \\
  \hline
  $Q$ & The mass ratio \\
  ${\cal C}$ & The compactness of the NS \\
  $a$ & The nondimensional spin parameter of the BH \\
  \hline
 \end{tabular}
 \label{table:notation}
\end{table}

\section{Initial condition} \label{sec:ini}

As in our previous works \cite{skyt2009,kst2010}, we employ BH-NS
binaries in quasiequilibrium states for initial conditions of our
numerical simulations. In this section, we summarize the formulation and
methods for the computation of a quasiequilibrium state, specifically
our method of estimating the spin angular momentum of the BH in a binary
and of determining the position of the rotation axis. The details of the
formulation and numerical methods, except for the issues on the BH spin,
are described in Ref.~\cite{kst2009}, to which the reader may
refer. Computations of the quasiequilibrium states are performed using
the spectral-method library LORENE \cite{LORENE}.

\subsection{Formulation and methods} \label{subsec:ini_method}

We compute a quasiequilibrium state of the BH-NS binary as a solution of
the initial value problem of general relativity \cite{cook}. As far as
the orbital separation is large enough, the time scale of the orbital
contraction due to the gravitational radiation reaction is much longer
than the orbital period, and, therefore, we may safely neglect the
gravitational radiation reaction in the calculation of the
quasiequilibrium state. In numerical simulations of the binary
coalescences, we have to track $\gtrsim 5$ orbits in order to calculate
accurate gravitational waveforms during the late inspiral and merger
phases, and hence, the orbital separation of the initial condition has
to be large enough. For such initial conditions, we can neglect the
gravitational radiation reaction. Thus, we give a BH-NS binary in a
quasicircular orbit, i.e., the binary in an approximate equilibrium
state in the corotating frame. To satisfy the quasiequilibrium
requirements described above, we assume the existence of a helical
Killing vector with the orbital angular velocity $\Omega$,
\begin{equation}
 \xi^\mu = ( \partial_t )^\mu + \Omega ( \partial_\varphi )^\mu .
\end{equation}
We also assume that the NS is in the hydrostatic equilibrium in the
corotating frame and has an irrotational velocity field, which is
believed to be a reliable approximation to an astrophysically realistic
configuration \cite{bildstencutler1992,kochanek1992}.

We compute the three-metric $\gamma_{ij}$, the extrinsic curvature
$K_{ij}$, the lapse function $\alpha$, and the shift vector $\beta^i$ by
a mixture of the extended conformal thin-sandwich approach
\cite{york1999,pfeifferyork2003} and the conformal transverse-traceless
decomposition \cite{cook} in the puncture framework
\cite{brandtbrugmann1997,clmz2006,bcckm2006}. In this formalism, we
assume the conformal flatness of the three-metric $\gamma_{ij} = \psi^4
\hat{\gamma}_{ij} = \psi^4 f_{ij}$, the stationarity of the conformal
three-metric $\pounds_\xi \hat{\gamma}_{ij} = 0$, the maximal slicing
condition $K = \gamma^{ij} K_{ij} = 0$, and its preservation in time
$\pounds_\xi K = 0$, where $f_{ij}$ is the flat spatial metric. Assuming
that the puncture is located at $x^i_{\rm P}$, we set the conformal
factor $\psi$ and a weighted lapse function $\Phi \equiv \alpha \psi$ as
\begin{equation}
 \psi = 1 + \frac{M_{\rm P}}{2 r_{\rm BH}} + \phi \; , \; \Phi = 1 -
  \frac{M_\Phi}{2 r_{\rm BH}} + \eta ,
\end{equation}
where $M_{\rm P}$ and $M_\Phi$ are positive constants of mass dimension,
and $r_{\rm BH} = | x^i - x^i_{\rm P} |$ is a coordinate distance from
the puncture. We adjust $M_{\rm P}$ to obtain a desired mass of the BH,
$M_{\rm BH}$, and determine $M_\Phi$ so as to satisfy the virial
relation, i.e., the equality of the Arnowitt-Deser-Misner (ADM) mass and
the Komar mass, which holds when the spacetime is stationary and
asymptotically flat \cite{beig1978,ashtekarashtekar1979}. $\phi$,
$\beta^i$, and $\eta$ are determined by solving elliptic equations
derived from the Hamiltonian constraint, momentum constraint, and
quasiequilibrium conditions, $\pounds_\xi \hat{\gamma}_{ij} = 0$ and
$\pounds_\xi K = 0$. We note that these quasiequilibrium conditions can
be replaced by $\partial_t \hat{\gamma}_{ij} = 0$ and $\partial_t K = 0$
in a conformal flatness approximation. In the puncture framework, we
also decompose a conformally weighted extrinsic curvature $\hat{A}_{ij}
= \psi^2 K_{ij}$ as
\begin{equation}
 \hat{A}_{ij} = \hat{\nabla}_i W_j + \hat{\nabla}_j W_i - \frac{2}{3}
  f_{ij} \hat{\nabla}^k W_k + K^{\rm P}_{ij} , \label{eq:extr}
\end{equation}
where $W_i$ is an auxiliary three-vector field and $\hat{\nabla}_i$ is
the covariant derivative associated with $f_{ij}$. $K^{\rm P}_{ij}$ is a
singular part of the extrinsic curvature, which is associated with the
linear and spin angular momenta of the BH \cite{bowenyork1980},
\begin{eqnarray}
 K_{ij}^{\rm P} &=& \frac{3}{2 r_{\rm BH}^2} [ l_i P^{\rm BH}_j + l_j
  P^{\rm BH}_i - ( f_{ij} - l_i l_j ) l^k P^{\rm BH}_k ] \nonumber \\
 &+& \frac{3}{r_{\rm BH}^3} [ \epsilon_{kil} S_{\rm P}^l l^k l_j +
  \epsilon_{kjl} S_{\rm P}^l l^k l_i] ,
\end{eqnarray}
where $l^i = x_{\rm BH}^i / r_{\rm BH}$ is a unit radial vector, $l_i =
f_{ij} l^j$, and $\epsilon_{ijk}$ is the Levi-Civita tensor associated
with the flat metric $f_{ij}$. $P^{\rm BH}_i$ and $S_{\rm P}^i$ are
parameters associated with the linear and spin angular momenta of the
BH, respectively. Here we determine $P^{\rm BH}_i$ by the condition in
which the total linear momentum of the binary vanishes, and we adjust
$S_{\rm P}^i$ to obtain a desired spin angular momentum of the BH,
$S_{\rm BH}^i$. The elliptic equation to determine $W_i$ is obtained by
taking the derivative of Eq.~(\ref{eq:extr}) and using the momentum
constraint.

The spin angular momentum of the BH, $S_{\rm BH}^i$, is evaluated on the
apparent horizon (hereafter AH), ${\cal S}$, according to the isolated
horizon framework (see
Refs.~\cite{ashtekarkrishnan,gourgoulhonjaramillo} for reviews). Because
we do not know the position of the AH in advance in the puncture
framework, we have to determine the location of the AH numerically
\cite{linnovak2007}. On the numerically determined AH, an approximate
rotational Killing vector may be defined using the method developed by
Cook and Whiting \cite{cookwhiting2007} with the normalization condition
proposed by Lovelace and his collaborators \cite{lopc2008}. We focus
only on the case in which the BH spin is aligned or antialigned with the
angular momentum of the binary in this work, and hence, the axis of the
BH spin is uniquely determined. Hereafter we set the rotational axis of
the binary, equivalently the axis of the BH spin, to be the {\it z} axis
and consider only the approximate Killing vector $\phi^i$ associated
with the rotation in this direction. Using $\phi^i$, we obtain the spin
angular momentum of the BH $S_{\rm BH}^{(\phi)} = S_{\rm BH}^z$ via the
surface integral at the AH,
\begin{equation}
 S_{\rm BH}^{(\phi)} = \frac{1}{8\pi} \int_{\cal S} K_{ij} \phi^i dS^j .
\end{equation}
We adjust $S_{\rm P}^z$ to obtain a desired value of $S_{\rm BH}^z$
(hereafter $S_{\rm BH}$). We note that $S_{\rm P}^i$ and $S_{\rm BH}^i$
do not agree exactly in the BH-NS binary spacetime due to the
contribution to the extrinsic curvature from the NS, associated with
$W_i$.

Because we adopt a conformal flatness approximation for the induced
metric, the Christodoulou mass of the BH evaluated on the AH, $M_{\cal
H} = \sqrt{M_{\rm irr}^2 + ( S_{\rm BH}^2 / 4 M_{\rm irr}^2 )}$, and the
gravitational mass evaluated at spatial infinity, $M_{\rm BH}$, do not
agree even for a single BH system due to the presence of so-called junk
waves. This difference leads to an ambiguity in defining the
nondimensional spin parameter of the BH. Here, we define the
nondimensional spin parameter of the BH with respect to the mass
evaluated {\it at spatial infinity}, i.e.,
\begin{equation}
 a \equiv \frac{S_{\rm BH}}{M_{\rm BH}^2} .
\end{equation}
The reason for this is that the mass and nondimensional spin parameter
of the BH evaluated at the AH quickly (in our simulations, within $\sim
1$ ms) relax to $M_{\rm BH}$ and $a$, defined at spatial infinity,
respectively, as the BH absorbs the junk radiation in the vicinity of
the BH \cite{dlz2008,lopc2008}. We note that these values show the
damping oscillation before the relaxation in the same manner as the
``scalar-curvature spin'' of Ref.~\cite{lopc2008} shows, because our
method of evaluating these values in the simulation is basically the
same as the method to define the scalar-curvature spin in
Ref.~\cite{lopc2008} (see Sec.~\ref{subsec:sim_diagnostics}).

To compute the equations of hydrostatic equilibrium for the NS matter,
we assume an ideal fluid, for which the energy-momentum tensor is
described by
\begin{equation}
 T^{\mu \nu} = \rho h u^\mu u^\nu + P g^{\mu \nu} ,
\end{equation}
where $\rho$ is the rest-mass density, $P$ is the pressure, $h \equiv 1
+ \varepsilon + ( P / \rho )$ is the specific enthalpy, $\varepsilon$ is
the specific internal energy, and $u^\mu$ is the four-velocity of the
fluid. Basic equations for the hydrostatics are derived from the
condition of irrotation, or the vanishing of the vorticity two-form
\cite{bgm1997,asada1998,shibata1998,teukolsky1998},
\begin{equation}
 \omega_{\mu \nu} = \nabla_\mu ( h u_\nu ) - \nabla_\nu ( h u_\mu ) = 0
  ,
\end{equation}
and the conservation of the specific momentum of the fluid along the
helical Killing vector field $\pounds_\xi ( h u_\mu ) = 0$. The specific
enthalpy is determined from the first integral of the relativistic Euler
equation (relativistic Bernoulli integral), and other thermodynamical
quantities are subsequently obtained by using the EOS, which is
described in Sec.~\ref{subsec:ini_pwp}. The four-velocity of the fluid
is calculated using the velocity potential $\Psi$, as $h u_i =
\partial_i \Psi$ in the assumption of the irrotational velocity
field. The elliptic equation for $\Psi$ is derived from the equation of
continuity, $\nabla_\mu ( \rho u^\mu ) = 0$.

We have no definite condition to determine the location of the center of
mass of the binary in the puncture framework and we use this ambiguity
to reduce an unphysical initial orbital eccentricity. We found in our
previous work \cite{skyt2009,kst2009} that orbits with a small
eccentricity could be obtained using the ``3PN-J method,'' i.e., a
phenomenological method to determine the location of the rotational axis
in which the total angular momentum of the binary for a given value of
$\Omega m_0$ agrees with that calculated from the third post-Newtonian
(3PN) approximation. In the present case in which the BH has a finite
amount of spin angular momentum, we extend the previous 3PN-J method to
include the contribution from the BH spin up to the 2.5PN order
\cite{fbb2006,bbf2006}. Specifically, the location of the rotational
axis is chosen from the condition that the orbital angular momentum of
the binary agrees with a sum of nonspin terms given by Eq.~(4) of
Ref.~\cite{blanchet2002} and of spin terms given by Eq.~(7.10) of
Ref.~\cite{bbf2006} for a given value of $\Omega m_0$. We show in
Sec.~\ref{subsec:res_merger} that this extended 3PN-J method again leads
to the initial condition with a small eccentricity.

\subsection{Piecewise polytropic equations of state}
\label{subsec:ini_pwp}

The matter inside the NS in the late inspiral phase is believed to be
well-approximated by a zero-temperature nuclear matter because the
cooling time scale of the NS in typical BH-NS binaries is shorter than
the time scale of the gravitational radiation reaction
\cite{lattimerprakash}. Hence, we employ a cold EOS, for which the
rest-mass density, $\rho$, determines all other thermodynamical
quantities, for calculating the quasiequilibrium state of the BH-NS
binary. To model nuclear-theory-based EOSs at high density with a small
number of parameters, we employ a piecewise polytropic EOS.  It is a
phenomenologically parametrized EOS of the form
\begin{equation}
 P (\rho) = \kappa_i \rho^{\Gamma_i} ~~ {\rm for} ~~ \rho_{i-1} \le
  \rho < \rho_i ~~ (1 \le i \le n) ,
\end{equation}
where $n$ is the number of the pieces used to parametrize an EOS,
$\rho_i$ is the rest-mass density at the boundary of two neighboring
$i$th and $(i+1)$th pieces, $\kappa_i$ is the polytropic constant for
the $i$th piece, and $\Gamma_i$ is the adiabatic index for the $i$th
piece. Here, $\rho_0 = 0$, $\rho_n \to \infty$, and other parameters
$(\rho_i , \kappa_i, \Gamma_i)$ are freely chosen. Requiring the
continuity of the pressure at each $\rho_i$, $2n$ free parameters---say
$(\kappa_i,\Gamma_i)$---determine the EOS completely. The specific
internal energy, $\varepsilon$, and hence the specific enthalpy, $h$,
are determined by the first law of thermodynamics and continuity of each
variable at boundary densities, $\rho_i$.

It was shown that piecewise polytropic EOSs with four pieces
approximately reproduce most properties of the nuclear-theory-based EOSs
at high density \cite{rlof2009}. If we focus on low mass NSs with
relatively low central density, the EOS at high density plays a minor
role. Thus, we adopt a simplified piecewise polytropic EOS composed of
two pieces, one of which models the crust EOS and the other of which the
core EOS. This simplification is based on the fact that NSs in the
observed binary NSs often have fairly small masses $\lesssim 1.4
M_\odot$ \cite{stairs2004} and the maximum rest-mass density in such NSs
may not be so high that the EOS at high density plays only a minor role
in determining their structure. Furthermore, the maximum rest-mass
density inside the NS should only decrease during the evolution of the
BH-NS binary due to tidal elongation of the NS by the companion BH.

Table \ref{table:EOS} lists the EOSs which we employ in this
study. Following Refs.~\cite{rmsucf2009,kst2010}, we always fix the EOS
for the crust region by the parameters below:
\begin{eqnarray}
 \Gamma_1 &=& 1.35692395 , \\
 \kappa_1 / c^2 &=& 3.99873692 \times 10^{-8} \; ( {\rm g} / {\rm cm}^3
  )^{1 - \Gamma_1} .
\end{eqnarray}
The EOS of the core region is determined by two parameters. One is the
adiabatic index of the core EOS, $\Gamma_2$. Whereas we find that
properties of gravitational waves and the merger remnants depend on
$\Gamma_2$ in our previous study \cite{kst2010}, we always fix $\Gamma_2
= 3$ in this work to focus on clarifying the effect of the BH spin aside
from the difference in the adiabatic index. The other parameter is
chosen to be the pressure $p$ at a fiducial density $\rho_{\rm fidu} =
10^{14.7} {\rm g/cm}^3$ because $p$ is closely related to the radius and
deformability of the NS \cite{lattimerprakash2001}. We vary the value of
$p$ systematically to investigate the effect of the stiffness of the EOS
\footnote{In this paper, we determine the stiffness simply as the
magnitude of the pressure for the nuclear-density region.}. With the
given values of $\Gamma_2$ and $p$, $\kappa_2$ and $\rho_1$ are
determined as
\begin{eqnarray}
 \kappa_2 &=& p \rho_{\rm fidu}^{-\Gamma_2} , \\
 \rho_1 &=& ( \kappa_1 / \kappa_2 )^{1 / ( \Gamma_2 - \Gamma_1 )} .
\end{eqnarray}

\begin{table*}[]
 \caption{Key ingredients of the adopted EOSs. $\Gamma_2 (=3.0)$ is the
 adiabatic index in the core region and $p$ is the pressure at the
 fiducial density $\rho_{\rm fidu} = 10^{14.7}$ ${\rm g/cm}^3$, which
 determines the polytropic constant $\kappa_2$ of the core region and
 $\rho_{1}$, the critical rest-mass density separating the crust and
 core regions. $M_{\rm max}$ is the maximum mass of the spherical NS for
 a given EOS. $R_{135} (R_{12},R_{145})$ and ${\cal C}_{135} ({\cal
 C}_{12},{\cal C}_{145})$ are the circumferential radius and the
 compactness of the NS with $M_{\rm NS} = 1.35M_\odot (1.2M_\odot,
 1.45M_\odot)$.}
 \begin{tabular}{ccc|cccccccc} \hline
 Model & $\Gamma_2$ & $\log_{10} p$ $({\rm g/cm}^3)$ & $\rho_1
  (10^{14}~{\rm g/cm}^3)$ & $M_{\rm max} [M_\odot]$ & $R_{135}$ (km) &
  ${\cal C}_{135}$ & $R_{12}$ (km) & ${\cal C}_{12}$ & $R_{145}$ (km) &
  ${\cal C}_{145}$ \\
 \hline \hline
  2H & 3.0 & 13.95 & 0.7033 & 2.835 & 15.23 & 0.1309 & 15.12 & 0.1172 &
  15.28 & 0.1401
  \\ \hline
  1.5H & 3.0 & 13.75 & 0.9308 & 2.525 & 13.69 & 0.1456 & 13.63 & 0.1300 &
  13.72 & 0.1561 \\ \hline
  H & 3.0 & 13.55 & 1.232 & 2.249 & 12.27 & 0.1624 & 12.25 & 0.1447 &
  12.27 & 0.1744
  \\ \hline
  HB & 3.0 & 13.45 & 1.417 & 2.122 & 11.61 & 0.1718 & 11.60 & 0.1527 &
  11.59 & 0.1848
  \\ \hline
  B & 3.0 & 13.35 & 1.630 & 2.003 & 10.96 & 0.1819 & 10.98 & 0.1614 &
  10.93 & 0.1960
  \\ \hline
 \end{tabular}
 \label{table:EOS}
\end{table*}

\subsection{Models} \label{sec:ini_model}

Numerical simulations are performed for a wide range of nondimensional
BH spin parameter, $a$, as well as for a variety of the mass ratios,
$Q$.  For nonspinning BH-NS binaries, we already found that the low mass
ratio of $Q \lesssim 3$ is required for tidal disruption of NSs to occur
sufficiently outside the innermost stable circular orbit (ISCO) of the
BH unless the EOS is extremely stiff \cite{kst2010}. If the tidal
disruption occurs inside or at an orbit very close to the ISCO, we do
not see strong effects of the tidal disruption. In such cases,
gravitational waveforms are similar to those of a BH-BH binary even in
the merger phase, and the mass of the remnant disk is negligible
\cite{lksbf}. However, the allowed range of the mass ratio for the tidal
disruption is modified drastically for a BH-NS binary with the prograde
BH spin \cite{shibata1996,wigginslai2000} because the ISCO radius
\footnote{In this paper, ``the ISCO radius'' always represents ``the
ISCO radius in the Boyer-Lindquist coordinates,'' which is physical in
the sense that it gives the proper circumferential length for the
equatorial circular orbit. It should be noted that the coordinate radius
of the ISCO in our simulation is different from the Boyer-Lindquist
one.} of the BH with a prograde spin becomes smaller by a factor of 1 --
6~\cite{bpt1972} than that of the nonspinning BH with the same
mass. Strong spin effects for the tidal disruption are also found in the
numerical-relativity simulation of the spinning BH-NS binary merger with
a simplified, $\Gamma$-law EOS \cite{elsb2009}. In this paper we perform
a more systematic study of the tidal disruption for different EOSs,
masses of each component, and BH spins.

Table \ref{table:model} summarizes several key quantities for the
initial conditions in our numerical simulations. The label for the model
denotes the EOS name, the mass ratio, the NS mass, and the
nondimensional spin parameter of the BH. Specifically, ``a75,'' ``a5,''
and ``a-5'' correspond to the spin parameters $a = 0.75$, $0.5$, and
$-0.5$, respectively. For example, HB-Q3M135a5 means that the EOS is HB
and $(Q,M_{\rm NS},a) = (3,1.35M_\odot,0.5)$. Although we vary the NS
mass systematically, the results of the merger remnant are reported only
for binaries with $M_{\rm NS} = 1.35 M_\odot$ in this paper because the
difference in the NS mass complicates the properties of the remnant,
such as the mass of the disk. Results for $M_{\rm NS} \neq 1.35 M_\odot$
are analyzed only for gravitational waves.

For the same value of the mass ratio, we basically prepare the initial
conditions with the same value of the initial angular velocity
$\Omega_0$ normalized by the total mass of the binary, $\Omega_0 m_0$.
For 2H EOS, in which the NS radius is the largest, we exceptionally
adopt a smaller value of $\Omega_0 m_0$ than for other EOSs to guarantee
$\gtrsim 5$ orbits before tidal disruption occurs. The reason for this
is that the tidal disruption occurs for a large orbital separation in 2H
EOS. When the BH has a prograde spin, the number of orbits to the merger
for a given value of $\Omega_0 m_0$ increases due to spin-orbit
repulsive interaction \cite{blanchet}, compared to the nonspinning BH
case. On the other hand, when the BH has a retrograde spin, the number
of orbits decreases due to spin-orbit attractive interaction. For
$a=-0.5$, the number of orbits is typically by $\sim 1$ orbit smaller
than for $a=0$. For this reason, we also prepare the initial condition
with a smaller value of $\Omega_0 m_0$ for H EOS and $a=-0.5$.

\begin{table*}[]
 \caption{Key parameters and quantities for the initial conditions
 adopted in numerical simulations. The adopted EOS, mass ratio ($Q$), NS
 mass in isolation ($M_{\rm NS}$), nondimensional spin parameter of the
 BH ($a$), initial angular velocity ($\Omega_0$) in units of $c^3/Gm_0$,
 baryon rest mass ($M_*$), compactness of the NS in isolation (${\cal
 C}$), maximum rest-mass density ($\rho_{\rm max}$), ADM mass of the
 system ($M_0$), and total angular momentum of the system ($J_0$),
 respectively. See also \cite{kst2010} for models of nonspinning BH-NS
 binaries.}
 \begin{tabular}{c|ccccc|ccccc} \hline
 Model & EOS & $Q$ & $M_{\rm NS} [M_\odot$] & $a$ & $G \Omega_0 m_0/c^3$
  & $M_* [M_\odot]$ & ${\cal C}$ & ~~$\rho_{\rm max} (10^{14} {\rm
  g/cm}^3)$~~ & $M_0 [M_\odot]$ & $J_0 [G M^2_\odot / c]$ \\ \hline
  \hline
  2H-Q2M135a75 & 2H & 2 & 1.35 & 0.75 & 0.025 & 1.455 & 0.1309 & 3.740 &
  4.014 & 13.83 \\
  1.5H-Q2M135a75 & 1.5H & 2  & 1.35 & 0.75 & 0.028 & 1.468 & 0.1456 &
  5.104 & 4.012 & 13.42 \\
  H-Q2M135a75 & H & 2 & 1.35 & 0.75 & 0.028 & 1.484 & 0.1624 & 7.019 &
  4.012 & 13.42 \\
  HB-Q2M135a75 & HB & 2 & 1.35 & 0.75 & 0.028 & 1.493 & 0.1718 & 8.263 &
  4.012 & 13.42 \\
  B-Q2M135a75 & B & 2 & 1.35 & 0.75 & 0.028 & 1.503 & 0.1819 & 9.762 &
  4.012 & 13.42 \\
  \hline
  2H-Q2M135a5 & 2H & 2 & 1.35 & 0.5 & 0.025 & 1.455 & 0.1309 & 3.740 &
  4.014 & 14.02 \\
  1.5H-Q2M135a5 & 1.5H & 2 & 1.35 & 0.5 & 0.028 & 1.468 & 0.1456 & 5.104
  & 4.012 & 13.63 \\
  H-Q2M135a5 & H & 2 & 1.35 & 0.5 & 0.028 & 1.484 & 0.1624 & 7.018 &
  4.012 & 13.63 \\
  HB-Q2M135a5 & HB & 2 & 1.35 & 0.5 & 0.028 & 1.493 & 0.1718 & 8.263 &
  4.012 & 13.63 \\
  B-Q2M135a5 & B & 2 & 1.35 & 0.5 & 0.028 & 1.503 & 0.1819 & 9.762 &
  4.012 & 13.63 \\
  \hline
  2H-Q2M135a-5 & 2H & 2 & 1.35 & $-0.5$ & 0.022 & 1.455 & 0.1309 & 3.740
				  & 4.019 & 15.15 \\
  H-Q2M135a-5 & H & 2 & 1.35 & $-0.5$ & 0.025 & 1.484 & 0.1624 & 7.018 &
				      4.017 & 14.74 \\
  HB-Q2M135a-5 & HB & 2 & 1.35 & $-0.5$ & 0.028 & 1.493 & 0.1718 & 8.262
				  & 4.015 & 14.41 \\
  B-Q2M135a-5 & B & 2 & 1.35 & $-0.5$ & 0.028 & 1.503 & 0.1819 & 9.760 &
				      4.015 & 14.41 \\
  \hline
  2H-Q2M12a75 & 2H & 2 & 1.2 & 0.75 & 0.025 & 1.282 & 0.1172 & 3.465 &
  3.568 & 10.93 \\
  H-Q2M12a75 & H & 2 & 1.2 & 0.75 & 0.028 & 1.303 & 0.1447 & 6.421 &
  3.566 & 10.60 \\
  HB-Q2M12a75 & HB & 2 & 1.2 & 0.75 & 0.028 & 1.310 & 0.1527 & 7.523 &
  3.566 & 10.60 \\
  B-Q2M12a75 & B & 2 & 1.2 & 0.75 & 0.028 & 1.317 & 0.1614 & 8.833 &
  3.566 & 10.60 \\
  \hline
  2H-Q2M145a75 & 2H & 2 & 1.45 & 0.75 & 0.025 & 1.572 & 0.1401 & 3.926 &
  4.312 & 15.96 \\
  H-Q2M145a75 & H & 2 & 1.45 & 0.75 & 0.028 & 1.607 & 0.1744 & 7.452 &
  4.309 & 15.48 \\
  HB-Q2M145a75 & HB & 2 & 1.45 & 0.75 & 0.028 & 1.617 & 0.1848 & 8.811 &
  4.309 & 15.48 \\
  B-Q2M145a75 & B & 2 & 1.45 & 0.75 & 0.028 & 1.629 & 0.1960 & 10.46 &
  4.309 & 15.48 \\
  \hline
  2H-Q3M135a75 & 2H & 3 & 1.35 & 0.75 & 0.028 & 1.455 & 0.1309 & 3.737 &
  5.357 & 20.00 \\
  1.5H-Q3M135a75 & 1.5H & 3 & 1.35 & 0.75 & 0.030 & 1.468 & 0.1456 &
  5.100 & 5.355 & 19.64 \\
  H-Q3M135a75 & H & 3 & 1.35 & 0.75 & 0.030 & 1.484 & 0.1624 & 7.013 &
  5.355 & 19.64 \\
  HB-Q3M135a75 & HB & 3 & 1.35 & 0.75 & 0.030 & 1.493 & 0.1718 & 8.256 &
  5.355 & 19.64 \\
  B-Q3M135a75 & B & 3 & 1.35 & 0.75 & 0.030 & 1.503 & 0.1819 & 9.753 &
  5.355 & 19.63 \\
  \hline
  2H-Q3M135a5 & 2H & 3 & 1.35 & 0.5 & 0.028 & 1.455 & 0.1309 & 3.737 &
  5.357 & 20.36 \\
  1.5H-Q3M135a5 & 1.5H & 3 & 1.35 & 0.5 & 0.030 & 1.468 & 0.1456 & 5.100
  & 5.356 & 20.02 \\
  H-Q3M135a5 & H & 3 & 1.35 & 0.5 & 0.030 & 1.484 & 0.1624 & 7.012 &
  5.356 & 20.01 \\
  HB-Q3M135a5 & HB & 3 & 1.35 & 0.5 & 0.030 & 1.493 & 0.1718 & 8.255 &
  5.356 & 20.01 \\
  B-Q3M135a5 & B & 3 & 1.35 & 0.5 & 0.030 & 1.503 & 0.1819 & 9.753 &
  5.356 & 20.01 \\
  \hline
  HB-Q3M135a-5 & HB & 3 & 1.35 & $-0.5$ & 0.030 & 1.493 & 0.1718 & 8.253
				  & 5.359 & 21.46 \\
  \hline
  2H-Q3M145a75 & 2H & 3 & 1.45 & 0.75 & 0.028 & 1.572 & 0.1401 & 3.923 &
  5.754 & 23.07 \\
  H-Q3M145a75 & H & 3 & 1.45 & 0.75 & 0.030 & 1.607 & 0.1744 & 7.445 &
  5.751 & 22.65 \\
  HB-Q3M145a75 & HB & 3 & 1.45 & 0.75 & 0.030 & 1.617 & 0.1848 & 8.803 &
  5.751 & 22.65 \\
  B-Q3M145a75 & B & 3 & 1.45 & 0.75 & 0.030 & 1.629 & 0.1960 & 10.45 &
  5.751 & 22.65 \\
  \hline
  2H-Q4M135a75 & 2H & 4 & 1.35 & 0.75 & 0.030 & 1.455 & 0.1309 & 3.735 &
  6.702 & 26.07 \\
  H-Q4M135a75 & H & 4 & 1.35 & 0.75 & 0.032 & 1.484 & 0.1624 & 7.007 &
  6.700 & 25.62 \\
  HB-Q4M135a75 & HB & 4 & 1.35 & 0.75 & 0.032 & 1.493 & 0.1718 & 8.249 &
  6.700 & 25.63 \\
  B-Q4M135a75 & B & 4 & 1.35 & 0.75 & 0.032 & 1.503 & 0.1819 & 9.746 &
  6.700 & 25.62 \\
  \hline
  2H-Q4M135a5 & 2H & 4 & 1.35 & 0.5 & 0.035 & 1.455 & 0.1309 & 3.732 &
  6.698 & 25.64 \\
  H-Q4M135a5 & H & 4 & 1.35 & 0.5 & 0.035 & 1.484 & 0.1624 & 7.004 &
  6.698 & 25.63 \\
  HB-Q4M135a5 & HB & 4 & 1.35 & 0.5 & 0.035 & 1.493 & 0.1718 & 8.244 &
  6.698 & 25.63 \\
  B-Q4M135a5 & B & 4 & 1.35 & 0.5 & 0.035 & 1.503 & 0.1819 & 9.740 &
  6.698 & 25.63 \\
  \hline
  2H-Q5M135a75 & 2H & 5 & 1.35 & 0.75 & 0.036 & 1.455 & 0.1309 & 3.730 &
  8.044 & 30.95 \\
  H-Q5M135a75 & H & 5 & 1.35 & 0.75 & 0.036 & 1.484 & 0.1624 & 7.000 &
  8.044 & 30.95 \\
  HB-Q5M135a75 & HB & 5 & 1.35 & 0.75 & 0.036 & 1.493 & 0.1718 & 8.241 &
  8.044 & 30.95 \\
  B-Q5M135a75 & B & 5 & 1.35 & 0.75 & 0.036 & 1.503 & 0.1819 & 9.736 &
  8.043 & 30.95 \\
  \hline
 \end{tabular}
 \label{table:model}
\end{table*}

\section{Methods of simulations} \label{sec:sim}

Numerical simulations are performed using an adaptive-mesh refinement
(AMR) code {\tt SACRA} \cite{yst2008}. The formulation, the gauge
conditions, the numerical scheme, and the methods of diagnostics are
basically the same as those described in Ref.~\cite{kst2010}, except for
the correction in the treatment of hydrodynamic equations in a far
region. Thus, we here only briefly review them and describe the present
setup of the computational domain for the AMR algorithm and grid
resolution.

\subsection{Formulation and numerical methods} \label{subsec:sim_method}

{\tt SACRA} solves the Einstein evolution equations in the
Baumgarte-Shapiro-Shibata-Nakamura (BSSN) formalism
\cite{shibatanakamura1995,baumgarteshapiro1998} with the moving-puncture
gauge \cite{brandtbrugmann1997,clmz2006,bcckm2006}. It evolves a
conformal factor $W \equiv \gamma^{-1/6}$, the conformal metric
$\tilde{\gamma}_{ij} \equiv \gamma^{-1/3} \gamma_{ij}$, the trace of the
extrinsic curvature $K$, a conformally weighted trace-free part of the
extrinsic curvature $\tilde{A}_{ij} \equiv \gamma^{-1/3} ( K_{ij} - K
\gamma_{ij} )$, and an auxiliary variable $\tilde{\Gamma}^i \equiv -
\partial_j \tilde{\gamma}^{ij}$. Introducing an auxiliary variable $B^i$
and a parameter $\eta_s$, which we typically set to be $\sim M_{\rm BH}
/ M_\odot$ in units of $c = G = M_\odot =1$, we employ a moving-puncture
gauge in the form \cite{bghhst2008}
\begin{eqnarray}
 ( \partial_t - \beta^j \partial_j ) \alpha &=& - 2 \alpha K , \\
 ( \partial_t - \beta^j \partial_j ) \beta^i &=& (3/4) B^i , \\
 ( \partial_t - \beta^j \partial_j ) B^i &=& ( \partial_t - \beta^j
  \partial_j ) \tilde{\Gamma}^i - \eta_s B^i .
\end{eqnarray}
We evaluate the spatial derivative by a fourth-order central finite
difference, except for the advection terms, which are evaluated by a
fourth-order noncentered, upwind finite difference, and employ a
fourth-order Runge-Kutta method for the time evolution.

To solve the hydrodynamic equations, we evolve $\rho_* \equiv \rho
\alpha u^t W^{-3}$, $\hat{u}_i \equiv h u_i$, and $e_* \equiv h \alpha
u^t - P / ( \rho \alpha u^t )$. The advection terms are handled with a
high-resolution central scheme by Kurganov and Tadmor
\cite{kurganovtadmor2000} with a third-order piecewise parabolic
interpolation for the cell reconstruction. For the EOS, we decompose the
pressure and specific internal energy into cold and thermal parts as
\begin{equation}
 P = P_{\rm cold} + P_{\rm th} \; , \; \varepsilon = \varepsilon_{\rm
  cold} + \varepsilon_{\rm th} .
\end{equation}
We calculate the cold parts of both variables using the piecewise
polytropic EOS from the primitive variable $\rho$, and then the thermal
part of the specific internal energy is defined from $\varepsilon$ as
$\varepsilon_{\rm th} = \varepsilon - \varepsilon_{\rm cold}$. Because
$\varepsilon_{\rm th}$ vanishes in the absence of shock heating,
$\varepsilon_{\rm th}$ is regarded as the finite-temperature part. In
this paper, we adopt a $\Gamma$-law ideal-gas EOS for the thermal part,
\begin{equation}
 P_{\rm th} = ( \Gamma_{\rm th} - 1 ) \rho \varepsilon_{\rm th} ,
\end{equation}
to determine the thermal part of the pressure, and choose $\Gamma_{\rm
th}$ equal to the adiabatic index in the crust region, $\Gamma_1$, for
simplicity.

Because the vacuum is not allowed in any conservative hydrodynamic
scheme, we put an artificial atmosphere of a small density outside the
NS in the same way as done in our previous work \cite{kst2010}. The
total rest mass of the atmosphere is always less than $10^{-4} M_\odot$,
and hence, we can safely neglect spurious effects by accretion of the
atmosphere onto the remnant disk as long as the disk mass is much larger
than $10^{-4} M_\odot$.

\subsection{Diagnostics} \label{subsec:sim_diagnostics}

We extract gravitational waves by calculating the outgoing part of the
Weyl scalar $\Psi_4$ at finite coordinate radii $r = 400$--$800 M_\odot$
and by integrating $\Psi_4$ twice in time as
\begin{equation}
 h_+ (t) - i h_\times (t) = - \int^t dt' \int^{t'} dt'' \Psi_4 ( t'' )
  . \label{eq:wave}
\end{equation}
In our previous works \cite{skyt2009,kst2010}, we directly perform this
integration of $\Psi_4 (t)$ and then subtract a quadratic function of
the form $a_2 t^2 + a_1 t + a_0$ to eliminate unphysical drift
components in the waveform, using the least-square fitting to obtain
constants $a_0$, $a_1$, and $a_2$. In this work, we adopt a
``fixed-frequency integration'' method proposed by Reisswig and Pollney
\cite{reisswigpollney2010} to obtain gravitational waveforms with less
unphysical components. In this method, we first perform a Fourier
transformation of $\Psi_4$ as
\begin{equation}
 \tilde{\Psi}_4 ( \omega ) = \int dt \Psi_4 (t) e^{i \omega t} .
\end{equation}
Using this, Eq.~(\ref{eq:wave}) is rewritten as
\begin{equation}
 h_+ (t) - i h_\times (t) = \frac{1}{2 \pi} \int \frac{\tilde{\Psi}_4 (
  \omega )}{\omega^2} e^{-i \omega t} d \omega .
\end{equation}
We then replace $1 / \omega^2$ of the integrand with $1 / \omega_0^2$
for $| \omega | < \omega_0$, where $\omega_0$ is a positive free
parameter in this method. By appropriately choosing $\omega_0$, this
procedure suppresses unphysical, low-frequency components of
gravitational waves. As proposed in Ref.~\cite{reisswigpollney2010}, we
choose $\omega_0$ to be $\sim 0.8 m \Omega_0$ for $m \neq 0$ mode
gravitational waves, where $m$ is the azimuthal quantum number. For the
$m=0$ mode gravitational waves, we adopt $\omega_0 \sim 0.8 \Omega_0$
and confirm that our results depend only very weakly on this choice. We
also adopt this method to calculate the energy $\Delta E$ and angular
momentum $\Delta J$ radiated by gravitational waves. Exceptionally, we
adopt the previous method of direct time integration to estimate the
orbital eccentricity in the inspiral phase because the fixed-frequency
integration method may change the modulation in the gravitational
waveform.

For comparisons between numerically calculated gravitational waveforms
and those calculated in the PN approximations, we use the Taylor-T4
formula for two-point masses in circular orbits \cite{bbkmpsct2007} with
an additional contribution from the BH spin angular momentum
\cite{santamariaetal2010}. In this formula, the time evolution of the
orbital angular velocity $\Omega (t)$ and orbital phase $\Theta (t)$ are
computed using a nondimensional angular velocity $X (t) \equiv [ m_0
\Omega (t) ]^{2/3}$ by
 \begin{widetext}
  \begin{eqnarray}
   \frac{dX}{dt} &=& \frac{64 \nu X^5}{5 m_0} \biggl[ 1 - \left(
							   \frac{743}{336}
							   +
							   \frac{11}{4}
							   \nu \right) X
   + \left( 4 \pi - \frac{113}{12} \chi + \frac{19}{6} \nu a \right)
   X^{3/2} \nonumber \\
   && + \left( \frac{34103}{18144} + 5 \chi^2 + \frac{13661}{2016} \nu +
	 \frac{59}{18} \nu^2 \right) X^2 \nonumber \\
   && - \left\{ \left( \frac{4159}{672} + \frac{189}{8} \nu \right) \pi
	 + \left( \frac{31571}{1008} - \frac{1165}{24} \nu \right) \chi
	 + \frac{3}{4} \chi^3 - \left( \frac{21863}{1008} \nu -
				 \frac{79}{6} \nu^2 \right) a \right\}
   X^{5/2} \nonumber \\
   && + \biggl\{ \frac{16447322263}{139708800} - \frac{1712}{105}
    \gamma_E + \frac{16}{3} \pi^2 - \left( \frac{56198689}{217728} -
				     \frac{451}{48} \pi^2 \right) \nu +
    \frac{541}{896} \nu^2 - \frac{5605}{2592} \nu^3 \nonumber \\
   && - \frac{856}{105} \ln (16X) - \frac{80 \pi}{3} \chi + \left(
							     \frac{64153}{1008}
							     -
							     \frac{457}{36}
							     \nu \right)
   \chi^2 + \left( \frac{20}{3} \pi - \frac{1135}{36} \chi \right) \nu a
   \biggr\} X^3 \nonumber \\
   && - \biggl\{ \left( \frac{4415}{4032} - \frac{358675}{6048} \nu -
		  \frac{91495}{1512} \nu^2 \right) \pi + \left(
		  \frac{2529407}{27216} - \frac{845827}{6048} \nu +
		  \frac{41551}{864} \nu^2 \right) \chi - 12 \pi \chi^2
		  \nonumber \\
   && + \left( \frac{1505}{24} + \frac{\nu}{8} \right) \chi^3 - \left(
	 \frac{1580239}{54432} - \frac{451597}{6048} \nu^2 +
	 \frac{2045}{432} \nu^3 + \frac{107}{6} \nu \chi^2 \right) a
   \biggr\} X^{7/2} \biggl] ,
  \end{eqnarray}
 \end{widetext}
 \begin{equation}
 \frac{d \Theta}{dt} = \frac{X^{3/2}}{m_0} ,
 \end{equation}
where $\nu \equiv Q / (1+Q)^2$, $\chi = a Q / (1+Q)$, and $\gamma_E
\approx 0.5772$ is the Euler constant. After $X (t)$ and $\Theta (t)$
are obtained, we calculate the complex gravitational-wave amplitude
$h^{22}$ of the $(l,m) = (2,2)$ mode and the spectrum up to the 3PN
order using the formula shown in
Refs.~\cite{kidder2008,santamariaetal2010}. Here, $h^{22}$ is
\begin{widetext}
 \begin{eqnarray}
  h^{22} &=& - 8 \sqrt{\frac{\pi}{5}} \frac{\nu m_0}{D} e^{-2 i \Theta}
   X \biggl[ 1 - \left( \frac{107}{42} - \frac{55}{42} \nu \right) X +
   \left( 2 \pi - \frac{4}{3} \chi + \frac{2}{3} \nu a \right) X^{3/2}
   \nonumber \\
  && - \left( \frac{2173}{1512} + \frac{1069}{216} \nu -
	\frac{2047}{1512} \nu^2 \right) X^2 - \left\{ \left(
					       \frac{107}{21} -
					       \frac{34}{21} \nu \right)
	\pi + 24 i \nu \right\} X^{5/2} \nonumber \\
  && + \biggl\{ \frac{27027409}{646800} - \frac{856}{105} \gamma_E +
   \frac{2}{3} \pi^2 - \frac{428}{105} \ln (16X) - \left(
						    \frac{278185}{33264}
						    - \frac{41}{96}
						    \pi^2 \right) \nu -
   \frac{20261}{2772} \nu^2 \nonumber \\ &&
   + \frac{114635}{99792} \nu^3 + \frac{428}{105} i \pi \biggr\} X^3 \biggr] ,
 \end{eqnarray}
\end{widetext}
where $D$ is the distance between the center of mass of the binary and
an observer. Hereafter, we simply refer to this formula as the Taylor-T4
formula irrespective of the presence of the BH spin. Another way for
deriving an approximate waveform is to employ an effective one-body
approach (see Ref.~\cite{damournagar} and references therein for
reviews). In accompanied papers \cite{lksbf}, comparisons between
numerical waveforms and those of the effective one body approach are
extensively performed.

To estimate the mass of the remnant disk, we calculate the total rest
mass outside the AH
\begin{equation}
 M_{r>r_{\rm AH}} \equiv \int_{r>r_{\rm AH}} \rho_* d^3 x,
  \label{eq:diskmass}
\end{equation}
where $r_{\rm AH} = r_{\rm AH} ( \theta , \varphi )$ is the radius of
the AH as a function of the angular coordinates. We note that we
systematically {\it underestimated} disk masses in our previous works
performed with an old version of {\tt SACRA} \cite{skyt2009,kst2010},
because we evolved hydrodynamic variables and estimated disk masses only
in the finer domains (described in Sec.~\ref{subsec:sim_grids}) of the
size $\sim 200^3 \; {\rm km}^3$. Such a domain size is insufficient for
the estimation of the disk mass if tidal disruption occurs at a distant
orbit, especially for the case in which the NS radius is large ($\sim
15$~km). In this study, we correct the treatment of hydrodynamics and
the estimation of disk masses: We follow the hydrodynamics for a wide
computational domain of the size $1000^3$--$2000^3 \; {\rm km}^3$. We
still possibly underestimated disk masses because some of the material
escapes from our computational domains and we cannot follow their return
which would occur if they are bounded.

We determine key quantities of the remnant BH, i.e., the mass $M_{\rm
BH,f}$ and nondimensional spin parameter $a_{\rm f}$, from the
circumferential radius of the AH, assuming that the deviation from the
Kerr spacetime is negligible in the vicinity of a BH horizon. We
estimate the remnant BH mass, $M_{\rm BH,f}$, from the circumferential
radius of the AH along the equatorial plane $C_e$ divided by $4\pi$,
i.e., $C_e / 4\pi$, which gives the BH mass in the stationary vacuum BH
spacetime. Similarly, the nondimensional spin parameter of the remnant
BH, $a_{\rm f}$, is estimated from the ratio of the circumferential
radius of the AH along the meridional plane $C_p$ to $C_e$ using the
relation
\begin{equation}
 \frac{C_p}{C_e} = \frac{\sqrt{2 \hat{r}_+}}{\pi} E \left( \frac{a_{\rm
						     f}^2}{2 \hat{r}_+}
						    \right) .
\end{equation}
This also holds for the stationary vacuum BH with the nondimensional
spin parameter $a_{\rm f}$. Here, $\hat{r}_+ = 1 + \sqrt{1 - a_{\rm
f}^2}$ is a normalized radius of the horizon, and $E (z)$ is an elliptic
integral
\begin{equation}
 E (z) = \int_{_0}^{\pi /2} \sqrt{1 - z \sin^2 \theta} d\theta .
\end{equation}
For comparison, the nondimensional spin parameter of the remnant BH is
also estimated from $C_e$ and the irreducible mass of the remnant BH
$M_{\rm irr,f}$ using the relation
\begin{equation}
 M_{\rm irr,f} = \frac{C_e}{4 \sqrt{2} \pi} \sqrt{1 + \sqrt{1-{a_{\rm
  f}}^2}} ,
\end{equation}
which holds for the stationary vacuum BH. The spin parameter obtained
using this relation is referred to as $a_{\rm f2}$ according to
Ref.~\cite{skyt2009}. Finally, we also estimate $a_{\rm f}$ from the
values of the remnant BH computed using approximate conservation laws
\begin{eqnarray}
 M_{\rm BH,c} & \equiv & M_0 - M_{r>r_{\rm AH}} - \Delta E , \\
 J_{\rm BH,c} & \equiv & J_0 - J_{r>r_{\rm AH}} - \Delta J ,
\end{eqnarray}
where the total angular momentum of the material located outside the AH,
$J_{r>r_{\rm AH}}$, is approximately defined by
\begin{equation}
 J_{r>r_{\rm AH}} \equiv \int_{r>r_{\rm AH}} \rho_* h u_\varphi d^3 x
  . \label{eq:jdisk}
\end{equation}
Here, we assume that the orbital angular momentum of the BH is
negligible. The nondimensional spin parameter of the remnant BH is
defined by $a_{\rm f1} \equiv J_{\rm BH,c} / M_{\rm BH,c}^2$, again
according to Ref.~\cite{skyt2009}.

\subsection{Setup of AMR grids} \label{subsec:sim_grids}

In {\tt SACRA}, an AMR algorithm is implemented so that both the radii
of compact objects in the near zone and the characteristic gravitational
wavelengths in the wave zone can be covered with sufficient grid
resolutions simultaneously. Our AMR grids consist of a number of
computational domains, each of which has the uniform, vertex-centered
Cartesian grids with $(2N+1, 2N+1, N+1)$ grid points for $(x,y,z)$ with
the equatorial plane symmetry at $z=0$. We always choose $N=50$ for the
best resolved runs in this work. We also perform simulations with $N=36$
and 42 for several arbitrary chosen models to check the convergence of
the results and find approximately the same level of convergence as that
found in the previous work (see the Appendix of Ref.~\cite{kst2010}). In
the Appendix \ref{app_conv} of this paper, we show the convergence of
gravitational waveforms and the masses of the remnant disks. The AMR
grids are classified into two categories: one is a coarser domain, which
covers a wide region, including both the BH and NS, with its origin
fixed at the approximate center of mass throughout the simulation. The
other is a finer domain, two sets of which comove with compact objects
and cover the region in the vicinity of these objects. We denote the
edge length of the largest domain, the number of the coarser domains,
and the number of the finer domains by $2L$, $l_c$, and $2 l_f$,
respectively. Namely, the total number of the domains is $l_c + 2
l_f$. The grid spacing for each domain is $h_l = L / (2^l N)$, where $l
= 0$--$(l_c + l_f - 1)$ is the depth of each domain.

Table \ref{table:grid} summarizes the parameters of the grid structure
for our simulations. The structure of the AMR grids depends primarily on
the mass ratio of the binary because the distances between two objects
and the center of mass depend strongly on the mass ratio for our initial
models. Specifically, we choose $(l_c , l_f) = (4,4)$ for all binaries
with $M_{\rm NS} = 1.35 M_\odot$ and $Q=2,3$, and 4. We choose $(l_c ,
l_f) = (3,5)$ for binaries with $Q=5$. For binaries with $M_{\rm NS}
\neq 1.35 M_\odot$, we choose $(l_c,l_f) = (3,4)$ because we do not
evaluate disk masses for them. In all the simulations, $L$ is chosen to
be larger than or comparable to the gravitational wavelengths at an
initial instant $\lambda_0 \equiv \pi / \Omega_0$. One of the two finest
regions covers the semimajor axis of the NS by $\sim 42$--45 grid
points. The other covers the coordinate radius of the AH typically by
$\sim 20$ grid points, depending on the mass ratio and the BH spin. For
the $Q=5$ runs, the total memory required is about 11 G bytes. We
perform numerical simulations with personal computers of 12 G bytes
memory and of core-i7X processors with clock speeds of 3.2 or 3.33
GHz. We use 2--6 processors to perform one job with an OPEN-MP
library. The typical computational time required to perform one
simulation (for $\sim 50$ ms in physical time of coalescence for the
$a=0.75$ case) is 4 weeks for the 6 processor case.

\begin{table*}
 \caption{Setup of the grid structure for the simulation with our AMR
 algorithm. $l_c$ and $l_f$ are the number of coarser domains and a half
 of finer domains, respectively. $\Delta x = h_l = L / (2^l N)$ ($l =
 l_c + l_f - 1$) is the grid spacing at the finest-resolution domain
 with $L$ being the location of the outer boundaries along each
 axis. $R_{\rm diam}/\Delta x$ denotes the grid number assigned inside
 the semimajor diameter of the NS. $\lambda_0$ is the gravitational
 wavelength of the initial configuration. (See Ref.~\cite{kst2010} for
 models with nonspinning BHs.)}
 \begin{tabular}{ccc|ccc} \hline
  Model & $l_c$ & $l_f$ & $\Delta x / M_0$ & $R_{\rm diam} / \Delta x$ &
  $L / \lambda_0$ \\ \hline \hline
  2H-Q2M135a75 & 4 & 4 & 0.0471 & 90.8 & 2.386 \\
  1.5H-Q2M135a75 & 4 & 4 & 0.0426 & 87.7 & 2.417 \\
  H-Q2M135a75 & 4 & 4 & 0.0377 & 86.2 & 2.138 \\
  HB-Q2M135a75 & 4 & 4 & 0.0347 & 87.1 & 1.968 \\
  B-Q2M135a75 & 4 & 4 & 0.0324 & 86.7 & 1.837 \\
  \hline
  2H-Q2M135a5 & 4 & 4 & 0.0471 & 90.8 & 2.378 \\
  1.5H-Q2M135a5 & 4 & 4 & 0.0426 & 87.7 & 2.410 \\
  H-Q2M135a5 & 4 & 4 & 0.0377 & 86.2 & 2.131 \\
  HB-Q2M135a5 & 4 & 4 & 0.0347 & 87.2 & 1.962 \\
  B-Q2M135a5 & 4 & 4 & 0.0324 & 86.7 & 1.831 \\
  \hline
  2H-Q2M135a-5 & 4 & 4 & 0.0470 & 90.7 & 2.092 \\
  H-Q2M135a-5 & 4 & 4 & 0.0376 & 86.4 & 1.902 \\
  HB-Q2M135a-5 & 4 & 4 & 0.0347 & 87.1 & 1.962 \\
  B-Q2M135a-5 & 4 & 4 & 0.0324 & 86.7 & 1.831 \\
  \hline
  2H-Q2M12a75 & 3 & 4 & 0.0583 & 84.7 & 1.476 \\
  H-Q2M12a75 & 3 & 4 & 0.0442 & 85.3 & 1.252 \\
  HB-Q2M12a75 & 3 & 4 & 0.0410 & 85.7 & 1.162 \\
  B-Q2M12a75 & 3 & 4 & 0.0389 & 84.2 & 1.102 \\
  \hline
  2H-Q2M145a75 & 3 & 4 & 0.0461 & 85.2 & 1.166 \\
  H-Q2M145a75 & 3 & 4 & 0.0347 & 85.3 & 0.985 \\
  HB-Q2M145a75 & 3 & 4 & 0.0316 & 87.1 & 0.896 \\
  B-Q2M145a75 & 3 & 4 & 0.0292 & 87.1 & 0.829 \\
  \hline
  2H-Q3M135a75 & 4 & 4 & 0.0367 & 85.5 & 2.084 \\
  1.5H-Q3M135a75 & 4 & 4 & 0.0326 & 84.0 & 1.986 \\
  H-Q3M135a75 & 4 & 4 & 0.0282 & 84.7 & 1.718 \\
  HB-Q3M135a75 & 4 & 4 & 0.0260 & 85.6 & 1.581 \\
  B-Q3M135a75 & 4 & 4 & 0.0235 & 87.9 & 1.431 \\
  \hline
  2H-Q3M135a5 & 4 & 4 & 0.0353 & 88.9 & 1.997 \\
  1.5H-Q3M135a5 & 4 & 4 & 0.0326 & 84.0 & 1.980 \\
  H-Q3M135a5 & 4 & 4 & 0.0282 & 84.7 & 1.712 \\
  HB-Q3M135a5 & 4 & 4 & 0.0260 & 85.7 & 1.576 \\
  B-Q3M135a5 & 4 & 4 & 0.0243 & 85.3 & 1.471 \\
  \hline
  HB-Q3M135a-5 & 4 & 4 & 0.0260 & 85.7 & 1.576 \\
  \hline
  2H-Q3M145a75 & 3 & 4 & 0.0328 & 87.7 & 0.933 \\
  H-Q3M145a75 & 3 & 4 & 0.0250 & 87.4 & 0.760 \\
  HB-Q3M145a75 & 3 & 4 & 0.0234 & 86.6 & 0.712 \\
  B-Q3M145a75 & 3 & 4 & 0.0214 & 87.7& 0.651 \\
  \hline
  2H-Q4M135a75 & 4 & 4 & 0.0296 & 83.4 & 1.804 \\
  H-Q4M135a75 & 4 & 4 & 0.0223 & 84.5 & 1.450 \\
  HB-Q4M135a75 & 4 & 4 & 0.0203 & 86.5 & 1.319 \\
  B-Q4M135a75 & 4 & 4 & 0.0190 & 85.8 & 1.237 \\
  \hline
  2H-Q4M135a5 & 4 & 4 & 0.0296 & 83.2 & 2.097 \\
  H-Q4M135a5 & 4 & 4 & 0.0219 & 85.9 & 1.548 \\
  HB-Q4M135a5 & 4 & 4 & 0.0205 & 85.5 & 1.448 \\
  B-Q4M135a5 & 4 & 4 & 0.0188 & 86.4 & 1.332 \\
  \hline
  2H-Q5M135a75 & 3 & 5 & 0.0235 & 86.3 & 1.718 \\
  H-Q5M135a75 & 3 & 5 & 0.0180 & 86.2 & 1.314 \\
  HB-Q5M135a75 & 3 & 5 & 0.0167 & 86.3 & 1.224 \\
  B-Q5M135a75 & 3 & 5 & 0.0159 & 84.6 & 1.159 \\
  \hline
 \end{tabular}
 \label{table:grid}
\end{table*}

\section{Numerical results} \label{sec:res}

We present numerical results of our simulations, focusing, in
particular, on their dependence on the BH spin and NS EOS. First, we
review general merger processes in
Sec.~\ref{subsec:res_merger}. Sections \ref{subsec:res_disk},
\ref{subsec:res_struc}, and \ref{subsec:res_BH} are devoted to the
analysis of properties of the remnant disk and BH formed after the
merger. Gravitational waveforms are shown in
Sec.~\ref{subsec:res_waveform}, their spectra in
Sec.~\ref{subsec:res_spectrum}, and the energy and angular momentum
radiated by gravitational waves in Sec.~\ref{subsec:res_energy}.

\subsection{Overview of the merger process} \label{subsec:res_merger}

\begin{figure*}[tbp]
 \begin{tabular}{ccc}
  \includegraphics[width=55mm,clip]{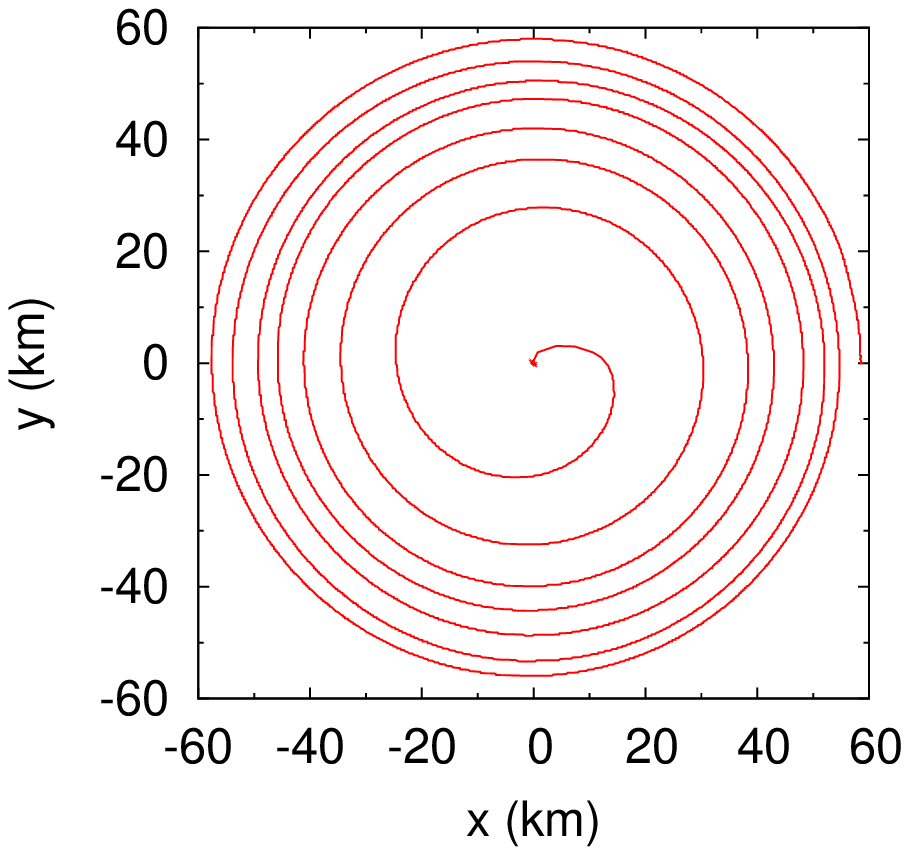} &
  \includegraphics[width=55mm,clip]{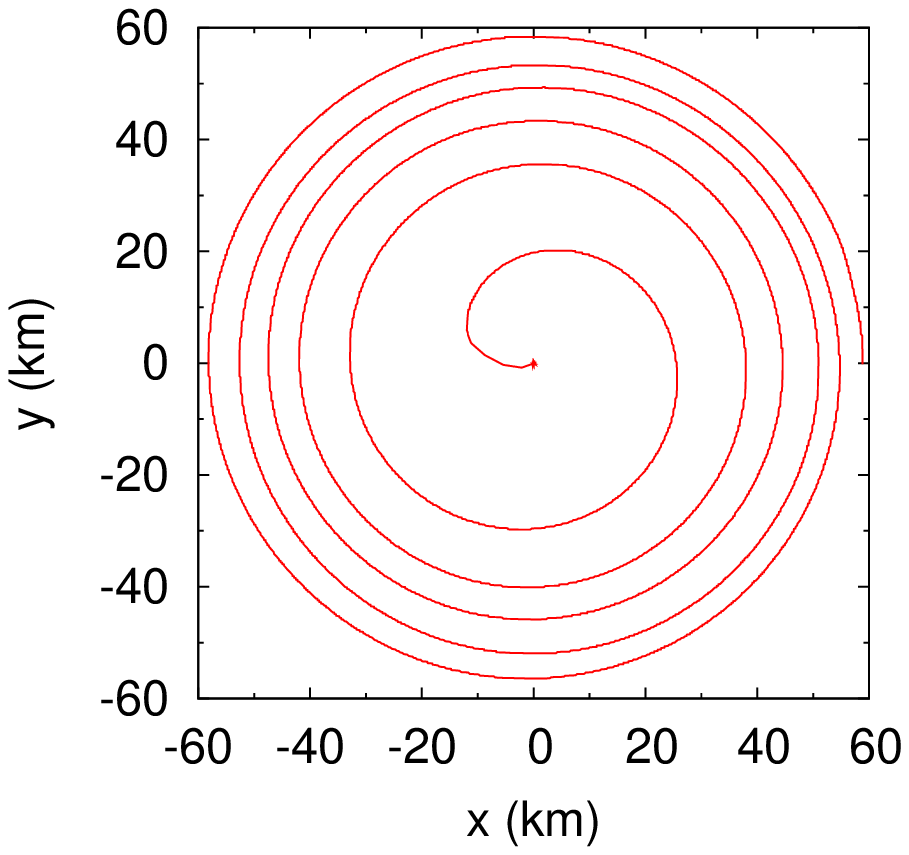} &
  \includegraphics[width=55mm,clip]{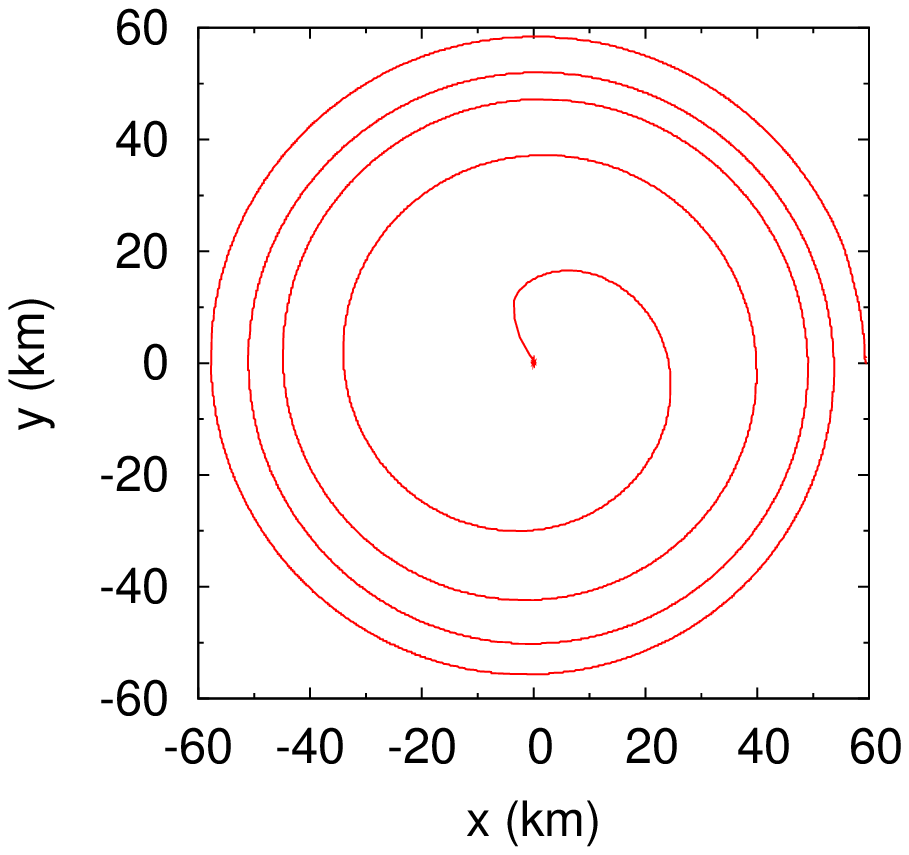} \\
 \end{tabular}
 \caption{Evolution of the orbital separation $x_{\rm sep}^i = x_{\rm
 NS}^i - x_{\rm BH}^i$ of binaries with $(Q,M_{\rm NS}) = (2,
 1.35M_\odot )$ and HB EOS. The left, middle, and right panels show the
 results with the prograde BH spin $a=0.5$, zero BH spin $a=0$, and
 retrograde BH spin $a=-0.5$, respectively.} \label{fig:orb}
\end{figure*}

Figure \ref{fig:orb} plots the evolution of the coordinate separation
defined by $x_{\rm sep}^i = x_{\rm NS}^i - x_{\rm BH}^i$ for models
HB-Q2M135a5, HB-Q2M135, and HB-Q2M135a-5, for which $\Omega_0 m_0$ takes
the same values. Here, $x_{\rm NS}^i$ is the position of the maximum
rest-mass density and $x_{\rm BH}^i$ is the location of the puncture,
$x_{\rm P}^i$. Figure \ref{fig:orb} shows that the numbers of orbits
increases as the BH spin increases from retrograde to prograde
\cite{elsb2009}. Specifically, the number of orbit are $\sim 7$, 5.5,
and 4 for $a=0.5$, 0, and $-0.5$, respectively. This difference comes
primarily from the spin-orbit interaction between these two angular
momenta \cite{kidder1995}; in the PN approximation, a force proportional
to the inner product of the orbital and spin angular momenta of two
objects appears at 1.5PN order. Here, we do not have to consider the NS
spin angular momentum in the assumption of the irrotational velocity
field and, therefore, we only consider the interaction between the
orbital and BH spin angular momenta throughout this paper. When these
two angular momenta are parallel and the inner product is positive
($a>0$), an additional repulsive force works between the BH and NS. This
repulsive force reduces the orbital angular velocity because the
centrifugal force associated with the orbital motion can be reduced, and
hence, the luminosity of gravitational radiation, which is proportional
to $\Omega^{10/3}$, is also reduced. This strong dependence of the
luminosity on $\Omega$ makes the approaching velocity smaller in the
late inspiral phase, and, therefore, the number of orbits
increases. Conversely, when these two angular momenta are antiparallel
($a<0$), an additional attractive force increases the angular velocity
and gravitational-wave luminosity in the late inspiral phase. In this
case, the orbital separation decreases faster due to a larger
approaching velocity, and the number of orbits becomes smaller as the
retrograde BH spin increases. All these results agree qualitatively with
those of Ref.~\cite{elsb2009}.

\begin{figure*}[t]
 \begin{tabular}{ccc}
  \includegraphics[width=55mm,clip]{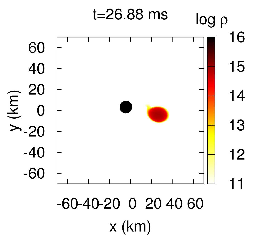} &
  \includegraphics[width=55mm,clip]{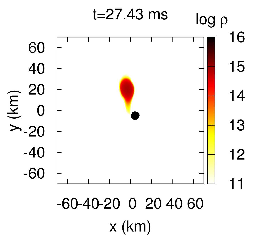} &
  \includegraphics[width=55mm,clip]{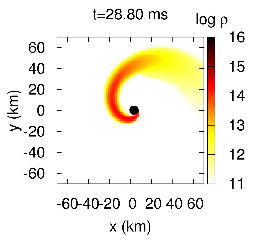} \\
  \includegraphics[width=55mm,clip]{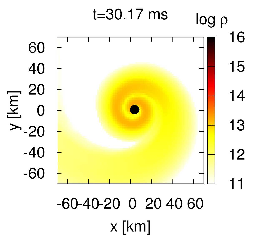} &
  \includegraphics[width=55mm,clip]{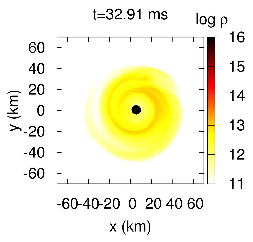} &
  \includegraphics[width=55mm,clip]{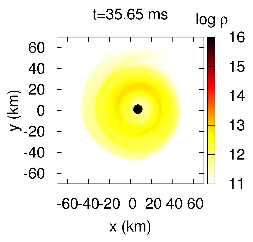} \\
 \end{tabular}
 \caption{Evolution of the rest-mass density profile in units of ${\rm
 g/cm^3}$ and the location of the AH on the equatorial plane for model
 HB-Q3M135a75. The filled circle denotes the region inside the AH. The
 color (gradational) panel on the right of each plot show $\log_{10} (
 \rho )$.}  \label{fig:snapshot1}
\end{figure*}

\begin{figure*}[t]
 \begin{tabular}{ccc}
  \includegraphics[width=55mm,clip]{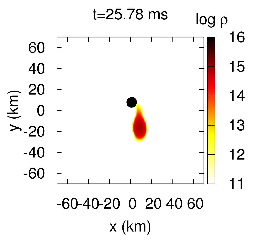} &
  \includegraphics[width=55mm,clip]{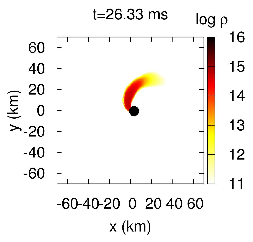} &
  \includegraphics[width=55mm,clip]{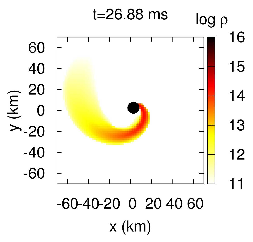} \\
  \includegraphics[width=55mm,clip]{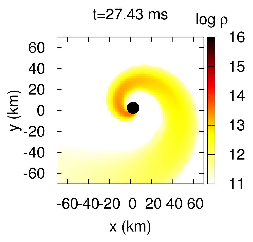} &
  \includegraphics[width=55mm,clip]{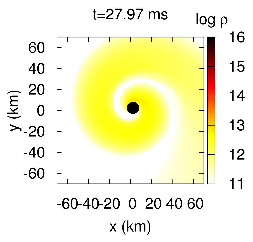} &
  \includegraphics[width=55mm,clip]{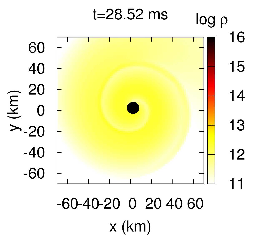} \\
 \end{tabular}
 \caption{The same as Fig.~\ref{fig:snapshot1} but for model
 HB-Q3M135a5.}  \label{fig:snapshot2}
\end{figure*}

\begin{figure*}[t]
 \begin{tabular}{ccc}
  \includegraphics[width=55mm,clip]{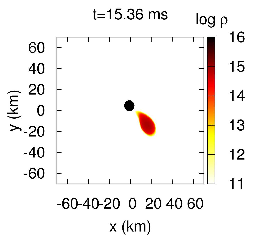} &
  \includegraphics[width=55mm,clip]{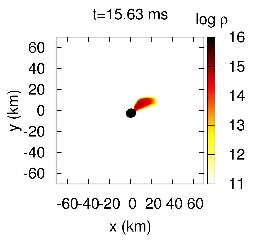} &
  \includegraphics[width=55mm,clip]{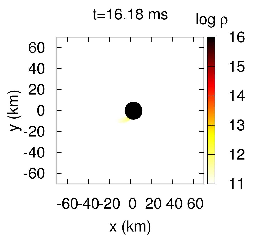} \\
 \end{tabular}
 \caption{The same as Fig.~\ref{fig:snapshot1} but for model
 HB-Q3M135a-5.}  \label{fig:snapshot3}
\end{figure*}

The fate of BH-NS binaries is classified into two categories. One is the
case in which the NS is disrupted by the BH tidal field before the BH
swallows the NS, and the other is the case in which the BH swallows the
NS without tidal disruption. In this paper, we focus mainly on the
former case. We plot snapshots of the rest-mass density profiles and the
location of the AH on the equatorial plane at selected time slices for
models HB-Q3M135a75, HB-Q3M135a5, and HB-Q3M135a-5 in
Figs.~\ref{fig:snapshot1}--\ref{fig:snapshot3}, respectively. The NS is
disrupted outside the ISCO in the $a>0$ cases (Figs.~\ref{fig:snapshot1}
and \ref{fig:snapshot2}) and forms a one-armed spiral structure with a
large angular momentum. The material in the inner part of the spiral arm
gradually falls onto the BH due to angular momentum transport via
hydrodynamic torque in the spiral arm. The material with a sufficiently
large specific angular momentum escapes the capture by the BH and forms
an accretion disk, which survives for a time much longer than the
dynamical time scale $\sim$ a few ms. We note that the prompt infall of
the one-armed spiral structure onto the BH occurs from a relatively
narrower region for $a=0.5$ than for $a=0.75$. The reason is that the
inner edge of the spiral arm contacts the AH well before the arm becomes
nearly axisymmetric due to a large radius of the AH and ISCO for
$a=0.5$. The infall of the disrupted material from a narrow region of
the BH frequently occurs when the NS is tidally disrupted in a binary
with a {\em high mass ratio}, whereas this is rare in a binary with a
nonspinning BH because the NS is not disrupted in a high mass-ratio
binary. This difference in the merger process is well-reflected in
gravitational waveforms (see Sec.~\ref{subsec:res_waveform}). By
contrast, the NS is swallowed by the BH without tidal disruption, and
essentially no material is left outside the ISCO for model HB-Q3M135a-5
(Fig.~\ref{fig:snapshot3}).

Note that the feature of the NS tidal disruption appears very weakly not
only for model HB-Q3M135a-5 but also for model HB-Q3M135 ($a=0$) because
the mass ratio $Q=3$ is so high that the tidal effect is less important
for the nonspinning BH with the typical NS radius $\sim 11$--12 km. The
enhancement of the tidal effect by a prograde BH spin results primarily
from the decrease of the BH ISCO radius \cite{bpt1972}. In the
Boyer-Lindquist coordinates, a Kerr BH has an ISCO with a smaller radius
than a Schwarzschild BH by a factor of $\ge 1/6$, depending on $a$ for a
prograde orbit: The ISCO radius approximately halves when the BH spin
increases from $a=0$ to 0.75. On the other hand, the orbital separation
at the onset of mass shedding depends only weakly on the BH spin in the
Boyer-Lindquist coordinates \cite{fishbone1973,marck1983,ism2005}. This
decrease of the ISCO radius enhances the possibility for the disrupted
material to escape capture by the BH and to form a more massive remnant
disk than in the nonspinning BH case. The retrograde BH spin plays an
opposite role; the ISCO radius of the Kerr BH increases by a factor of
1--1.5 for a retrograde orbit, and hence, the tidal effect is less
important in the merger process.

Before closing this subsection, we estimate the degree of (undesired)
orbital eccentricity in our simulations to assess the circularity of the
orbital motion. For this purpose, we compute the evolution of the
gauge-invariant orbital angular velocity $\Omega (t)$, which is defined
from the $(l,m) = (2,2)$ mode of $\Psi_4$ by
\begin{equation}
 \Omega (t) = \frac{1}{2} \frac{| \Psi_4 (l=m=2) |}{| \int \Psi_4
  (l=m=2) dt |} . \label{eq:omega}
\end{equation}
The evolution of the orbital angular velocity in our simulation agrees
with that derived from the Taylor-T4 formula in the inspiral phase
within a small modulation, typically $\Delta \Omega / \Omega \lesssim
5\%$, which is equivalent to the orbital eccentricity of $\lesssim
3\%$. This amount of orbital eccentricity is as small as that observed
in the nonspinning BH case with a low mass ratio $Q=2$ \cite{kst2010}.

\subsection{Global properties of the disk} \label{subsec:res_disk}

The mass of the remnant disk reflects the significance of NS tidal
disruption in a clear way because the disk formation is a result of
tidal disruption. A massive disk is formed if tidal disruption of the NS
occurs far outside the ISCO. If the mass shedding starts in the vicinity
of or inside the ISCO, only a small portion of the mass is left outside
the AH. The material is not left outside the AH when the mass shedding
does not occur before the BH swallows the NS, and the merger of a BH-NS
binary may be indistinguishable from that of a BH-BH binary except for
very small tidal corrections to the inspiral. Thus, the mass of a
remnant disk is a reliable indicator of the degree of tidal disruption.

\begin{figure*}[tbp]
 \begin{tabular}{cc}
  \includegraphics[width=80mm,clip]{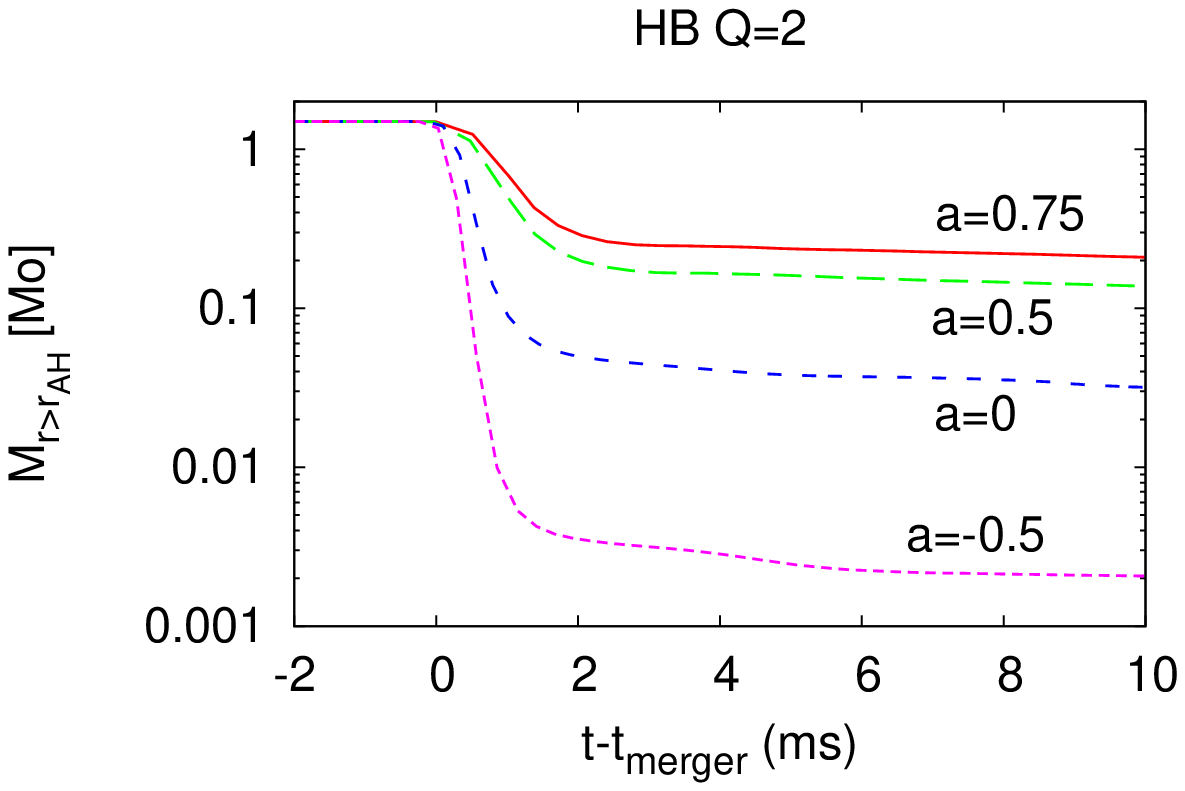} &
  \includegraphics[width=80mm,clip]{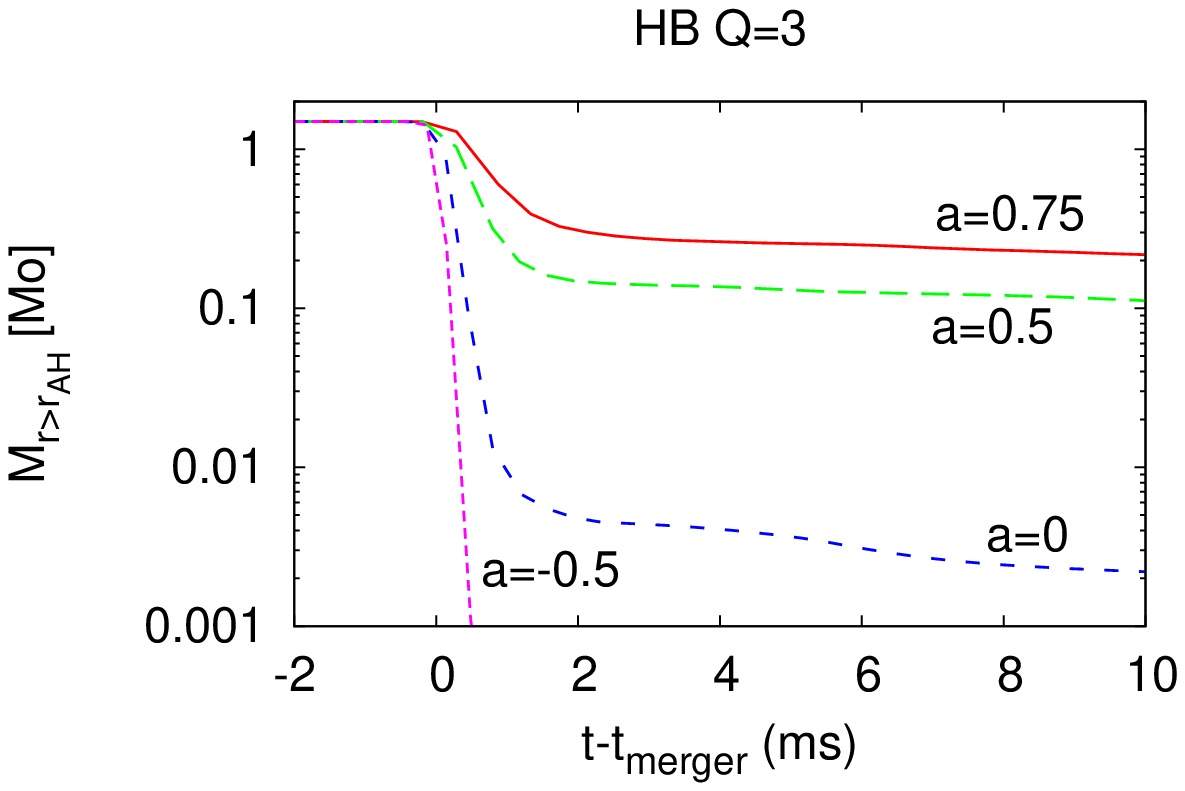}
 \end{tabular} 
 \caption{Evolution of the rest mass of the material located outside an
 AH of the BH, $M_{r>r_{\rm AH}}$. The left and right panels show the
 results for $Q=2$ and 3, respectively. In both plots, $M_{\rm NS} =
 1.35 M_\odot$ and HB EOS are adopted. The results revised from the
 previous paper~\cite{kst2010} are plotted for $a=0$.}
 \label{fig:disk1}
\end{figure*}

Figure \ref{fig:disk1} plots the time evolution of the rest mass located
outside the AH, $M_{r>r_{\rm AH}}$, for $Q=2$ and 3 with different
nondimensional BH spin parameters $a=0.75$, 0.5, 0, and $-0.5$. In both
plots, $M_{\rm NS} = 1.35 M_\odot$ and HB EOS are adopted. We note that
the results revised from the previous work~\cite{kst2010} are plotted
for $a=0$. The dependence of $M_{r>r_{\rm AH}}$ on $a$ for HB EOS found
here is similar to those for other EOSs. We set the time origin to be an
approximate merger time $t_{\rm merger}$. These plots indicate that the
mass of the material left outside the AH relaxes to a quasisteady value
for $t-t_{\rm merger} \agt 3$--4~ms, and the relaxed value increases
monotonically as the BH spin increases from retrograde to prograde. This
is consistent with the decrease of the BH ISCO radius with the increase
of its spin, as described in Sec.~\ref{subsec:res_merger}. In
particular, the remnant disk mass at $\approx 10$ ms after the merger is
$\gtrsim 0.1 M_\odot$ for all the EOSs with $(Q,a) = (2, \ge 0.5)$ and
$(\le 4,0.75)$, as shown in Table \ref{table:remnant}, and $\gtrsim 0.05
M_\odot$ for $(Q,a) = (3,0.5)$, irrespective of the EOS. The formation
of such a massive disk may be encouraging for the BH-NS binary merger
hypothesis of a short-hard GRB. For the $a=-0.5$ cases, by contrast,
massive accretion disks of $\gtrsim 0.01 M_\odot$ are not expected to be
formed as merger remnants even for $Q=2$ unless the EOS is extremely
stiff (the NS radius is $\approx$ 15 km). This fact indicates that the
retrograde BH spin is unfavorable for producing a central engine of a
short-hard GRB.

\begin{figure*}[tbp]
 \begin{tabular}{cc}
  \includegraphics[width=80mm,clip]{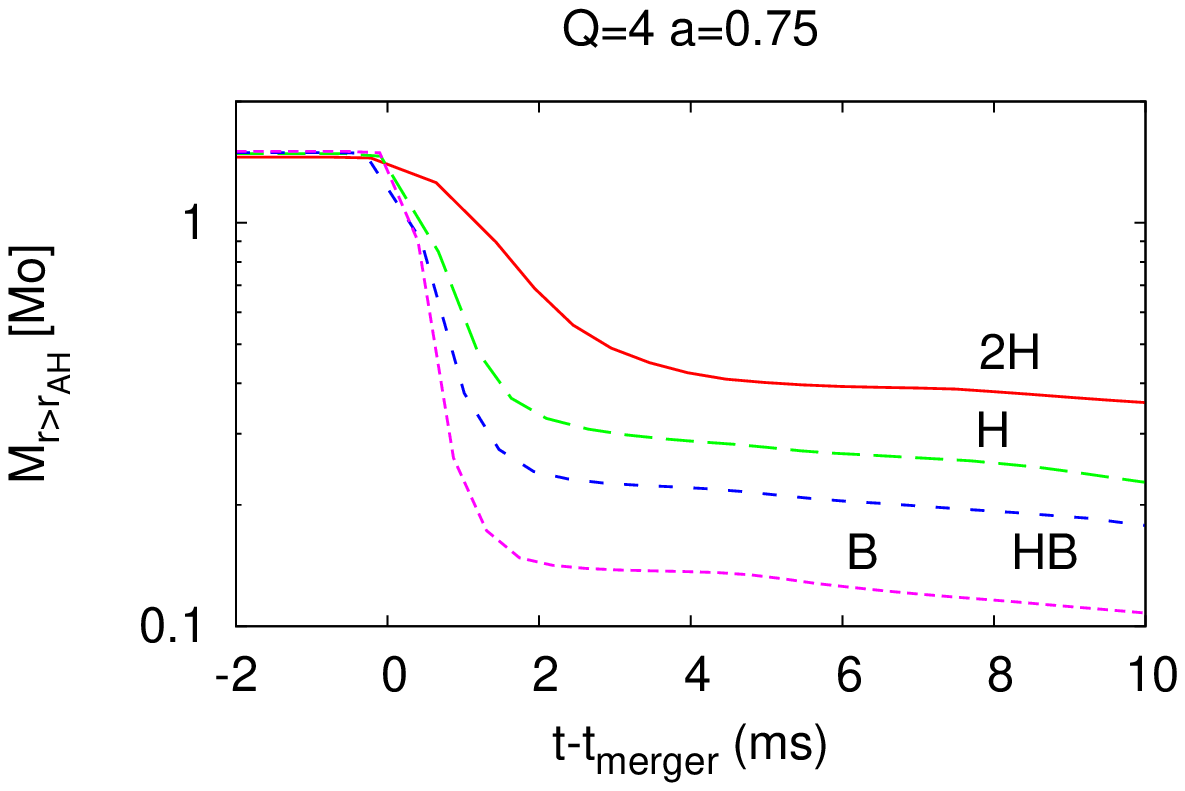} &
  \includegraphics[width=80mm,clip]{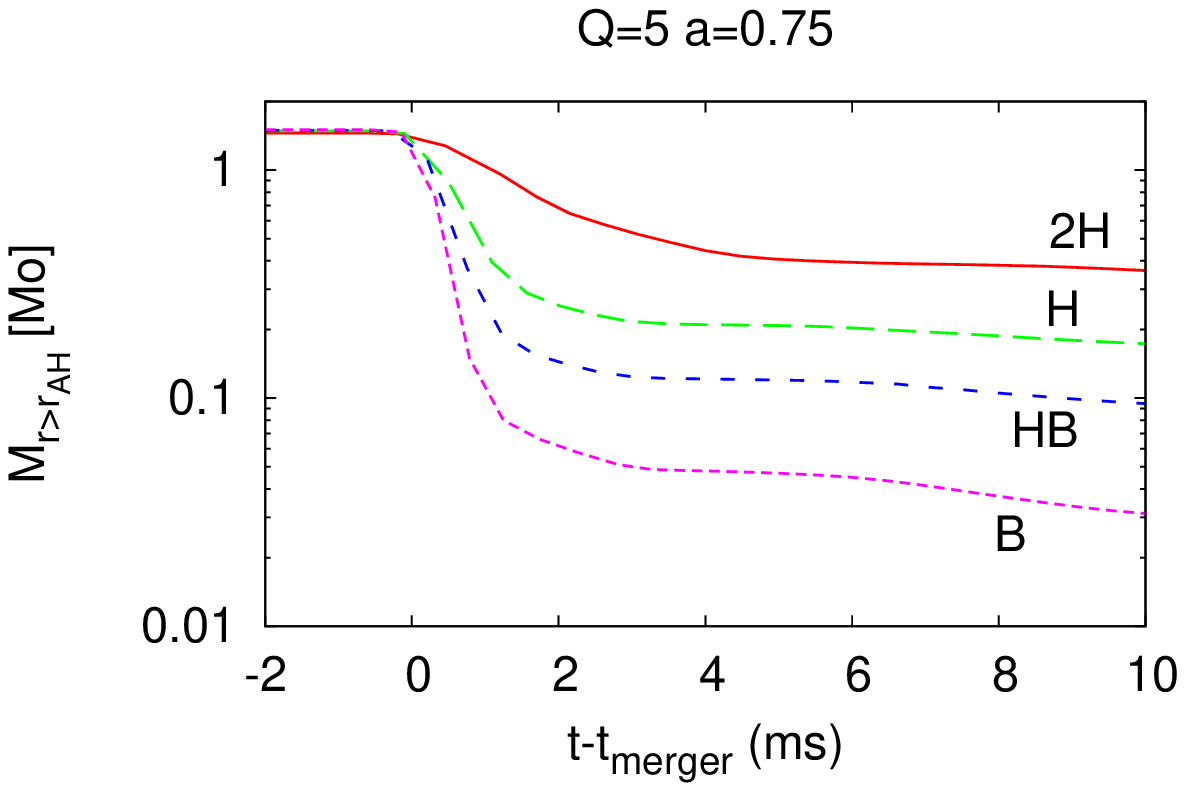}
 \end{tabular} 
 \caption{The same as Fig.~\ref{fig:disk1} but for different models $Q =
 4$ (left) and 5 (right). In both plots, $M_{\rm NS} = 1.35 M_\odot$ and
 $a = 0.75$ are adopted.} \label{fig:disk2}
\end{figure*}

The prograde BH spin enhances the disk formation dramatically for a
BH-NS binary with {\it a high mass ratio}, for which the disk mass is
very low when the BH is nonspinning. We plot the time evolution of
$M_{r>r_{\rm AH}}$ for $Q=4$ and 5 with different EOSs in
Fig.~\ref{fig:disk2}. In both plots, $M_{\rm NS} = 1.35 M_\odot$ and
$a=0.75$ are adopted. Figure \ref{fig:disk2} clearly shows that a
massive accretion disk is formed for $Q=4$ and 5 if the BH has a
prograde spin of $a=0.75$. Namely, the formation of a massive accretion
disk is universal for the merger of a BH-NS binary with a mass ratio of
$Q \lesssim 5$ as far as $a \sim 0.75$ and $M_{\rm NS} = 1.35 M_\odot$
(equivalently, $M_{\rm BH} \lesssim 6.75 M_\odot$). Note that a heavy BH
of $M_{\rm BH} \gtrsim 5 M_\odot$ is predicted to be realistic as an
astrophysical consequence of the stellar evolution with solar
metallicity \cite{mcclintockremillard} (see, e.g.,
Ref.~\cite{bbfrvvh2010} for a population synthesis study) and hence as a
possible progenitor of the short-hard GRB.

\begin{figure*}[tbp]
 \begin{tabular}{cc}
  \includegraphics[width=90mm,clip]{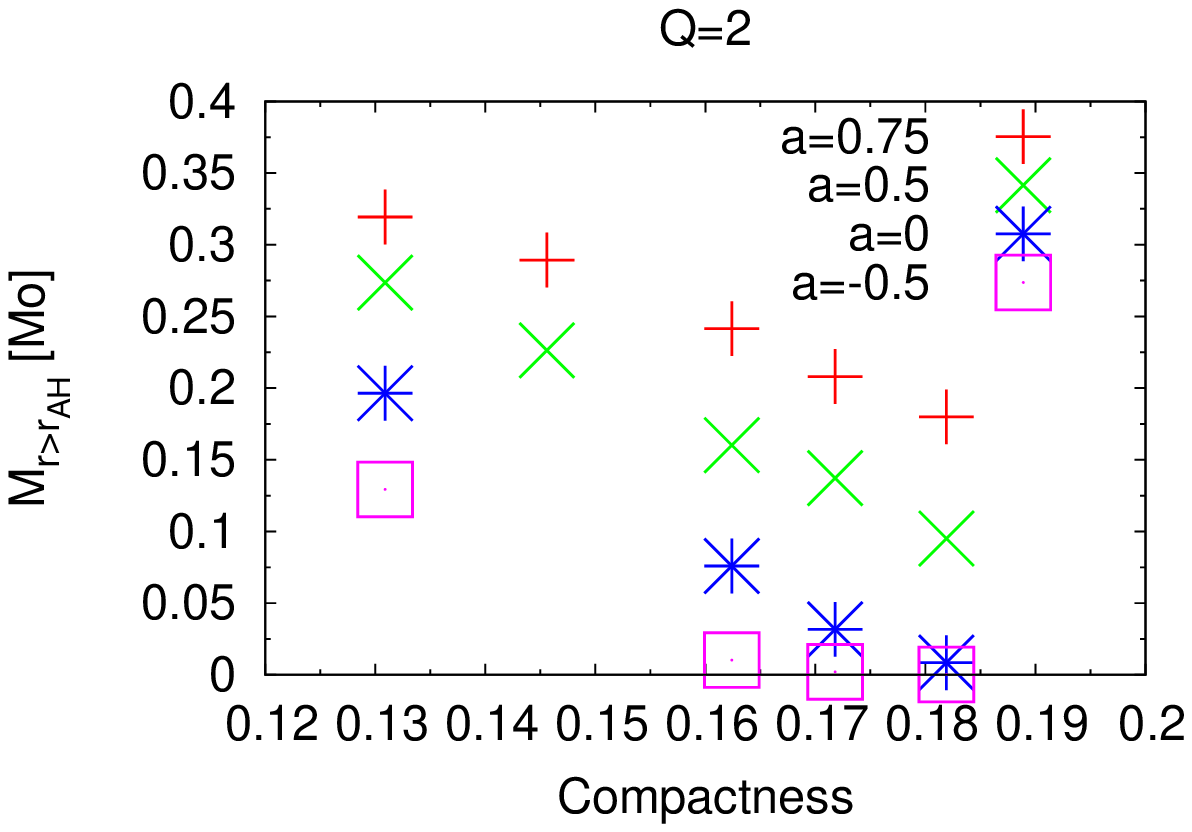} &
  \includegraphics[width=90mm,clip]{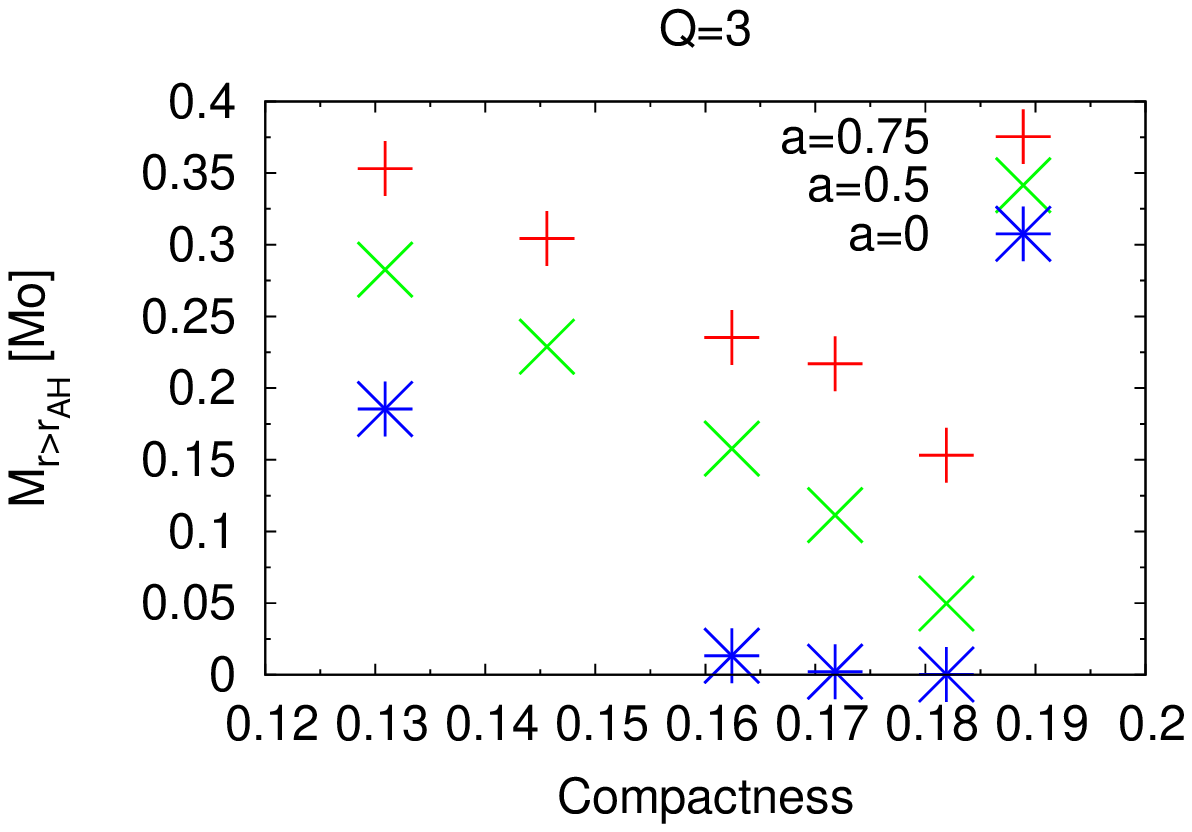}
 \end{tabular}
 \caption{The disk mass $M_{r>r_{\rm AH}}$ at $\approx 10$ ms after the
 onset of the merger as a function of the NS compactness ${\cal C}$. The
 left and right panels show the results for $Q=2$ (left) and $Q=3$
 (right). The results revised from the previous work~\cite{kst2010} are
 plotted for a=0.}  \label{fig:ctom}
\end{figure*}

For more quantitative discussion, we plot the disk mass estimated at
$\approx 10$ ms after the merger for all the models with $Q=2$ and for
models with $(Q,a) = (3, \ge 0)$ as a function of the NS compactness,
${\cal C}$, in Fig.~\ref{fig:ctom}. Numerical values of $M_{r>r_{\rm
AH}}$ are shown in Table \ref{table:remnant}, as well as other
quantities associated with the merger remnants. For any fixed value of
$a$, a negative correlation between $M_{r>r_{\rm AH}}$ and ${\cal C}$ is
found to hold in Fig.~\ref{fig:ctom}. This correlation indicates that
the NS with a larger compactness is less subject to tidal deformation
and disruption than the NS with a smaller compactness for any fixed
value of $a$.  This correlation is expected from the nature of a tidal
force as a finite-size effect, as found in the study of nonspinning
BH-NS binaries~\cite{kst2010}. On the other hand, Fig.~\ref{fig:ctom}
again shows that the prograde BH spin increases the disk mass for any
fixed value of ${\cal C}$. A remarkable fact is that the disk mass does
not decrease steeply to a value of $\ll 0.1 M_\odot$ as the compactness
increases for binaries with $(Q,a)=( \le 3, \ge 0.5)$. We expect that
the coalescence of a BH-NS binary with $(Q,a) = ( \le 3, \ge 0.5)$ may
always produce a remnant disk of $\gtrsim 0.01 M_\odot$ within a
plausible range of the NS compactness, ${\cal C} \lesssim 0.2$, although
it is possible only if ${\cal C} \lesssim 0.18$ for $(Q,a)=(2,0)$ and
${\cal C} \lesssim 0.16$ for $(Q,a)=(2,-0.5)$ or $(3,0)$.

\begin{figure*}[tbp]
 \begin{tabular}{cc}
  \includegraphics[width=90mm,clip]{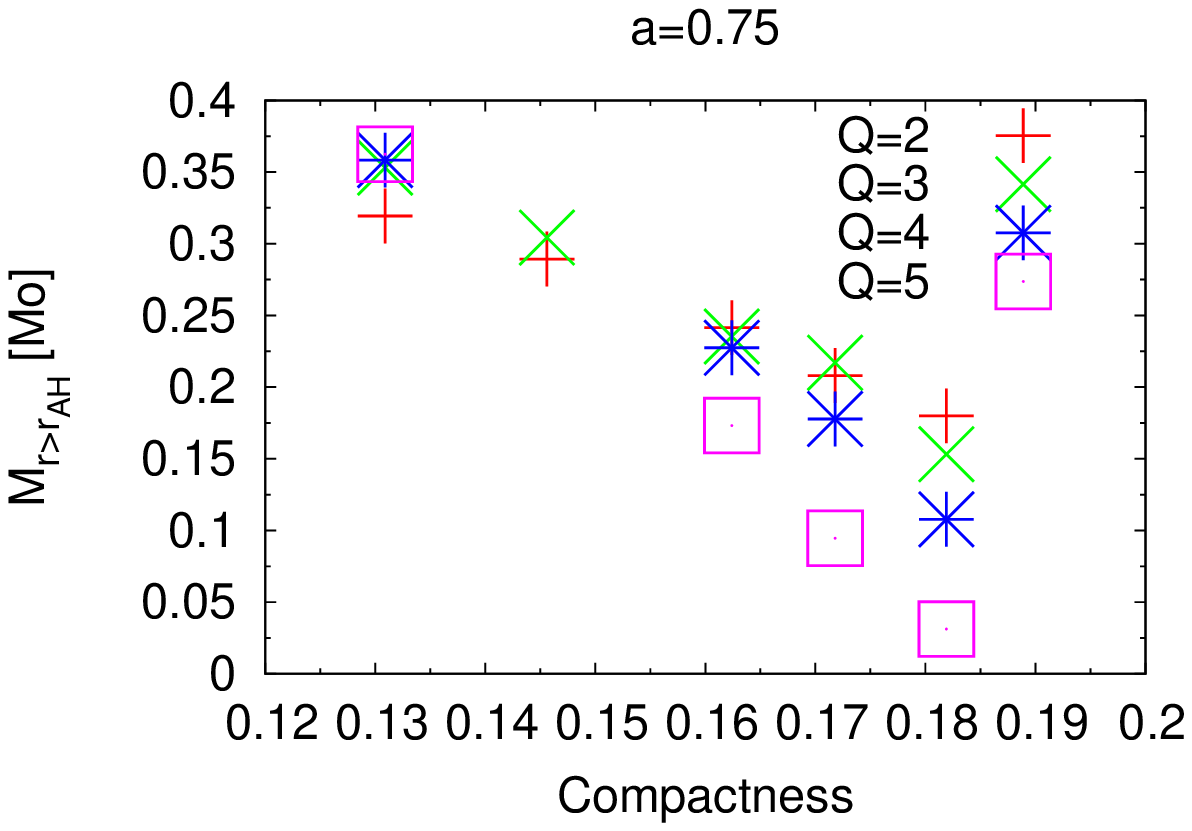} &
  \includegraphics[width=90mm,clip]{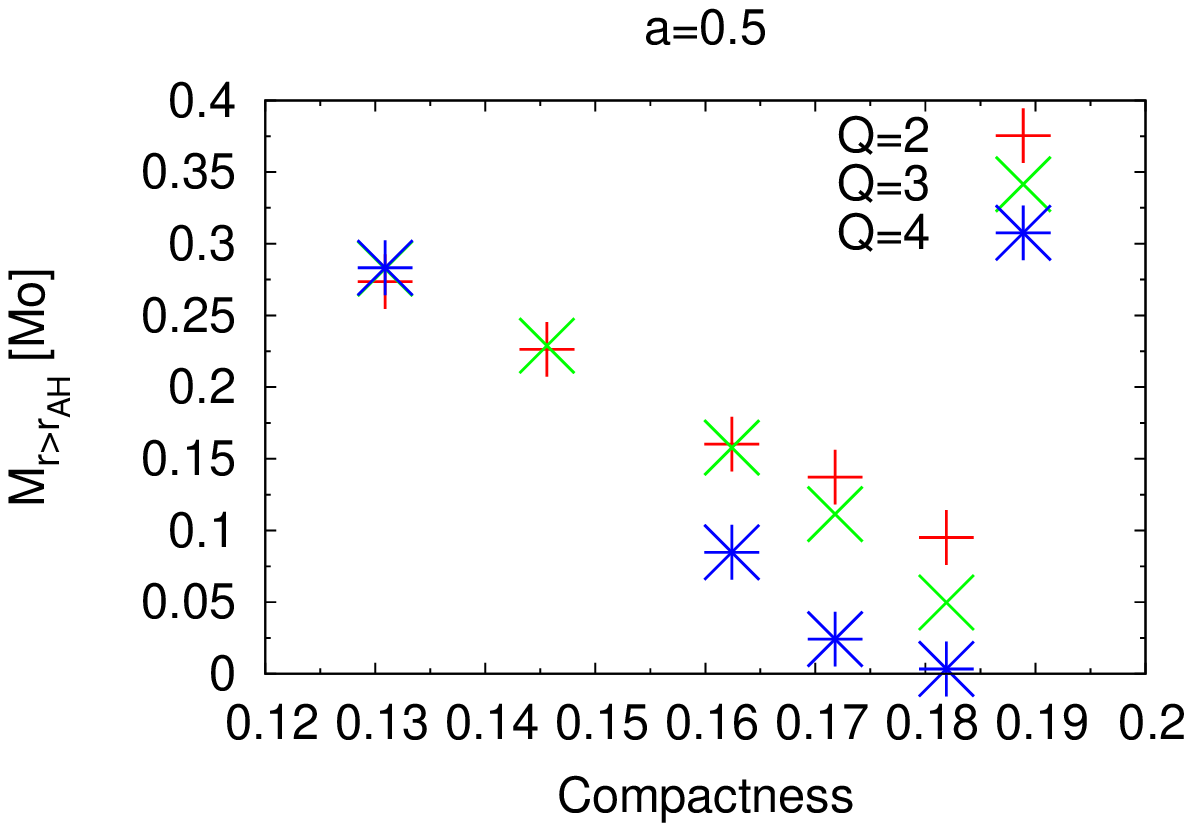} 
 \end{tabular}
 \caption{The same as Fig.~\ref{fig:ctom} but for $a=0.75$ and $a=0.5$
 with $Q=2$--5.} \label{fig:ctom2}
\end{figure*}

The dependence of the disk mass on the NS compactness is different for
different values of the mass ratio. We plot in Fig.~\ref{fig:ctom2} the
disk mass as a function of the NS compactness as in Fig.~\ref{fig:ctom},
but for $a=0.75$ and 0.5. This figure shows that the disk mass depends
more strongly on ${\cal C}$ when the mass ratio, $Q$, is larger. The
disk mass is larger for smaller values of $Q$ when the EOS is soft and
${\cal C} \gtrsim 0.16$, except for HB-Q2M135a75 and HB-Q3M135a75, for
which the disk masses depend only weakly on $Q$. This dependence on $Q$
is expected from the comparison between the mass-shedding radius,
$r_{\rm shed}$, and the ISCO radius, $r_{\rm ISCO}$,
\begin{equation}
 \frac{r_{\rm shed}}{r_{\rm ISCO}} \propto {\cal C}^{-1} Q^{-2/3} ,
\end{equation}
where we assume Newtonian gravity for simplicity. This relation states
that a larger amount of mass can escape the capture by the BH and can
form an accretion disk when $Q$ is small because the mass shedding sets
in at relatively more distant orbit. However, the disk mass may be
larger for larger values of $Q$ when the EOS is stiff as ${\cal C}
\lesssim 0.15$ for $a \ge 0.5$ and $2 \lesssim Q \lesssim 5$. This
should be ascribed to the redistribution process of the specific angular
momentum of the NS to the disrupted material and to subsequent behavior
of the material (such as collision of the fluid elements in spiral
arms). This feature suggests that a binary with a larger value of $Q$,
say $Q \gtrsim 6$, possibly form a massive remnant disk of $\gtrsim 0.1
M_\odot$ and could be a progenitor of a short-hard GRB if the EOS is
stiff and the BH has a large spin $\gtrsim 0.5$.

\begin{figure*}[tbp]
 \begin{tabular}{cc}
  \includegraphics[width=80mm,clip]{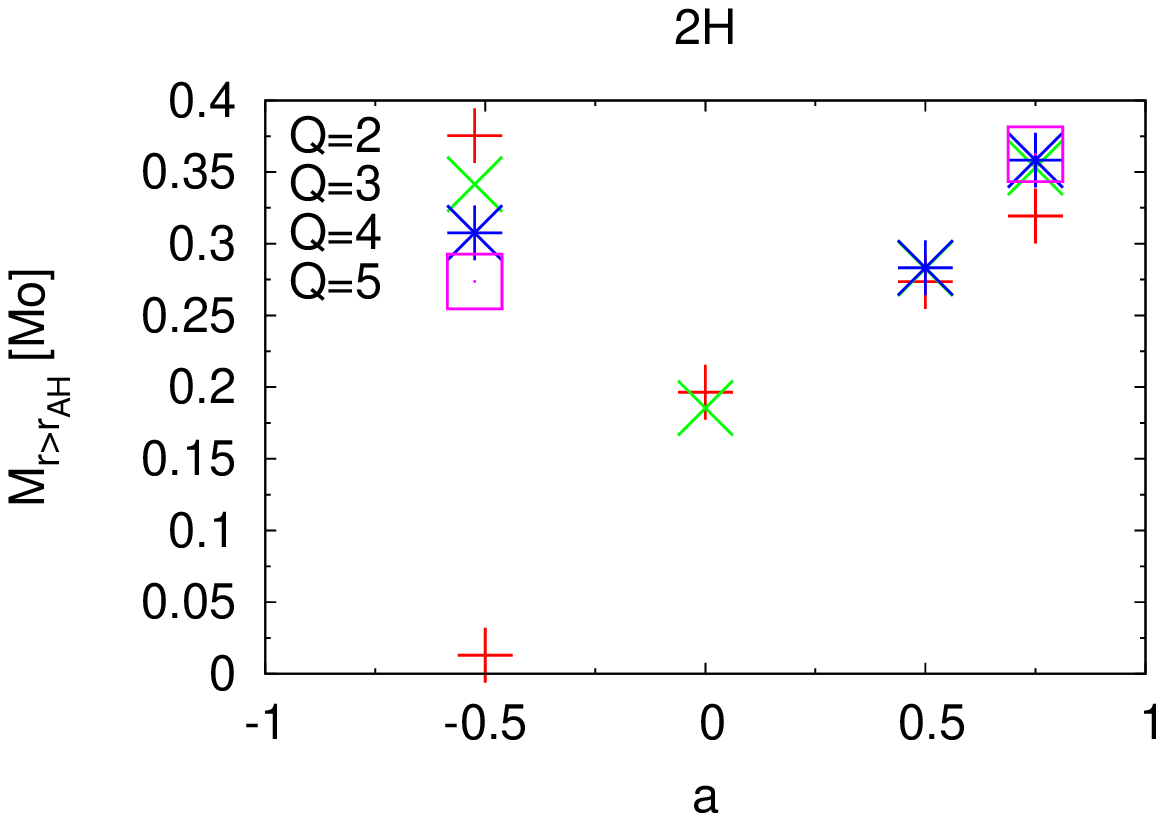} &
  \includegraphics[width=80mm,clip]{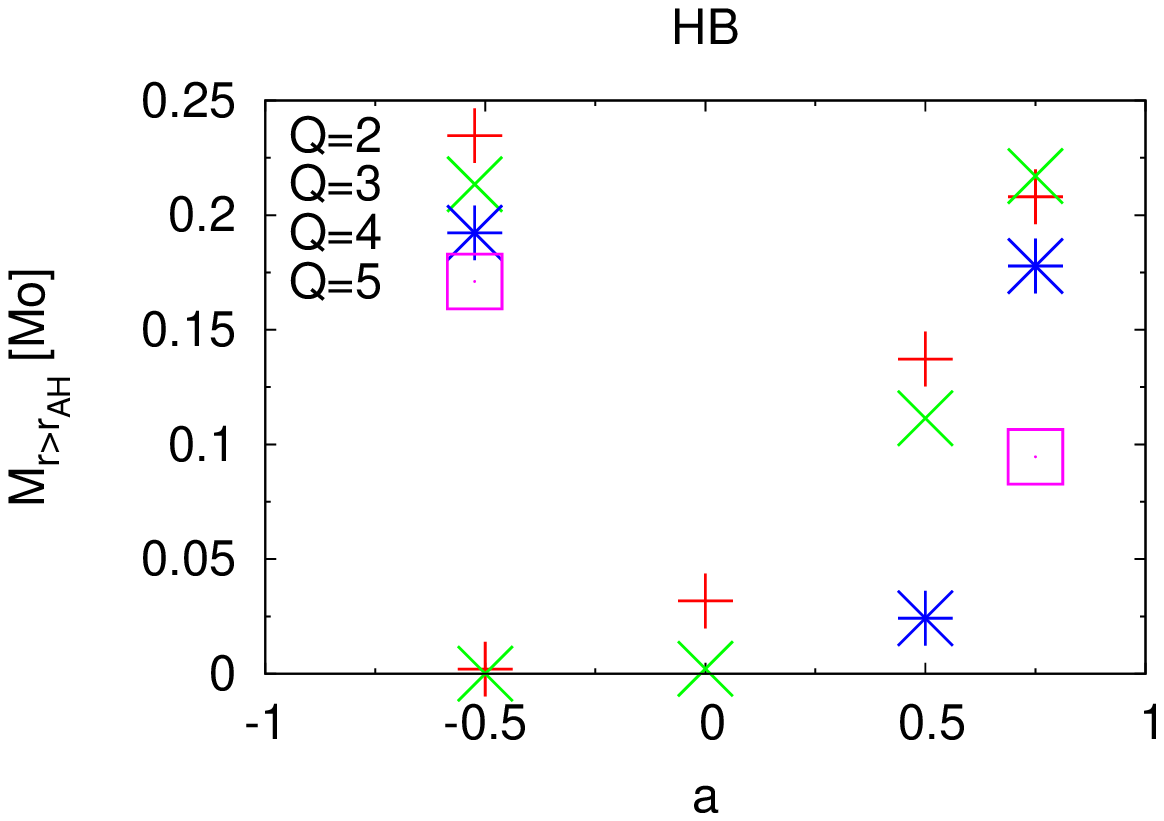}
 \end{tabular}
 \caption{The disk mass, $M_{r>r_{\rm AH}}$, at $\approx 10$ ms after
 the onset of the merger as a function of the nondimensional spin
 parameter of the BH $a$ for models with $M_{\rm NS} = 1.35
 M_\odot$. The left and right panels show the results for models with
 the 2H and HB EOSs, respectively. The results revised from the previous
 work~\cite{kst2010} are plotted for $a=0$.}  \label{fig:atom}
\end{figure*}

To clarify the dependence of the disk mass on the BH spin, we plot the
disk mass as a function of $a$ in Fig.~\ref{fig:atom}. The EOS (and,
equivalently, ${\cal C}$) is the same for each plot. Again, we find a
monotonic and steep increase of the disk mass as the increase of $a$ for
the fixed EOS and mass ratio. The enhancement of the disk mass by a
prograde spin is more dramatic for the compact NS (for the soft
EOS). For example, the difference in the disk mass between the cases of
$a = 0.75$ and $-0.5$ is only by a factor of $\sim 3$ when $Q=2$ and 2H
EOS is adopted. This low amplification is natural because tidal
disruption of a large NS occurs at an orbit far enough from the ISCO for
a substantial amount of the disrupted material to escape the capture by
the BH irrespective of $a$ and because at such a large orbital
separation the spin-orbit coupling effect is relatively weak. On the
other hand, a few-orders-of-magnitude amplification of the disk mass is
seen when $Q=2$ and HB EOS is adopted.

Finally, we comment on a possible unbound outflow. To estimate the rest
mass of unbound material, we compute
\begin{equation}
 M_{\rm ub} \equiv \int_{r>r_{\rm AH}} \rho_* H ( - u_t - 1 )d^3 x ,
\end{equation}
where $H(x)$ is a step function. Here, the material with $u_t < -1$
should be considered to have an unbound orbit. We find that $M_{\rm ub}$
can be larger than $0.01 M_\odot$ at $\approx 10$ ms after the merger
for the stiff EOS like 2H and H, and $a \ge 0$. However, $M_{\rm ub}$
does not approach a constant value and rather continues to
decrease. Therefore, it is unclear whether $M_{\rm ub}$ estimated at 10
ms after the merger can really become unbound or not, and we do not show
the precise values of $M_{\rm ub}$. When the EOS is not stiff, $M_{\rm
ub}$ is negligible within the accuracy of our simulations.

\begin{table*}
 \caption{Several key quantities for the merger remnants for $M_{\rm NS}
 = 1.35 M_\odot$. All the quantities are estimated at $\approx 10$ ms
 after the approximate merger time $t=t_{\rm merger}$. $M_{r>r_{\rm
 AH}}$ is the rest mass of the disk surrounding the BH; because the
 accretion is still ongoing due to the hydrodynamic angular momentum
 transport process, the listed values only give approximate masses of
 the long-lived accretion disks, which survive for a time longer than
 the dynamical time scale $\sim$ a few ms. $C_e$ and $C_p$ are the
 circumferential radii of the AH along the equatorial and meridional
 planes, respectively. $C_e/4\pi$ and $M_{\rm BH,c}$ denote approximate
 masses of the remnant BH. $M_{\rm irr,f}$ is the irreducible mass of
 the remnant BH, and $a_{\rm f}$ is the nondimensional spin parameter of
 the remnant BH estimated from $C_p/C_e$. $a_{\rm f2}$ and $a_{\rm f1}$
 are also the nondimensional spin parameters, estimated from the
 quantities on the AH and approximate conservation laws, respectively.
 We note that the values associated with the remnant BH for model
 B-Q2M135a75 (with an asterisk) are evaluated at $\approx 5$ ms after
 the onset of the merger because the BH area decreases by $\gtrsim 1 \%$
 at $\approx 10$ ms after the onset of the merger and the error becomes
 large.}
 \begin{tabular}{c|cccccccc} \hline
  Model & $M_{r>r_{\rm AH}} [M_\odot]$ & $C_e / 4 \pi M_0$ & $M_{\rm
  BH,c} / M_0$ & $M_{\rm irr,f} / M_0$ & $C_p / C_e$ & $a_{\rm f}$ &
  $a_{\rm f2}$ & $a_{\rm f1}$ \\ \hline
  \hline
  2H-Q2M135a75 & 0.32 & 0.913 & 0.915 & 0.789 & 0.807 & 0.87 & 0.87 &
				  0.95 \\
  1.5H-Q2M135a75 & 0.29 & 0.918 & 0.920 & 0.785 & 0.794 & 0.89 & 0.89 &
				  0.95 \\
  H-Q2M135a75 & 0.24 & 0.927 & 0.929 & 0.783 & 0.780 & 0.91 & 0.90 &
				  0.94 \\
  HB-Q2M135a75 & 0.21 & 0.933 & 0.934 & 0.783 & 0.772 & 0.91 & 0.91 &
				  0.94 \\
  B-Q2M135a75 & 0.18 & 0.937$^*$ & 0.938 & 0.790$^*$ & 0.778$^*$ &
			  0.91$^*$ & 0.91$^*$ & 0.93 \\
  \hline
  2H-Q2M135a5 & 0.27 & 0.925 & 0.926 & 0.825 & 0.843 & 0.81 & 0.81 &
				  0.84 \\
  1.5H-Q2M135a5 & 0.23 & 0.935 & 0.936 & 0.831 & 0.840 & 0.82 & 0.81 &
				  0.84 \\
  H-Q2M135a5 & 0.17 & 0.945 & 0.946 & 0.837 & 0.836 & 0.82 & 0.82 &
				  0.84 \\
  HB-Q2M135a5 & 0.14 & 0.951 & 0.952 & 0.840 & 0.832 & 0.83 & 0.83 &
				  0.84 \\
  B-Q2M135a5 & 0.095 & 0.959 & 0.960 & 0.846 & 0.830 & 0.83 & 0.83 &
				  0.84 \\
  \hline
  2H-Q2M135a-5 & 0.13 & 0.961 & 0.962 & 0.931 & 0.954 & 0.48 & 0.48 &
				  0.50 \\
  H-Q2M135a-5 & 0.010 & 0.985 & 0.986 & 0.950 & 0.948 & 0.51 & 0.51 &
				  0.52 \\
  HB-Q2M135a-5 & 0.0021 & 0.985 & 0.986 & 0.952 & 0.950 & 0.50 & 0.50 &
				  0.51 \\
  B-Q2M135a-5 & $2 \times 10^{-4} $ & 0.983 & 0.984 & 0.952 & 0.952 &
			  0.49 & 0.49 & 0.50 \\
  \hline
  2H-Q3M135a75 & 0.35 & 0.927 & 0.927 & 0.807 & 0.815 & 0.86 & 0.86 &
				  0.90 \\
  1.5H-Q3M135a75 & 0.30 & 0.931 & 0.934 & 0.811 & 0.815 & 0.86 & 0.86 &
				  0.90 \\
  H-Q3M135a75 & 0.24 & 0.939 & 0.943 & 0.820 & 0.818 & 0.85 & 0.85 &
				  0.91 \\
  HB-Q3M135a75 & 0.22 & 0.941 & 0.943 & 0.812 & 0.805 & 0.87 & 0.87 &
				  0.90 \\
  B-Q3M135a75 & 0.15 & 0.949 & 0.951 & 0.824 & 0.812 & 0.86 & 0.86 &
				  0.89 \\
  \hline
  2H-Q3M135a5 & 0.28 & 0.939 & 0.940 & 0.858 & 0.874 & 0.74 & 0.74 &
				  0.77 \\
  1.5H-Q3M135a5 & 0.23 & 0.946 & 0.948 & 0.862 & 0.871 & 0.75 & 0.75 &
				  0.78 \\
  H-Q3M135a5 & 0.16 & 0.955 & 0.957 & 0.867 & 0.866 & 0.76 & 0.76 &
				  0.78 \\
  HB-Q3M135a5 & 0.11 & 0.961 & 0.963 & 0.871 & 0.864 & 0.77 & 0.77 &
				  0.78 \\
  B-Q3M135a5 & 0.050 & 0.969 & 0.971 & 0.877 & 0.862 & 0.77 & 0.77 &
				  0.79 \\
  \hline
  HB-Q3M135a-5 & $< 10^{-4}$ & 0.986 & 0.987 & 0.973 & 0.980 & 0.32 &
			      0.32 & 0.33 \\
  \hline
  2H-Q4M135a75 & 0.36 & 0.937 & 0.938 & 0.825 & 0.828 & 0.84 & 0.84 &
				  0.87 \\
  H-Q4M135a75 & 0.23 & 0.948 & 0.951 & 0.831 & 0.823 & 0.84 & 0.84 &
				  0.88 \\
  HB-Q4M135a75 & 0.18 & 0.953 & 0.956 & 0.833 & 0.821 & 0.85 & 0.85 &
				  0.88 \\
  B-Q4M135a75 & 0.11 & 0.960 & 0.963 & 0.837 & 0.817 & 0.85 & 0.85 &
				  0.88 \\
  \hline
  2H-Q4M135a5 & 0.28 & 0.950 & 0.951 & 0.879 & 0.891 & 0.70 & 0.70 &
				  0.72 \\
  H-Q4M135a5 & 0.085 & 0.970 & 0.973 & 0.890 & 0.880 & 0.73 & 0.73 &
				  0.74 \\
  HB-Q4M135a5 & 0.024 & 0.976 & 0.979 & 0.894 & 0.878 & 0.74 & 0.74 &
				  0.75 \\
  B-Q4M135a5 & 0.0034 & 0.978 & 0.980 & 0.896 & 0.878 & 0.74 & 0.74 &
				  0.75 \\
  \hline
  2H-Q5M135a75 & 0.36 & 0.946 & 0.947 & 0.838 & 0.835 & 0.82 & 0.82 &
				  0.85 \\
  H-Q5M135a75 & 0.17 & 0.960 & 0.963 & 0.844 & 0.827 & 0.84 & 0.84 &
				  0.86 \\
  HB-Q5M135a75 & 0.095 & 0.966 & 0.970 & 0.848 & 0.824 & 0.84 & 0.84 &
				  0.86 \\
  B-Q5M135a75 & 0.031 & 0.972 & 0.975 & 0.851 & 0.821 & 0.85 & 0.85 &
				  0.87 \\
  \hline
 \end{tabular}
 \label{table:remnant}
\end{table*}

\subsection{Structure of the remnant disk} \label{subsec:res_struc}

\begin{figure*}[t]
 \begin{tabular}{ccc}
  \includegraphics[width=55mm,clip]{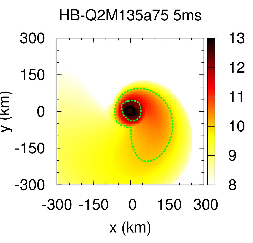} &
  \includegraphics[width=55mm,clip]{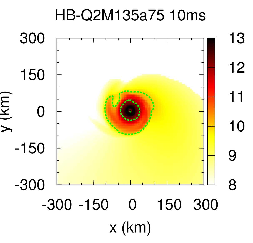} &
  \includegraphics[width=55mm,clip]{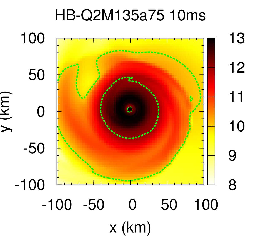} \\
  \includegraphics[width=55mm,clip]{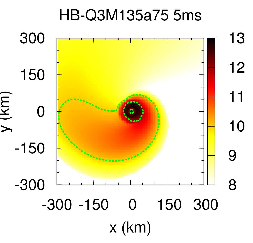} &
  \includegraphics[width=55mm,clip]{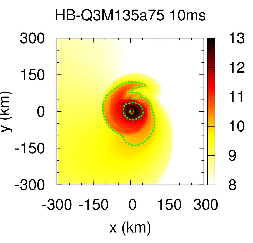} &
  \includegraphics[width=55mm,clip]{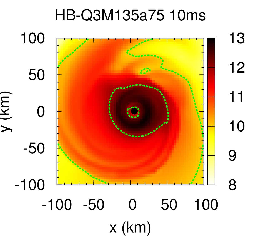} \\
  \includegraphics[width=55mm,clip]{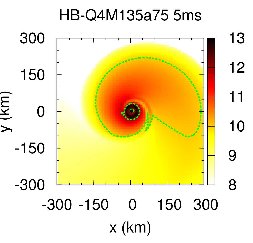} &
  \includegraphics[width=55mm,clip]{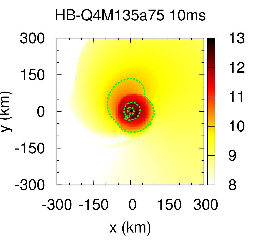} &
  \includegraphics[width=55mm,clip]{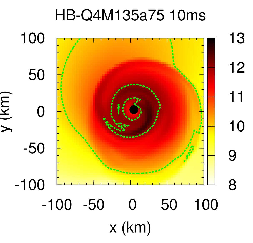} \\
  \includegraphics[width=55mm,clip]{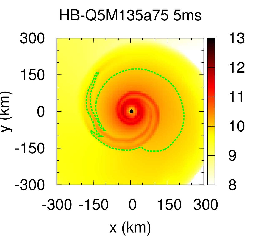} &
  \includegraphics[width=55mm,clip]{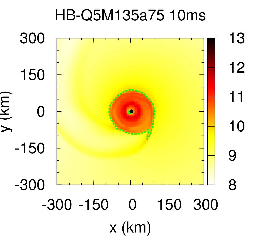} &
  \includegraphics[width=55mm,clip]{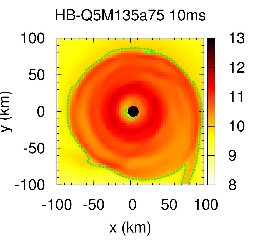} \\
 \end{tabular}
 \caption{The same as Fig.~\ref{fig:snapshot1} with contour curves for
 $\rho = 10^{10}$ and $10^{12} {\rm g/cm}^3$ plotted. In all the plots,
 HB EOS and $a=0.75$ are adopted. The first, second, third, and fourth
 rows are for $Q=2,3,4,$ and 5, respectively. The left column plots the
 snapshots at 5 ms after the onset of the merger. The middle column
 plots the snapshots at 10 ms after the onset of the merger, and the
 right column plots close-ups of the middle column.}
 \label{fig:structure}
\end{figure*}

The structure of the remnant disk and its time evolution process depend
on the mass ratio of the binary. We plot the rest-mass density profile
at $\approx$ 5 and 10 ms after the onset of the merger for binaries with
$a=0.75$, HB EOS, and different values of $Q$ in
Fig.~\ref{fig:structure}. The left column of Fig.~\ref{fig:structure} is
plotted for $\approx$ 5 ms after the onset of the merger and shows that
the dense material of $\rho \gtrsim 10^9 \; {\rm g/cm}^3$ always extends
to $\gtrsim 400$~km. The spiral arm always spreads to a far region
irrespective of EOSs, as far as the tidal disruption results in a
massive disk. These plots also suggest that the accretion disk for a
large value of $Q$---say, $Q=5$---keeps a nonaxisymmetric structure in
the vicinity of the remnant BH at this time. This feature is
qualitatively the same for binaries with other EOSs. When $Q$ is small
as $\sim 2$, the accretion disk becomes nearly axisymmetric in $\approx
5$ ms after tidal disruption because the dynamical time scale of the
system (which is proportional to the BH mass) is shorter for a smaller
value of $Q$. Also, because the ISCO radius of the BH is smaller, the
maximum rest-mass density, $\rho_{\rm max}$, of the disk (which should
be approximately proportional to the inverse square of the BH mass)
reaches a higher value on average in time for a smaller value of $Q$.
It should be noted that the difference in $\rho_{\rm max}$ comes
primarily from the difference in the radius and not from the difference
in the disk mass, which do not vary by an order of magnitude for
$a=0.75$ and $Q=2$--5. This difference in the nonaxisymmetric structure
results in different features of gravitational waves (see
Sec.~\ref{subsec:res_waveform}).

The middle and right columns of Fig.~\ref{fig:structure} plot snapshots
at $\approx$ 10 ms after the onset of the merger. At this time,
nonaxisymmetric structures are not as significant as those at $\approx$
5 ms after the onset of the merger because the accretion disk settles
toward an approximately stationary state in the vicinity of the BH. The
maximum values of the rest-mass density, $\rho_{\rm max}$, in the
accretion disk are still higher for a smaller value of $Q$. Indeed, the
right column of Fig.~\ref{fig:structure} shows that smaller values of
$Q$ result in producing a wider region with $\rho > 10^{12} \; {\rm
g/cm}^3$. By contrast, the disk for $Q=5$ does not have such a
high-density region. The smaller density may be unfavorable to be the
short-hard GRB model.

\begin{figure*}[tbp]
 \begin{tabular}{cc}
  \includegraphics[width=80mm,clip]{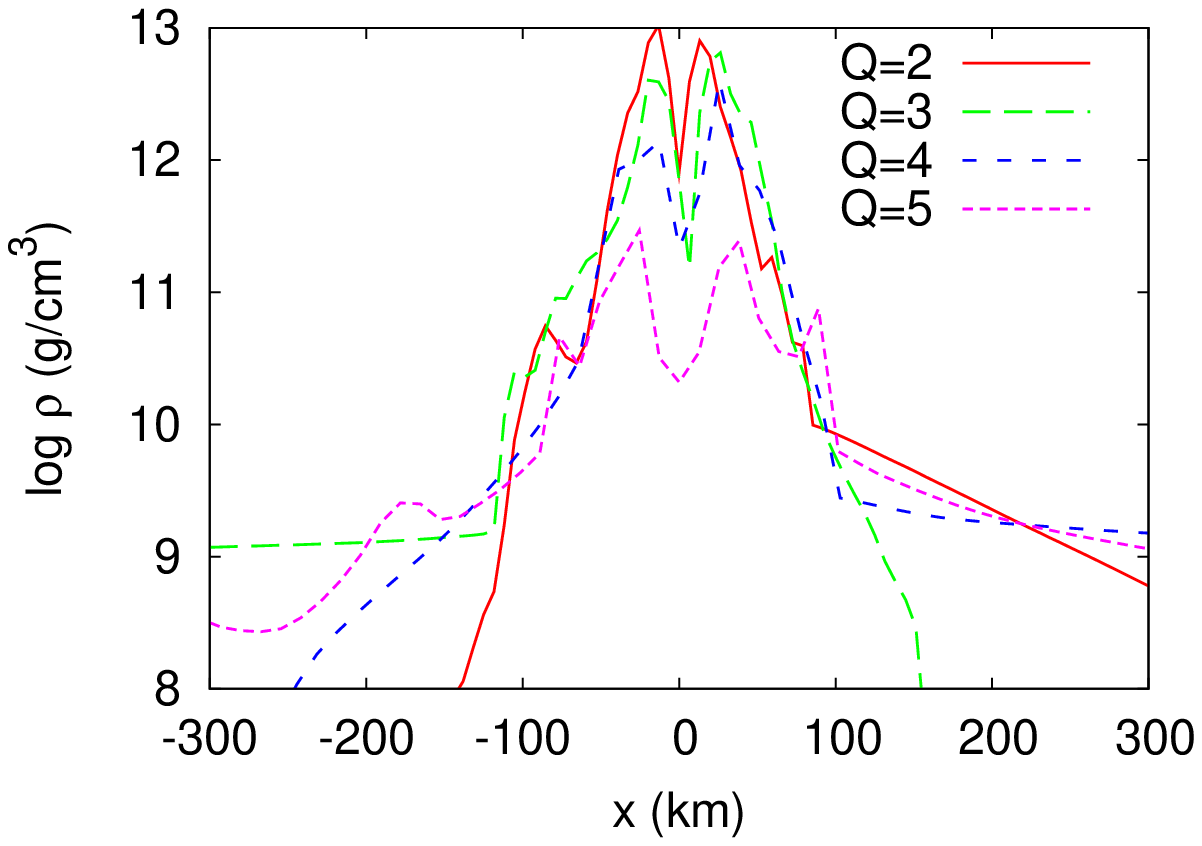} &
  \includegraphics[width=80mm,clip]{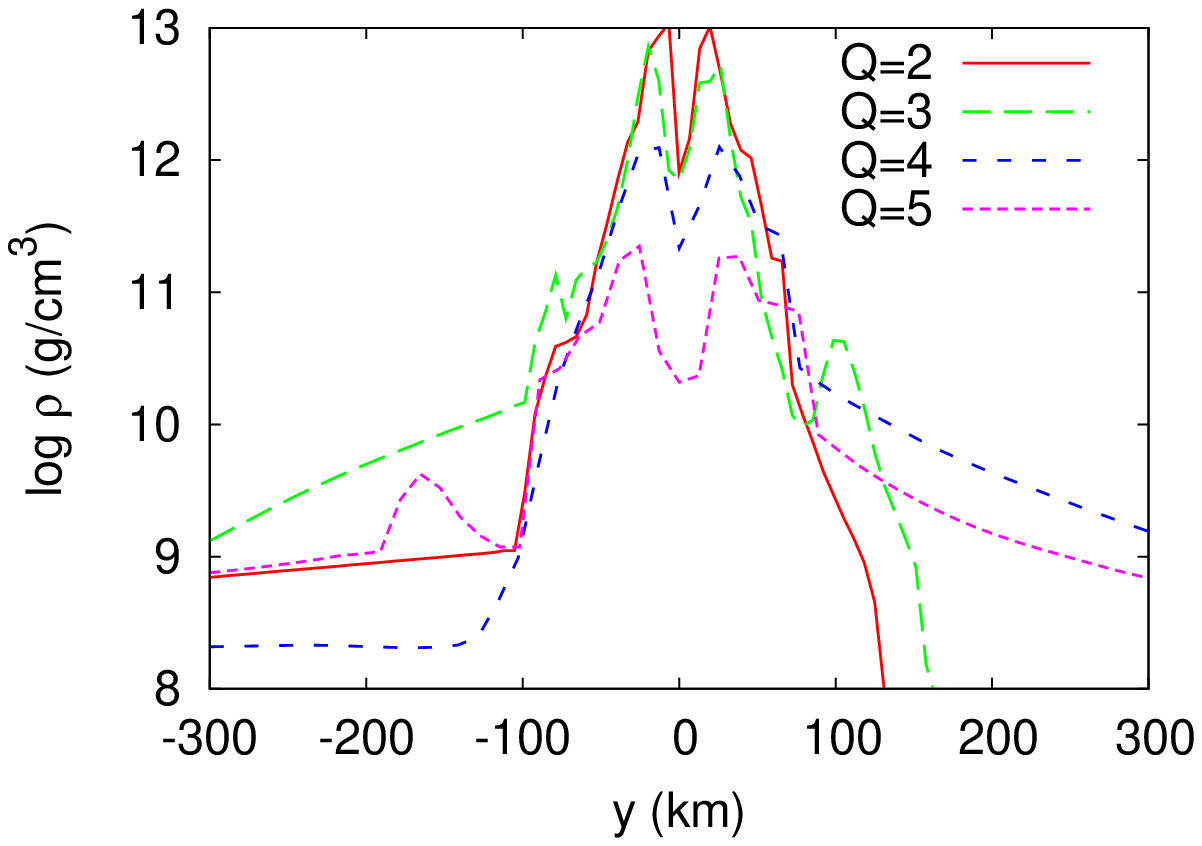}
 \end{tabular} 
 \caption{Radial distribution of the rest-mass density at 10 ms after
 the onset of the merger for different values of $Q$. The left and right
 panels show the distribution along the {\it x} and {\it y} axes,
 respectively. In both plots, $(M_{\rm NS},a) = (1.35M_\odot , 0.75)$
 and HB EOS are adopted.}  \label{fig:rhotor}
\end{figure*}

The size of a region where $\rho > 10^{10} \; {\rm g/cm}^3$ coincides
approximately among four models with different values of $Q$ and is
always $\sim 100$ km. Furthermore, the middle column suggests that the
region of $\rho > 10^8 \; {\rm g/cm}^3$ extends to larger distances when
$Q$ is larger. We plot the radial distribution of $\rho$ along {\it x}
and {\it y} axes for these models in Fig.~\ref{fig:rhotor}. Note that
low-density regions near the origin are inside the BH. These plots show
that $\rho_{\rm max}$ is systematically higher for the binary with a
smaller value of $Q$. These also show that the location of the
isodensity surface of $\rho = 10^{10} \; {\rm g/cm}^3$ approximately
coincides among different values of $Q$. Taking these facts into
account, we conclude that a typical profile of $\rho (r)$ is steeper for
smaller values of $Q$ in the vicinity of the BH. A region of $\gtrsim
100$ km away from the BH, where the profile $\rho (r)$ shows relatively
shallow decrease and $\lesssim 10^{10} \; {\rm g/cm}^3$, corresponds to
the tail component, as is seen in the middle column of
Fig.~\ref{fig:structure}.

\subsection{Properties of the remnant BH} \label{subsec:res_BH}

Table \ref{table:remnant} shows that masses and nondimensional spin
parameters of the remnant BHs depend weakly on the adopted EOSs. The
mass of the remnant BH tends to become large as the EOS softens for
fixed values of $(Q, M_{\rm NS}, a)$ for the case in which tidal
disruption of the NS occurs. The reason for this is that the tidal
disruption occurs near the ISCO, and then the BH swallows a large amount
of the NS mass when the EOS is soft.  Exceptionally, the mass of the BH
becomes slightly larger for H and HB EOSs than for B EOS for binaries
with $(Q,a) = (2,-0.5)$. The reason for this is that the remnant disk
masses are small as $\lesssim 0.01 M_\odot$ for these cases and the
amount of the energy radiated by gravitational waves primarily
determines the final state (for more compact NSs, the radiated energy is
larger because a closer inspiral orbit is achieved). The spin angular
momentum of the remnant BH $S_{\rm BH,f}$ shows similar behavior to that
of the BH mass. The situation becomes complicated for a spin parameter
of the remnant BH defined by $a_{\rm f} = S_{\rm BH,f} / M_{\rm
BH,f}^2$; the competition between the mass and angular momentum losses
from the system makes the dependence of the nondimensional spin
parameter of the remnant BH on the EOS very weak. For comparison,
$a_{\rm f1}$ and $a_{\rm f2}$ defined in
Sec.~\ref{subsec:sim_diagnostics} are also shown in Table
\ref{table:remnant}. As is found in our previous work \cite{kst2010},
$a_{\rm f}$ and $a_{\rm f2}$ agree with each other within the error of
$\Delta a = 0.003$. By contrast, $a_{\rm f1}$ does not agree well with
the other two estimates, as is found in another previous work of ours
\cite{skyt2009}, particularly when the massive remnant disk is formed
and/or the mass of the BH is small: the maximum error is $\Delta a
\approx 0.08$. Taking into account the fact that the agreement between
$C_e / 4 \pi$ and $M_{\rm BH,f}$ is always better than 0.5\%, a possible
reason for this discrepancy is that Eq.~(\ref{eq:jdisk}) systematically
underestimates the angular momentum of the disk. Hereafter, we only
refer to $a_{\rm f}$ as the nondimensional spin parameter of the remnant
BH.

\begin{figure}[tbp]
 \includegraphics[width=80mm,clip]{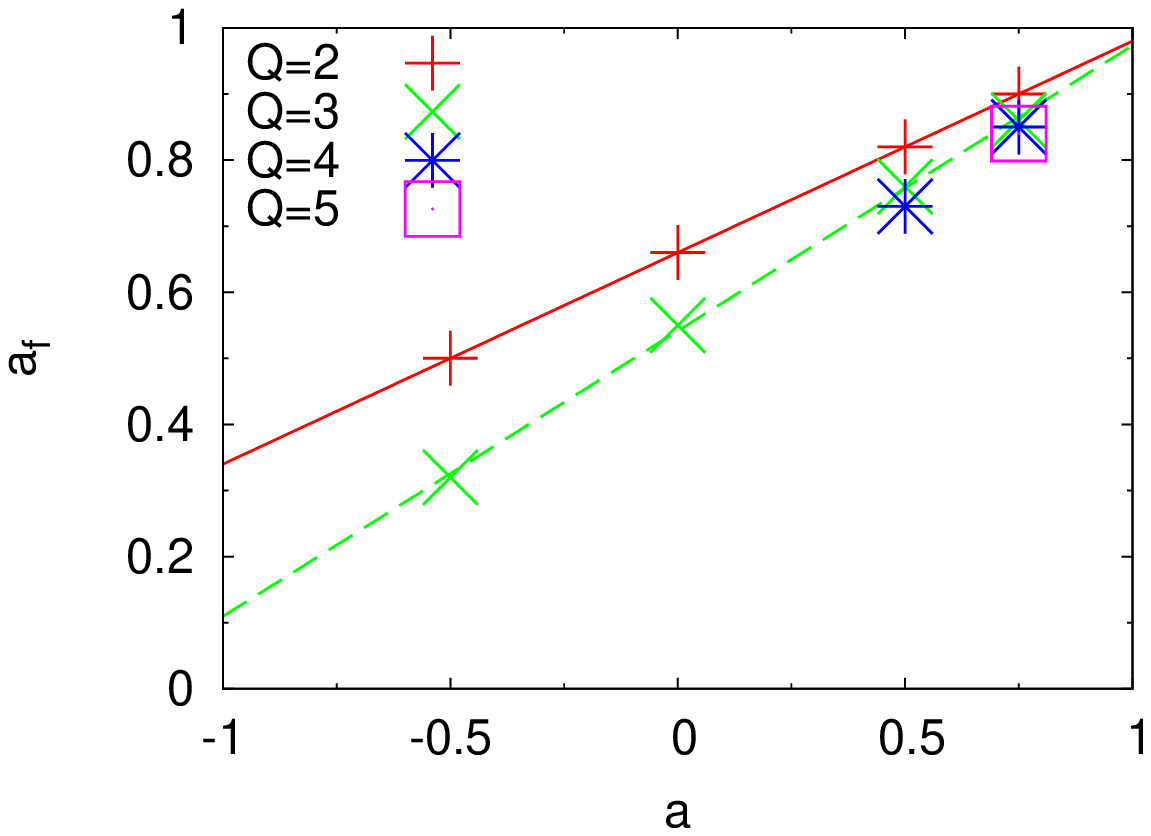} \caption{The typical
 nondimensional spin parameters of the remnant BH $a_{\rm f}$ as a
 function of the initial BH spin parameter $a$. The solid lines are
 obtained by a linear fitting of the data for $Q=2$ and $Q=3$.}
 \label{fig:spininifin}
\end{figure}

The nondimensional spin parameter of the remnant BH depends strongly on
the initial spin parameter, $a$, and the mass ratio, $Q$. Approximate
values of the nondimensional spin parameter of the remnant BH, $a_{\rm
f}$, are shown in Fig.~\ref{fig:spininifin} as a function of the initial
BH spin parameter, $a$. We also plot lines obtained by a linear fitting
of data for $Q=2$ and 3 of the following form,
\begin{eqnarray}
 a_{\rm f} &=& 0.32 a + 0.66 \; (Q=2) , \\
 a_{\rm f} &=& 0.43 a + 0.54 \; (Q=3) .
\end{eqnarray}
The relation for $Q=3$ agrees approximately with the results reported in
Refs.~\cite{elsb2009,fdkt2011} within an error of $\lesssim 5\%$, and
the agreement becomes better for a larger value of $a$. Figure
\ref{fig:spininifin} and these relations show that $a_{\rm f}$ is an
approximately linear function of $a$. In a zeroth approximation, the
slope and intercept of the linear relation denote the contribution from
the initial BH spin angular momentum, $S_{\rm BH}$, and the orbital
angular momentum of the binary, $J_0$, respectively. The larger slope
for a larger value of $Q$ is explained by a larger contribution from the
spin of the initial BH of mass $M_{\rm BH} = Q M_{\rm NS}$ to the spin
of the remnant BH of mass $M_{\rm BH,f} \sim (1+Q) M_{\rm NS}$. These
predict the value of the slope to be $Q^2 / (1+Q)^2$. However, the slope
obtained by numerical simulations is smaller by $\sim 25$--30 \% than
this predicted slope, because the amount of angular momenta
redistributed to the remnant disk and extracted by gravitational waves
become larger for a larger value of $a$. The fitting function also
suggests that the merger of an extremely spinning BH of $a=1$ and a NS
with an irrotational velocity field results in a remnant BH with $a_{\rm
f} \approx 0.98$ for BH-NS binaries with $Q=2$ and 3 and hence never
forms an overspinning BH, i.e., a BH with $a_{\rm f} > 1$. Furthermore,
the results for $Q=4$ shown in Table \ref{table:remnant} also suggest
$a_{\rm f} \approx 0.97$ for the merger of an extremely spinning BH and
an irrotational NS. These results give a circumstantial support for
cosmic censorship conjecture \cite{penrose}. Whether $a_{\rm f} \lesssim
0.98 (< 1)$ is an universal consequence of a general BH-NS binary merger
or not should be confirmed by simulations of higher mass-ratio binary
mergers, in particular, with (nearly) extremal BH spin.

\begin{figure}[tbp]
 \includegraphics[width=80mm,clip]{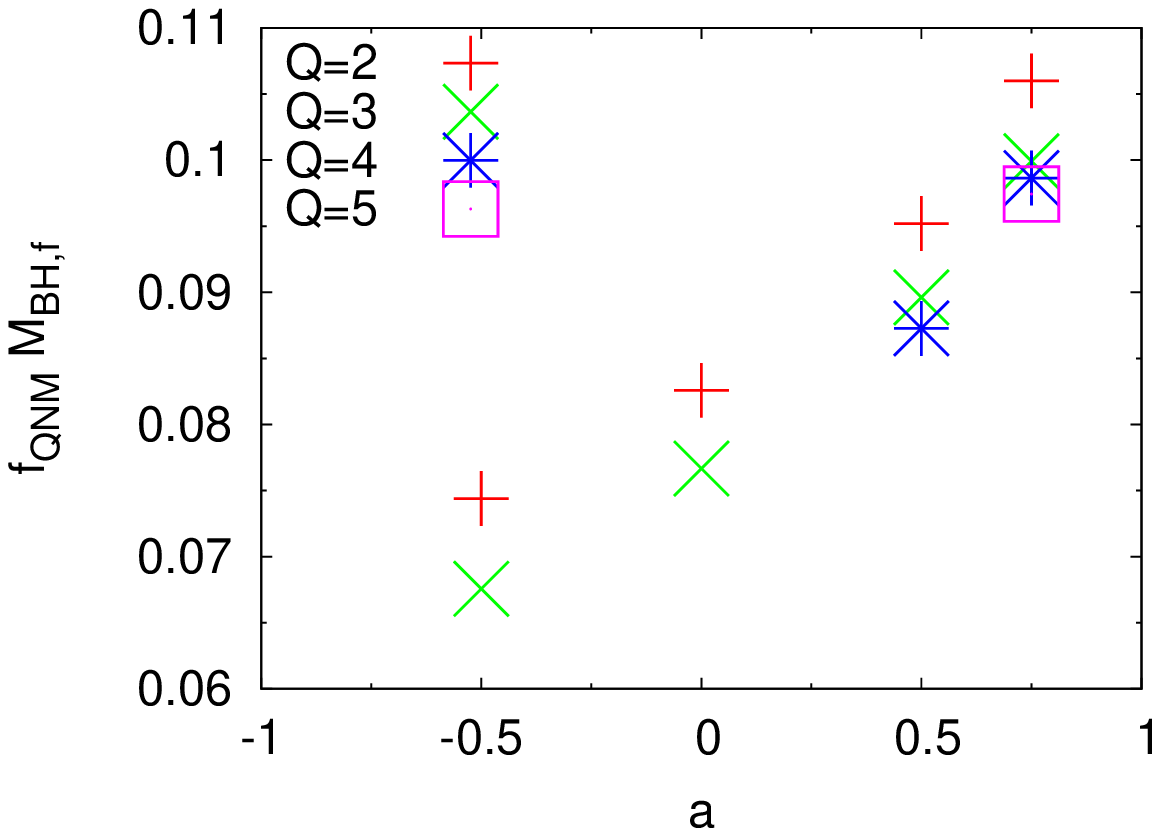} \caption{The typical QNM
 frequency of the remnant BH normalized by its mass $f_{\rm QNM} M_{\rm
 BH,f}$.} \label{fig:qnm}
\end{figure}

From these typical values of $a_{\rm f}$ and $M_{\rm BH,f}$, we can
estimate typical frequencies of quasinormal modes (hereafter QNM)
$f_{\rm QNM}$ of the remnant BH by a fitting formula \cite{bcs2009}
\begin{equation}
 f_{\rm QNM} M_{\rm BH,f} \approx \frac{1}{2\pi} [ 1.5251 - 1.1568
  (1-a)^{0.1292} ] .
\end{equation}
We plot these values in Fig.~\ref{fig:qnm}. They are in good agreement
with those of the ringdown waveforms, if the QNM is excited after the
merger.

\subsection{Gravitational waveforms} \label{subsec:res_waveform}

In this section, we show $(l,m)=(2,2)$, plus-mode gravitational
waveforms $h_+$ for selected models obtained in this study, as well as
the waveform for models obtained in our previous simulations
\cite{kst2010}. We plot all the waveforms for an observer along the {\it
z} axis as a function of the approximate retarded time
\begin{equation}
 t_{\rm ret} = t - D - 2 M_0 \ln ( D / M_0 ) .
\end{equation}
The amplitude of the waveforms is normalized as $D h_+ / m_0$ or we show
physical amplitude observed at a hypothetical distance $D = 100$ Mpc
along the {\it z} axis. Gravitational waveforms calculated in the
Taylor-T4 formula are plotted together in the figures to validate the
waveforms obtained in our numerical simulations during the inspiral
phase. Numerical waveforms during 2--3 initial cycles deviate from ones
obtained from the Taylor-T4 formula in all the cases due to the lack of
an approaching velocity in the initial data. This deficit is ascribed to
insufficient modeling of the quasiequilibrium state and improvement in
the future is important to obtain more accurate gravitational-wave
templates \cite{ulfgs2006,ulfgs2009}. Our waveforms also deviate from
the Taylor-T4 waveforms in the late inspiral phase due to a physical
reason, which we describe below. Comparisons between waveforms obtained
from simulations with different grid resolutions are shown in the
Appendix \ref{app_conv}.

\begin{figure*}[tbp]
 \begin{tabular}{cc}
  \includegraphics[width=80mm,clip]{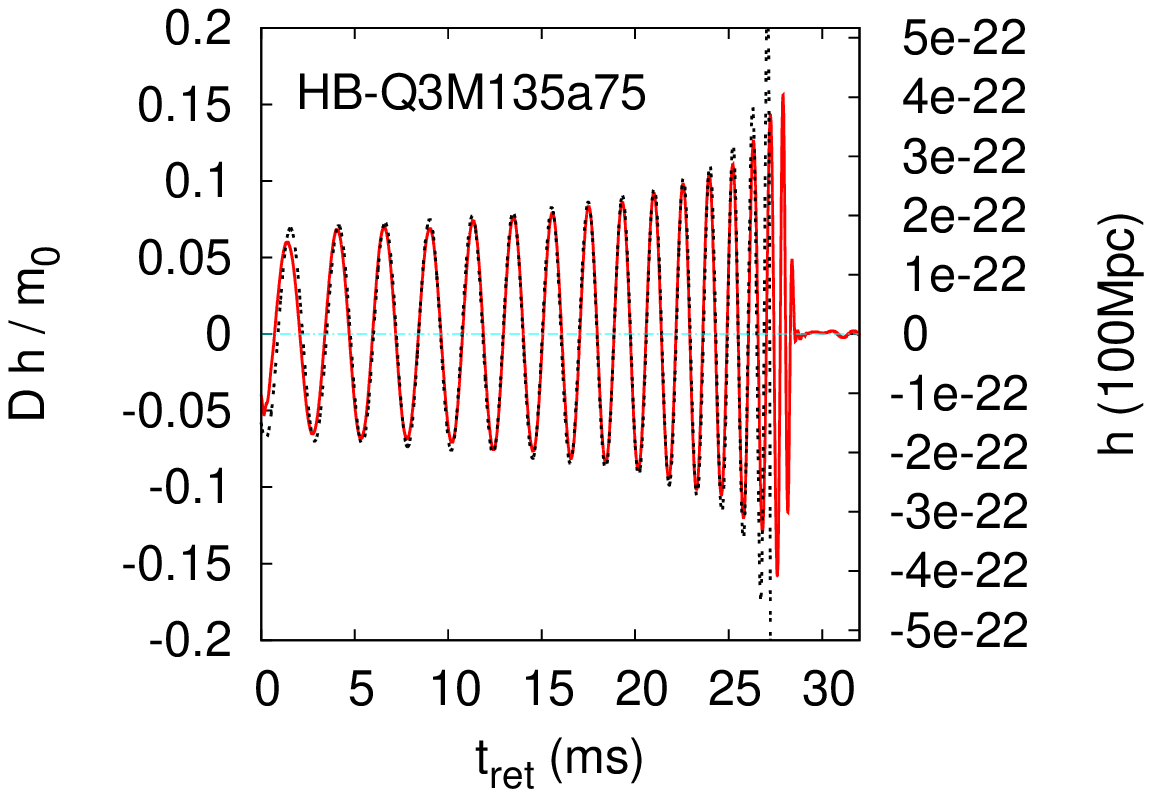} &
  \includegraphics[width=80mm,clip]{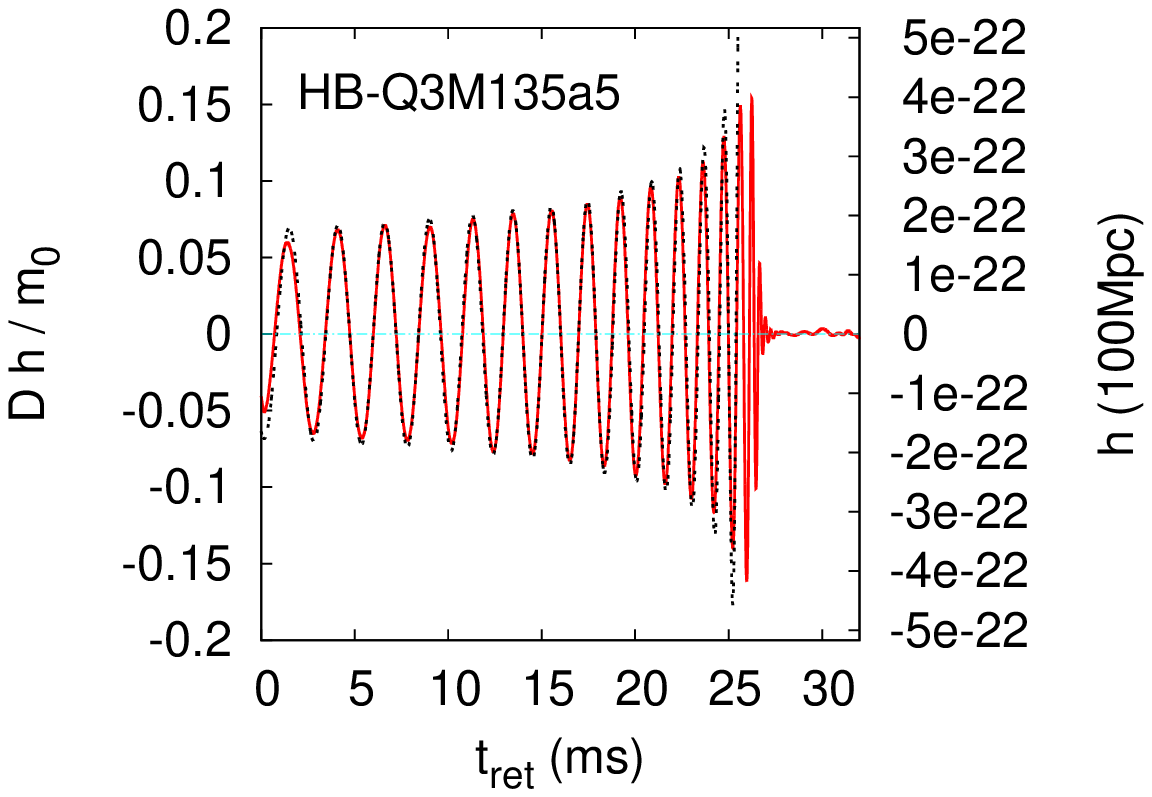} \\
  \includegraphics[width=80mm,clip]{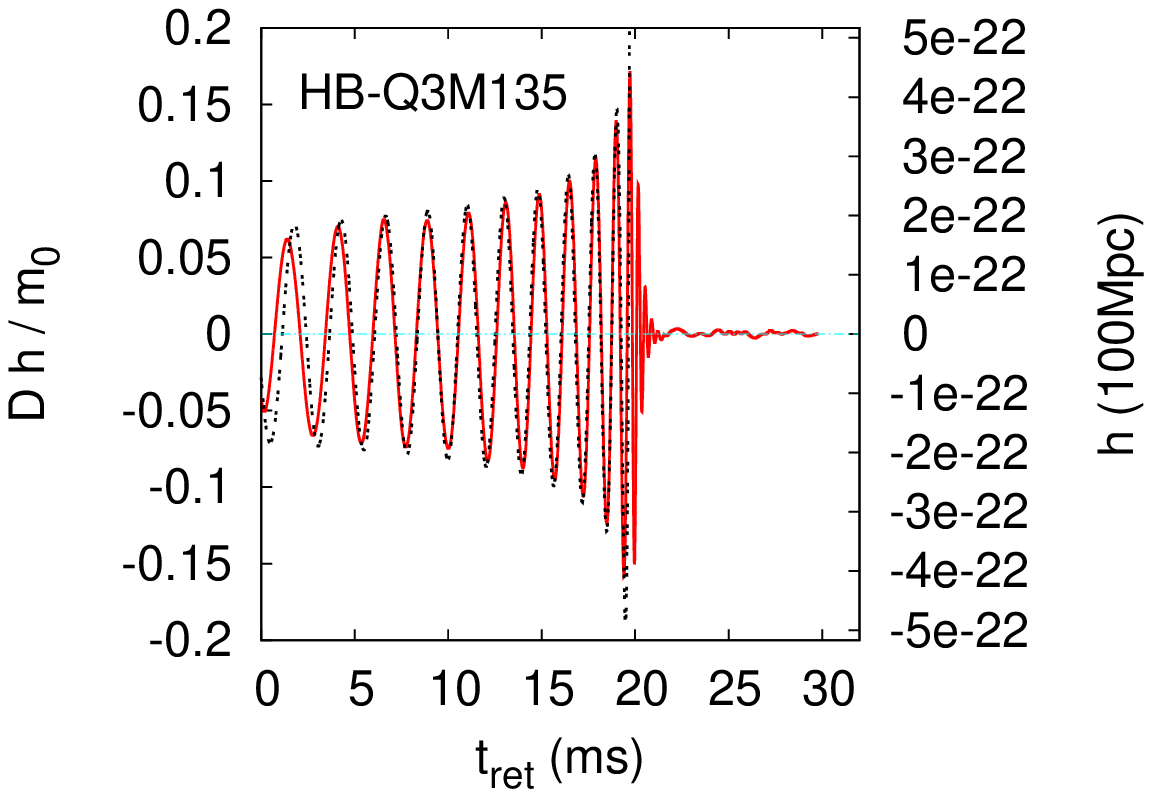} &
  \includegraphics[width=80mm,clip]{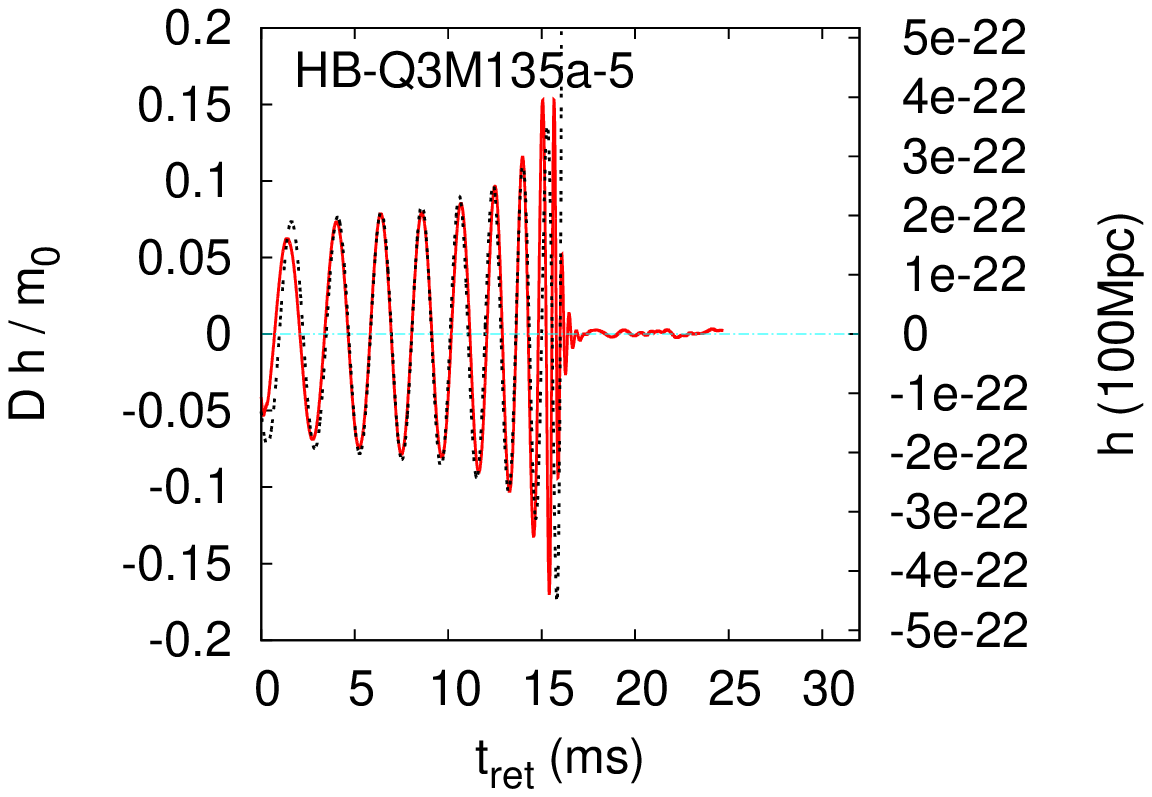}
 \end{tabular} 
 \caption{$(l,m)=(2,2)$, plus-mode gravitational waves for models
 HB-Q3M135a75, HB-Q3M135a5, HB-Q3M135, and HB-Q3M135a-5. All the
 waveforms are shown for an observer located along the {\it z} axis (the
 axis of rotation) and plotted as a function of a retarded time. The
 left axis denotes the amplitude normalized by the distance from the
 binary $D$ and the total mass $m_0$. The right axis denotes the
 physical amplitude of gravitational waves observed at a hypothetical
 distance 100 Mpc. The dotted curves denote the waveform calculated by
 the Taylor-T4 formula.} \label{fig:gwq3}
\end{figure*}

Figure \ref{fig:gwq3} shows the gravitational waveforms for binaries
with HB EOS, $(Q,M_{\rm NS}) = (3,1.35 M_\odot )$ but with different BH
spin parameters, $a=0.75$, 0.5, 0, and $-0.5$. This figure shows that
the time to the merger, to which we refer approximately as the time at
which the maximum gravitational-wave amplitude is achieved, for
$\Omega_0 m_0 = 0.030$ becomes longer by $\approx 10$ ms as the increase
of the BH spin within the range concerned here. This difference in the
merger time owes primarily to the spin-orbit interaction described in
Sec.~\ref{subsec:res_merger}, and this behavior is qualitatively the
same for binaries with any EOS. The numerical and Taylor-T4 waveforms
agree well with each other during an inspiral phase for all the cases.

\begin{figure*}[tbp]
 \begin{tabular}{cc}
  \includegraphics[width=80mm,clip]{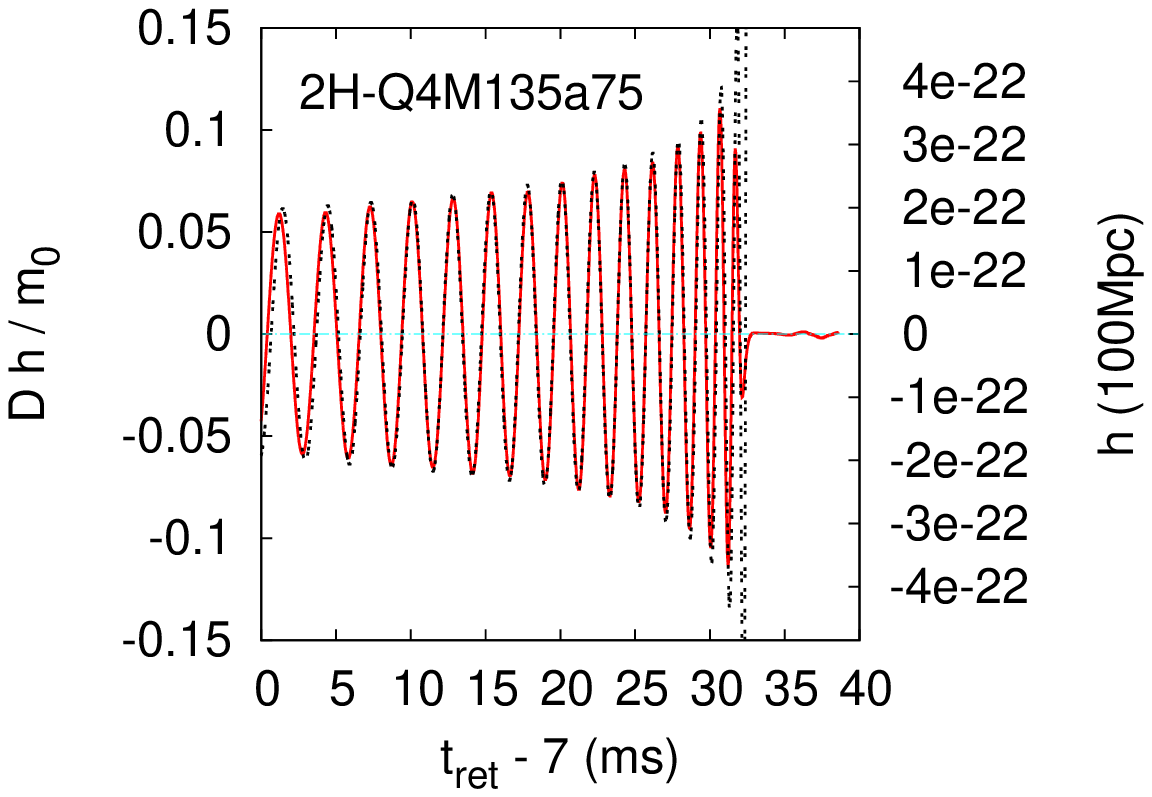} &
  \includegraphics[width=80mm,clip]{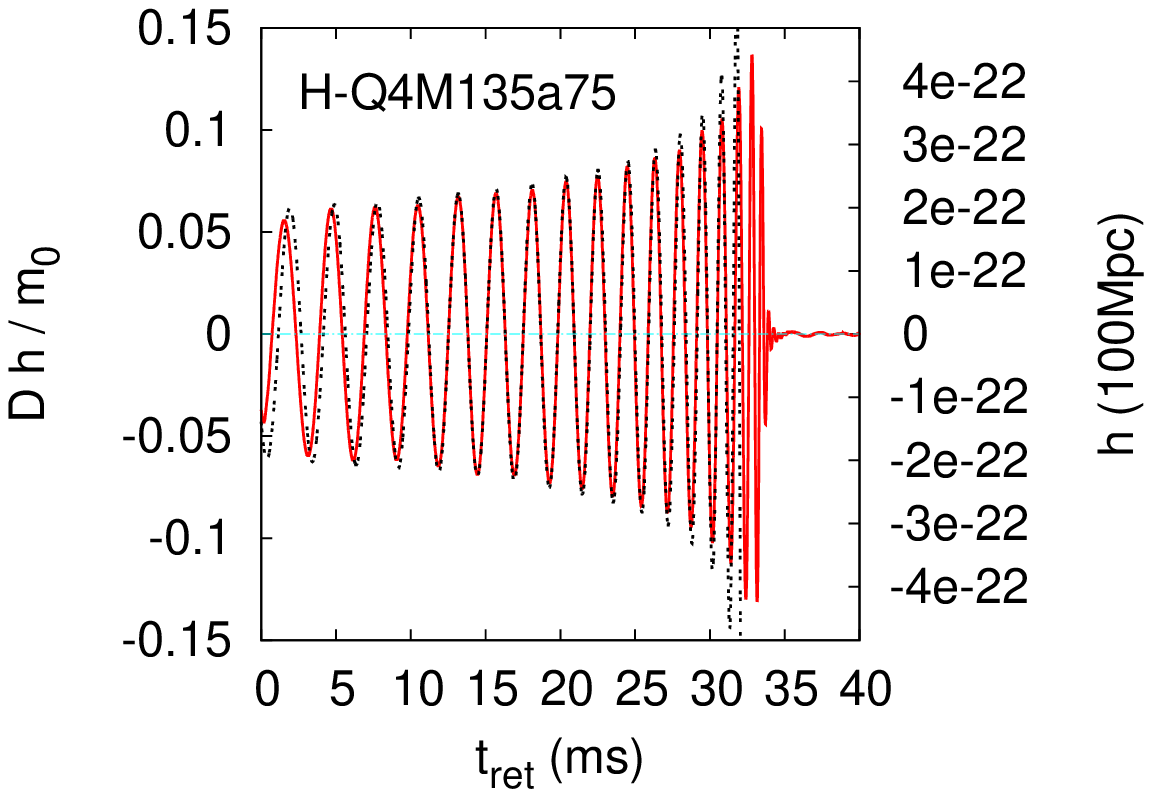} \\
  \includegraphics[width=80mm,clip]{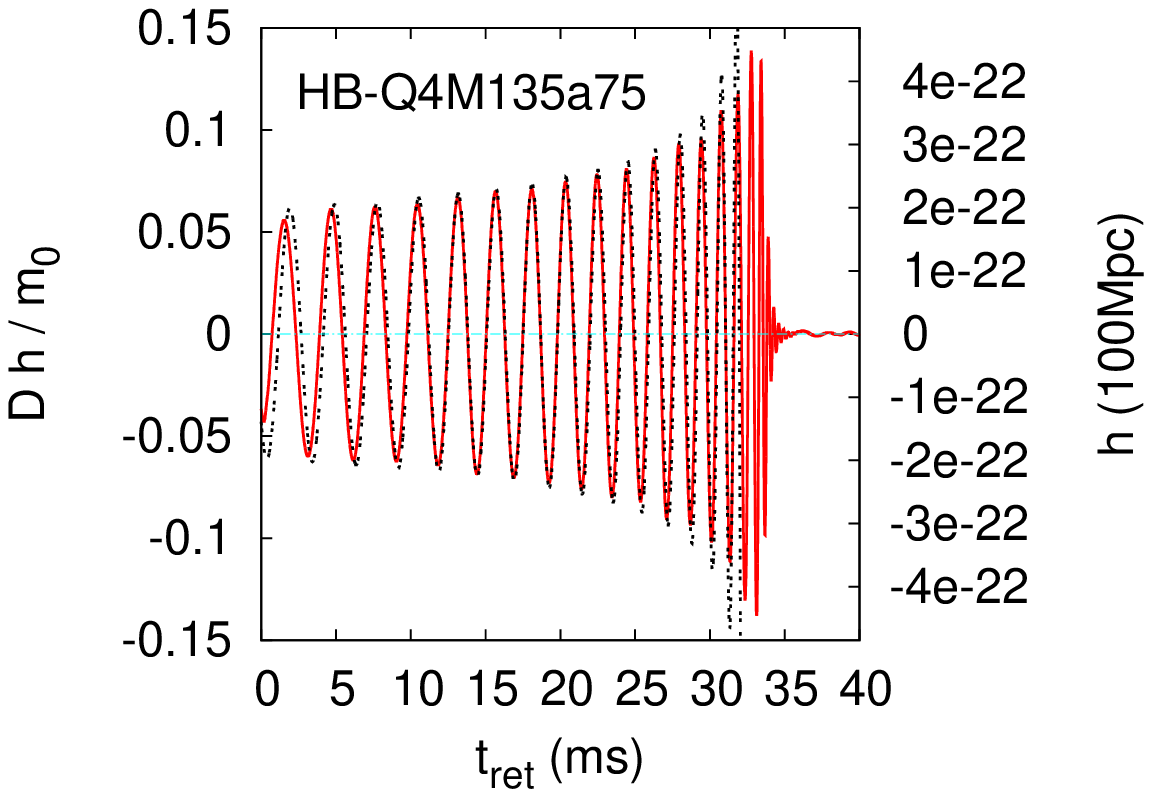} &
  \includegraphics[width=80mm,clip]{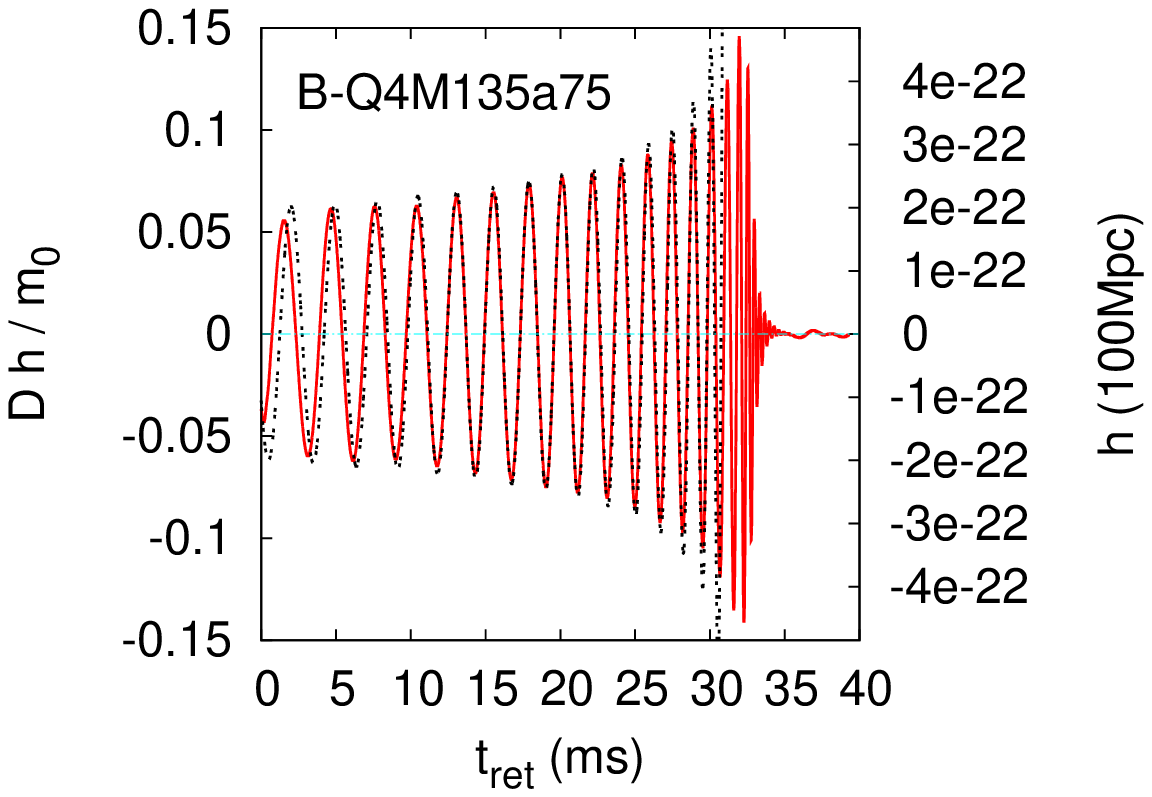}
 \end{tabular} 
 \caption{The same as Fig.~\ref{fig:gwq3} but for models 2H-Q4M135a75,
 H-Q4M135a75, HB-Q4M135a75, and B-Q4M135a75.} \label{fig:gwq4}
\end{figure*}

\begin{figure*}[tbp]
 \begin{tabular}{cc}
  \includegraphics[width=80mm,clip]{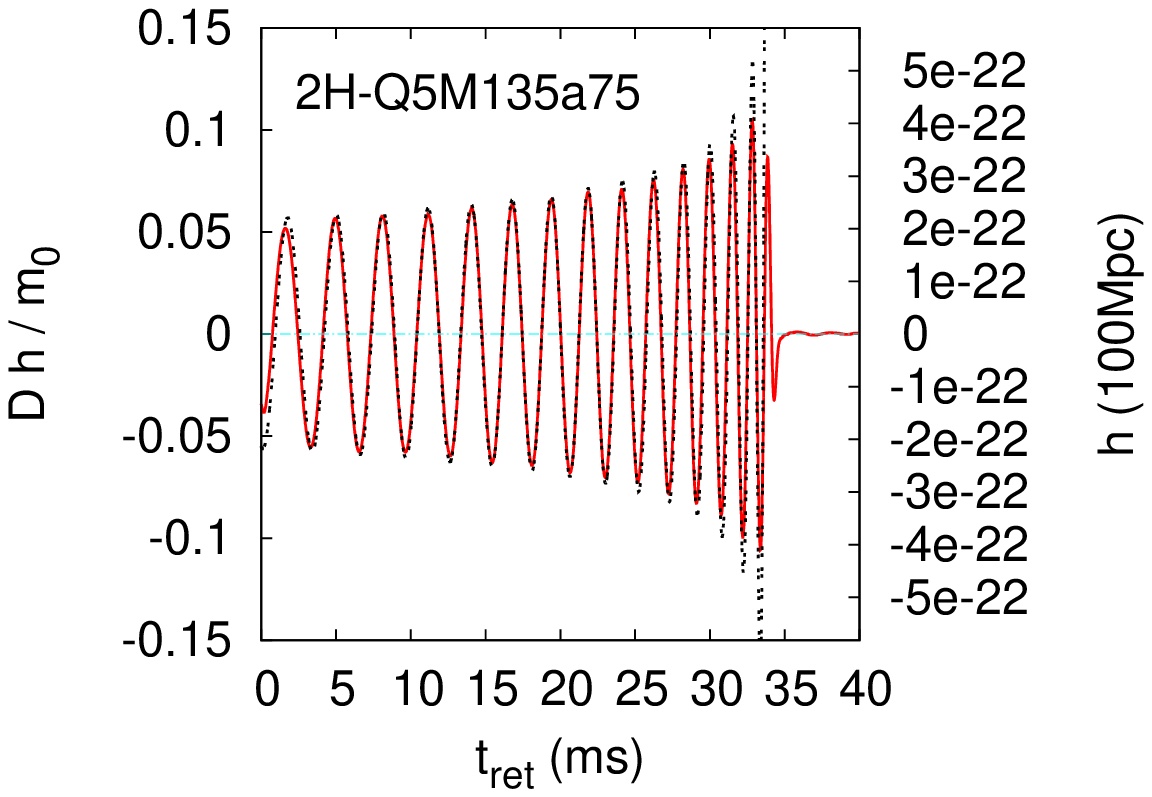} &
  \includegraphics[width=80mm,clip]{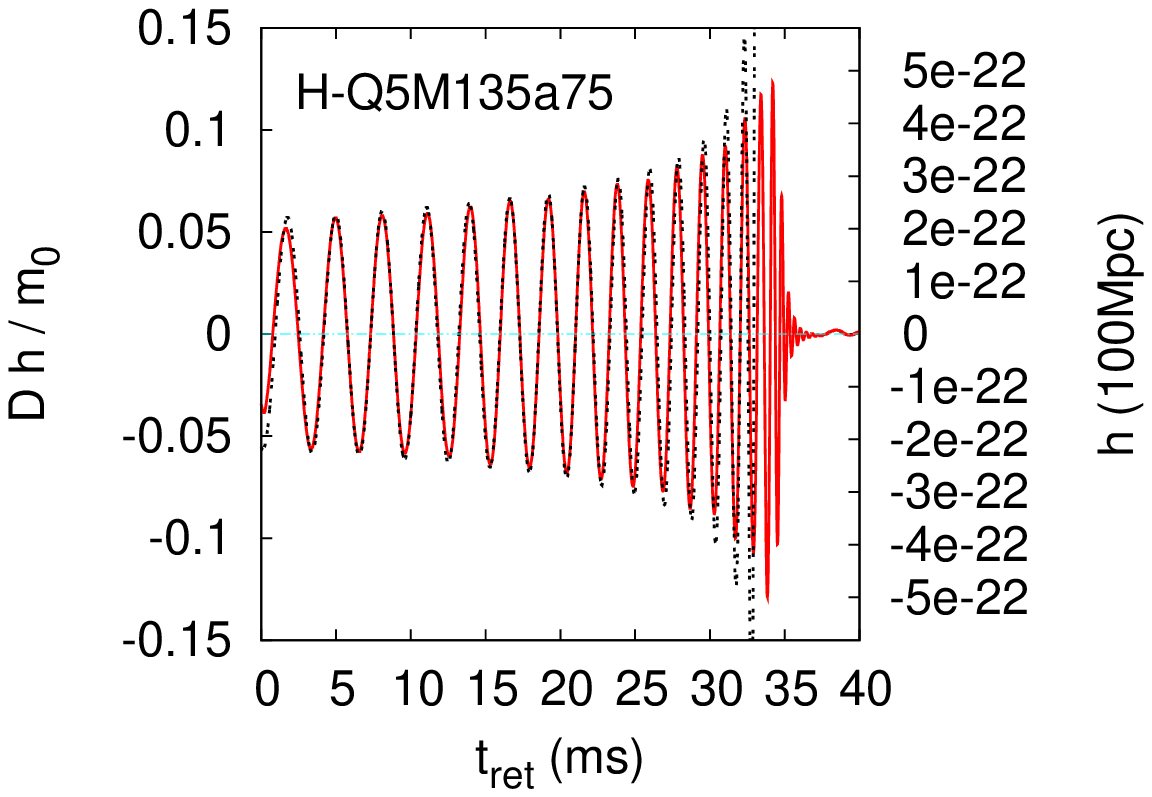} \\
  \includegraphics[width=80mm,clip]{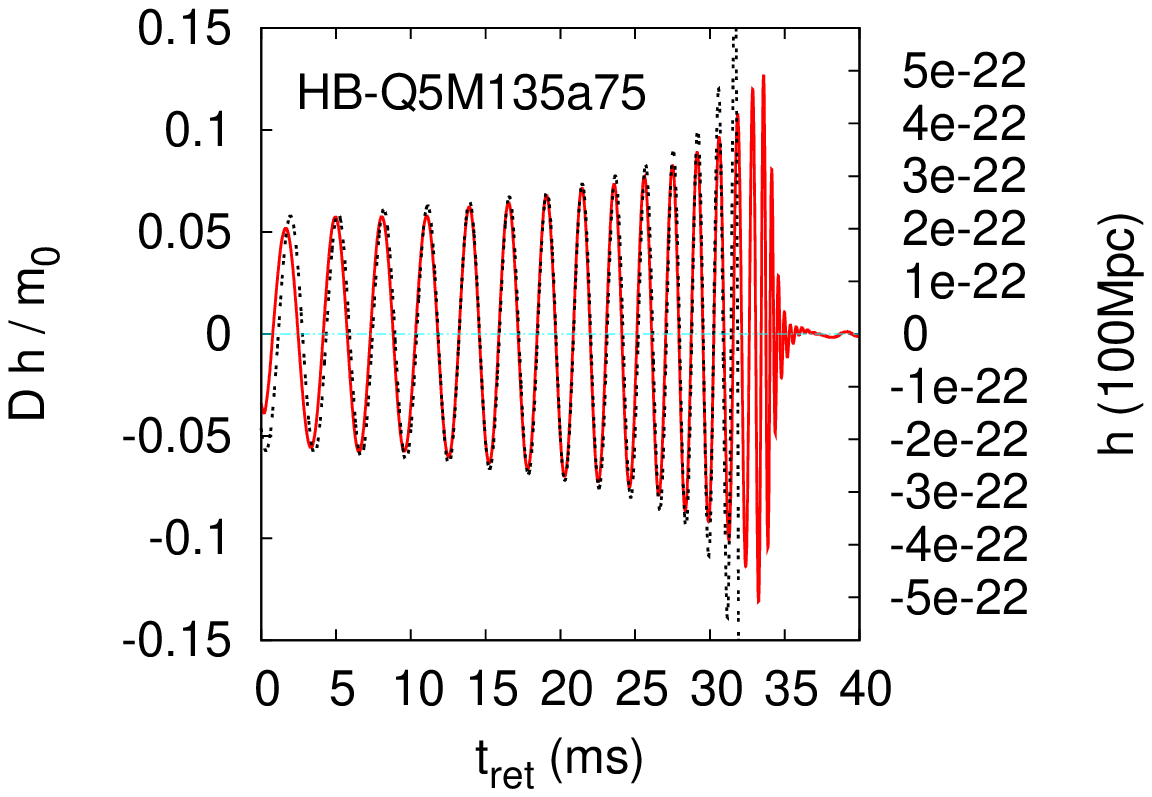} &
  \includegraphics[width=80mm,clip]{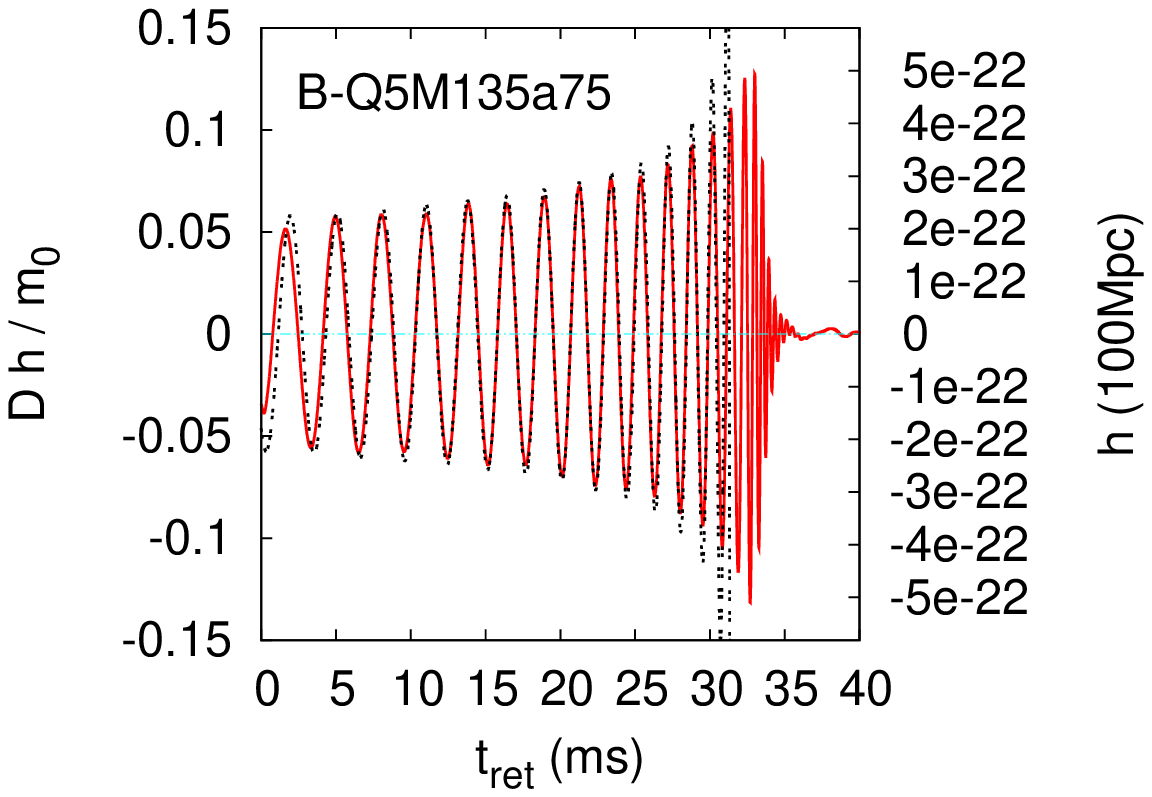}
 \end{tabular} 
 \caption{The same as Fig.~\ref{fig:gwq3} but for models 2H-Q5M135a75,
 H-Q5M135a75, HB-Q5M135a75, and B-Q5M135a75.} \label{fig:gwq5}
\end{figure*}

For the prograde BH spin cases, the Taylor-T4 formula does not track the
evolution for $\sim 0.5$ inspiral orbit just before the merger. The
Taylor-T4 amplitude departs from that of numerical relativity and even
diverges. Accordingly, the number of gravitational-wave cycles differs
by as much as unity between the numerical and Taylor-T4 waveforms. The
difference in the number of cycles is larger for higher mass-ratio
binaries with prograde BH spins. We show the waveforms for binaries with
$(Q, M_{\rm NS}, a) = (4, 1.35 M_\odot, 0.75)$ and with $(Q, M_{\rm NS},
a) = (5, 1.35 M_\odot, 0.75)$ for 2H, H, HB, and B EOSs in
Figs.~\ref{fig:gwq4} and \ref{fig:gwq5}, respectively. The deviation is
clear for H, HB, and B EOSs in both figures. This difference indicates
that the phase evolution predicted by the Taylor-T4 formula is not
sufficient to model the last inspiral phase of a coalescing binary with
the high mass ratio of $Q \gtrsim 3$ and the prograde BH spin of $a
\gtrsim 0.5$.

For the retrograde BH spin case (the bottom right panel of
Fig.~\ref{fig:gwq3}), the phase evolution deviates between the numerical
and Taylor-T4 waveforms in the last orbit before the merger. This
deviation may be partly ascribed to the small number of orbits in our
simulation but appears to be primarily ascribed to a larger angular
velocity, or equivalently a larger PN parameter, $\Omega m_0$, at the
last orbit for a retrograde BH spin. Thus, the Taylor-T4 formula seems
to be again insufficient for modeling the retrograde BH spin cases.

Figure \ref{fig:gwq3} also shows that the gravitational waveform in the
merger stage depends strongly on the BH spin. For a binary with $(Q,a) =
(3,0.75)$, gravitational waves show a sudden decrease in the amplitude
at $t_{\rm ret} \approx 27$~ms, which is a clear signature of tidal
disruption. Gravitational waves associated with the ringdown of a
remnant BH are absent due to the phase cancellation by nearly
axisymmetric accretion of the disrupted material. This feature is
consistent with the formation of a massive remnant disk, which is
described in Sec.~\ref{subsec:res_disk}, for the prograde BH spin. For
binaries with $(Q,a) = (3, \le 0)$, on the other hand, gravitational
waves end up with ringdown waveforms associated with the remnant BHs
because the tidal effect is very weak throughout the merger. In these
circumstances, gravitational waves do not show strong signatures of
tidal deformation and disruption of the NS.

Gravitational waves for a binary with $(Q,a) = (3,0.5)$ show a
qualitatively new feature (the top right panel of
Fig.~\ref{fig:gwq3}). In this case, a ringdown waveform of the remnant
BH is seen in the final stage, although the NS is tidally disrupted and
the disk mass is larger than $0.1 M_\odot$. Namely, both tidal
disruption of the NS and excitation of a QNM of the remnant BH occur in
a compatible manner. The same feature is also found for a binary with a
high mass ratio and a prograde BH spin, i.e., $(Q,a) = (\ge 4,0.75)$,
shown in Figs.~\ref{fig:gwq4} and \ref{fig:gwq5} except for 2H EOS, with
which the NS is disrupted at a fairly distant orbit. These waveforms are
often seen for BH-NS binaries with a heavy BH (or a high mass ratio)
with the prograde BH spin, which results in the NS tidal disruption, and
is never seen for BH-NS binaries with $Q=2$ or high mass-ratio binaries
with nonspinning BHs.

\begin{figure*}[tbp]
 \includegraphics[width=170mm,clip]{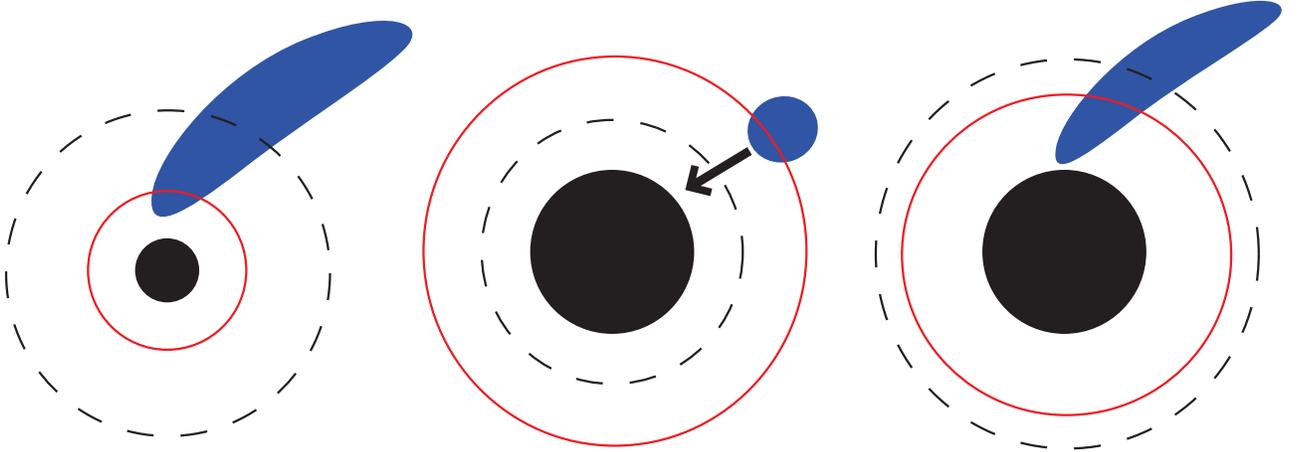} \caption{Schematic
 pictures for three types of the merger process. The solid filled circle
 denotes the BH, the distorted ellipsoid denotes the NS, the solid
 circle is the location of the ISCO, and the dashed circle is the
 location of the radius at which the tidal disruption occurs. Left: the
 NS is tidally disrupted, and the spatial extent of the disrupted
 material is larger than or as large as that of the BH. Middle: the NS
 is not tidally disrupted. Right: the NS is tidally disrupted, and the
 spatial extent of the disrupted material is smaller than that of the
 BH.}  \label{fig:scheme}
\end{figure*}

The ratio of the areal radius of the remnant BH to the NS radius,
$R_{\rm NS}$, is intimately related to the different excitation degree
of the QNM between low and high mass-ratio binaries in the presence of
NS tidal disruption~\cite{saijonakamura2000,saijonakamura2001}.
Schematic pictures of merger processes are depicted in
Fig.~\ref{fig:scheme}. If tidal disruption does not occur, the NS is
simply swallowed by the BH and excites a QNM, as shown in the middle
panel of Fig.~\ref{fig:scheme}. If tidal disruption occurs in a binary
with a low mass ratio, the disrupted material spreads around the BH to
soon form a nearly axisymmetric disk. Approximately speaking, this
occurs if the BH areal radius is smaller than the NS radius, as is shown
in the left panel of Fig.~\ref{fig:scheme}. Thus, the NS tidal
disruption has a strong effect to suppress the excitation of a QNM
through the phase cancellation in the low mass-ratio binary. However,
the situation is different in a high mass-ratio binary. Whereas the
disrupted material forms an axisymmetric accretion disk around the BH in
a sufficiently long time duration, the accretion just after the merger
does not proceed in an axisymmetric way in high mass-ratio binaries,
such as $Q=4$, except for the extremely stiff EOS. This is because the
BH radius for $Q=4$ approximately doubles that for $Q=2$, and hence, the
disrupted material takes longer time to spread around the BH. In other
words, the NS material accretes onto the BH coherently even after the
tidal disruption, as is shown in the right panel of
Fig.~\ref{fig:scheme}, because the BH radius is so large that the
disrupted material cannot fully cover the BH surface before the BH
swallows a large portion of the material. In the exceptional 2H EOS
case, tidal disruption occurs sufficiently far outside the BH due to the
large radius of the NS, and hence, the disrupted material is able to
spread around the BH to form a nearly axisymmetric accretion disk before
the prompt infall. Therefore, the QNM of a remnant BH is not excited for
2H EOS.

\subsection{Gravitational-wave spectrum} \label{subsec:res_spectrum}

Key features of gravitational waves are reflected in the Fourier
spectrum. In this paper, we define the Fourier spectrum as a sum of each
Fourier component of two independent polarizations of the $(l,|m|) =
(2,2)$ mode as
\begin{eqnarray}
 \tilde{h} (f) &=& \sqrt{\frac{|\tilde{h}_+ (f)|^2 + |\tilde{h}_\times
  (f)|^2}{2}} , \\
 \tilde{h}_A (f) &=& \int e^{2 \pi ift} h_A (t) dt ,
\end{eqnarray}
where {\it A} denotes two polarization modes, + or $\times$. We show a
nondimensional spectrum, $f \tilde{h} (f)$, observed at a hypothetical
distance of 100~Mpc as a function of the gravitational-wave frequency,
$f$ (Hz), or a normalized amplitude, $f \tilde{h} (f) D / m_0$, as a
function of a nondimensional frequency, $f m_0$. The amplitude of
gravitational waves, $h_A$, is given as the amplitude observed along the
{\it z} axis, which is the most optimistic direction for the
gravitational-wave detection. We note that the actual amplitude of
gravitational waves depends on an angle locating the source in the sky
and on an angle specifying the orientation of the orbital plane of the
binary. The angular average of the effective amplitude is $\approx 0.4 f
\tilde h(f)$. We always exclude spurious radiation components for
$t_{\rm ret} \lesssim 0$ ms, using a step function of the retarded time
as a window function.

\begin{figure*}[tbp]
 \begin{tabular}{cc}
  \includegraphics[width=88mm,clip]{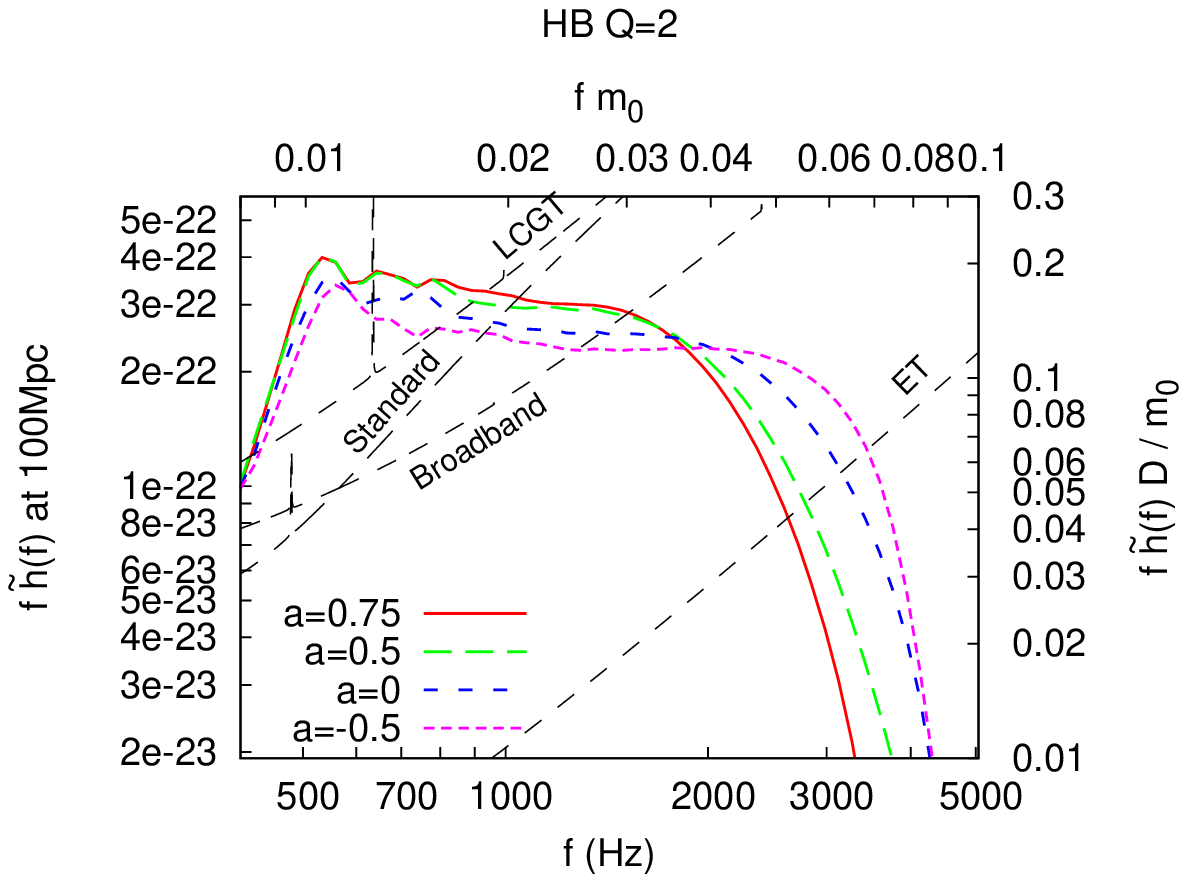} &
  \includegraphics[width=88mm,clip]{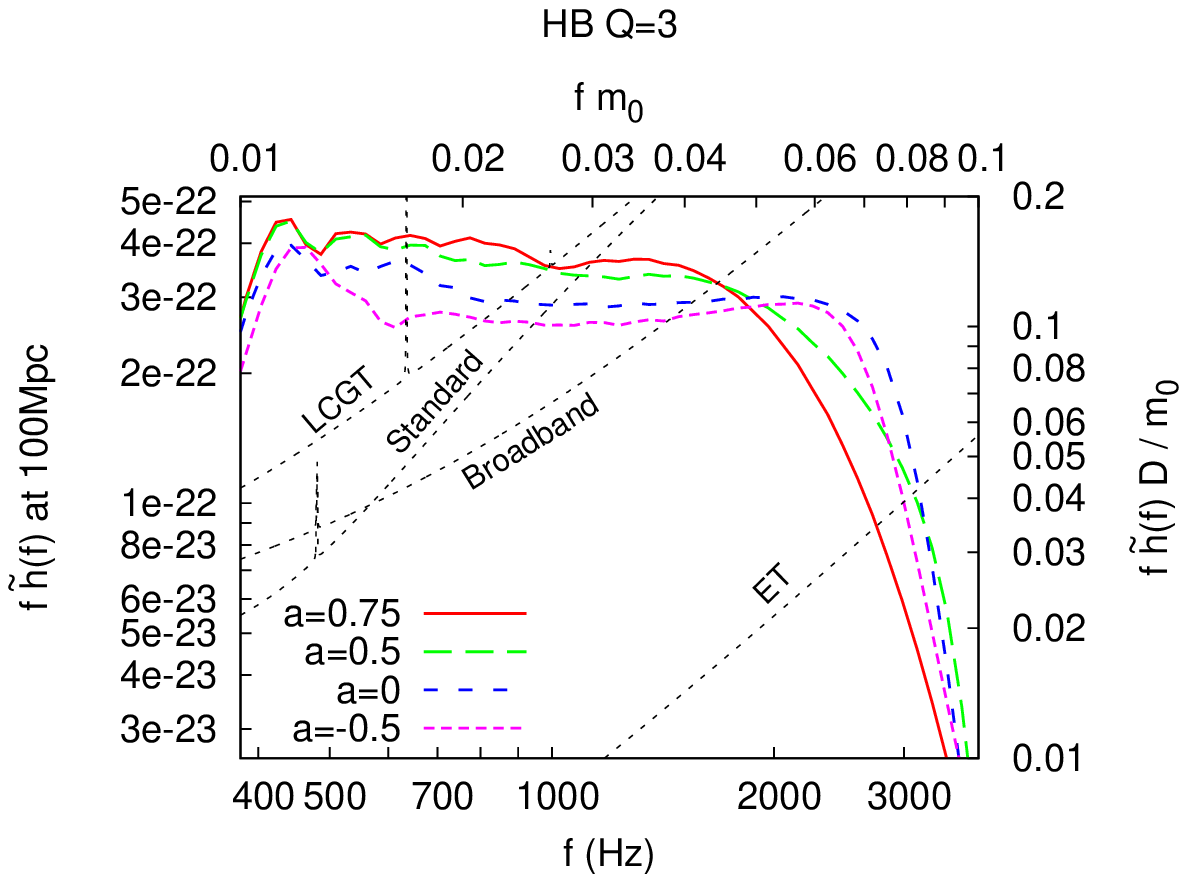}
 \end{tabular}
 \caption{Gravitational-wave spectra for BH-NS binaries with HB EOS,
 $M_{\rm NS} = 1.35 M_\odot$ and $a=0.75$, 0.5, 0, and $-0.5$. The left
 and right panels show the spectra for $Q=2$ and 3, respectively. The
 upper axis denotes the normalized frequency, $f m_0$, and the right
 axis denotes the normalized amplitude, $f \tilde{h} (f) D / m_0$. The
 bottom axis denotes the frequency, $f$, in Hz, and the left axis
 denotes the nondimensional amplitude of gravitational waves, $f
 \tilde{h} (f)$, observed at a hypothetical distance 100 Mpc from the
 binary along the {\it z} axis. The dashed curves are planned noise
 curves of the LCGT (``LCGT''), the Advanced LIGO optimized for $1.4
 M_\odot$ NS-NS detection (``Standard''), the Advanced LIGO optimized
 for the burst detection (``Broadband''), and the Einstein Telescope
 (``ET'') \cite{et2010}.} \label{fig:spec1}
\end{figure*}

\begin{figure}
 \includegraphics[width=88mm,clip]{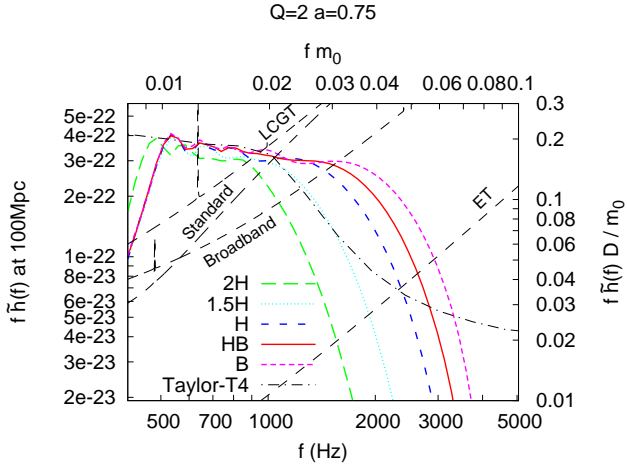} \caption{The
 same as Fig.~\ref{fig:spec1} but for $(Q,M_{\rm NS},a) = (2,1.35
 M_\odot,0.75)$ with 2H, H, HB, and B EOSs. The spectrum derived by the
 Taylor-T4 formula is also included.} \label{fig:spec2}
\end{figure}

To show the dependence of the gravitational-wave spectra on the BH spin
parameter, we plot the spectra for models HB-Q2M135a75, HB-Q2M135a5,
HB-Q2M135, and HB-Q2M135a-5 in the left panel of Fig.~\ref{fig:spec1}
and for models HB-Q3M135a75, HB-Q3M135a5, HB-Q3M135, and HB-Q3M135a-5 in
the right panel of Fig.~\ref{fig:spec1}. In the early inspiral phase of
$f \lesssim 1$ kHz, where the point-particle approximation works well,
the amplitude of the gravitational-wave spectrum for a given frequency
increases monotonically as $a$ increases. This is a feature expected
from the PN calculation and is explained by the spin-orbit interaction
as follows: The power spectrum of gravitational radiation is written as
\begin{equation}
 \frac{dE}{df} \propto [ f \tilde{h} (f) ]^2 .
\end{equation}
On the other hand, retaining only 1.5PN, the lowest-order spin-orbit
interaction terms, Eqs.~(4.10) and (4.14) of Ref.~\cite{kidder1995},
derive the expression for this quantity as
\begin{equation}
 \frac{dE}{df} = \frac{Q}{3 (1+Q)^2} \frac{X^{5/2}}{\pi f^2} \left[ 1 +
							      a X^{3/2}
							      \left\{
							       \frac{5 (
							       4Q + 3
							       )}{3
							       (1+Q)^2}
							      \right\}
							     \right]
 . \label{eq:dedf}
\end{equation}
Thus, the effective amplitude, $f \tilde{h} (f)$, for a given frequency
$f$ increases monotonically as the BH spin parameter, $a$, increases in
the inspiral phase.

Figure \ref{fig:spec2} plots the spectra for models 2H-Q2M135a75,
H-Q2M135a75, HB-Q2M135a75, and B-Q2M135a75, for which only the EOS is
different, and indicates that the amplitude in the early inspiral phase
does not depend strongly on ${\cal C}$. This is because the finite-size
effect of the NS does not play an important role in the early inspiral
phase (but, see Refs.~\cite{bdgnr2010,bdgnr2011}), as already found for
nonspinning BH-NS binaries~\cite{kst2010}.

In the late inspiral phase of 1~kHz $\lesssim f \lesssim f_{\rm cut}$,
where $f_{\rm cut}$ is a characteristic frequency at which the spectrum
starts damping exponentially (see below), the amplitude is significantly
larger than the Taylor-T4 formula for the cases in which the NS is not
disrupted. This is because the binaries in the inspiral and plunge after
the NS enters the BH's ISCO emit gravitational waves in reality, whereas
the Taylor-T4 formula does not take into account the motion inside the
ISCO. In contrast to the spectrum calculated by the Taylor-T4 formula,
which decreases steeply after the last inspiral phase, the amplitude
obtained from the simulation depends only weakly on the
gravitational-wave frequency in that phase, as far as the tidal
disruption does not occur.

The most fruitful information of the NS comes from the
gravitational-wave spectrum in the merger phase through the ``cutoff
frequency,'' $f_{\rm cut}$, which depends on the BH spin as well as the
NS compactness \cite{skyt2009,kst2010}. If the NS tidal disruption
occurs, the spectrum damps at $f = f_{\rm tidal} \sim 2$--4 kHz, which
denotes the frequency at the tidal disruption and depends sensitively on
physical parameters of the binary. In that case, gravitational waves for
a higher frequency, $f \gtrsim f_{\rm cut} \approx f_{\rm tidal}$, are
not emitted by the binary in the inspiral motion but only weakly by
disrupted material. Because the disrupted material gradually spreads
around the BH to form a nearly axisymmetric disk, the emission of
gravitational waves is suppressed at the high frequency. Thus, the
spectrum shows a relatively moderate damping around $f \approx f_{\rm
cut}$, which is closely related to the NS compactness through the tidal
disruption. The spectra for binaries with $(Q,a) = (2,\ge 0)$ and $(3,
\ge 0.5)$ in the left panel of Fig.~\ref{fig:spec1} correspond to these
cases. We see that the cutoff frequency, $f_{\rm cut}$, for these models
decreases as the BH spin parameter increases. This is ascribed to the
decrease of the orbital frequency at the tidal disruption for a binary
with the prograde BH spin. The enhancement of the effective centrifugal
force by the spin-orbit interaction reduces the orbital frequency at the
tidal disruption, $f_{\rm tidal}$, although the orbital separation at
the tidal disruption itself does not vary much even in the presence of
the BH spin. If the tidal disruption does not occur during the merger,
however, inspiral-like motion continues at higher frequencies near and
even inside the ISCO until the BH swallows the NS. In this case, the
spectrum amplitude depends only weakly on $f$ in the frequency range $f
\lesssim f_{\rm cut}$ and damps for $f \gtrsim f_{\rm cut}$, which is
closely related to the QNM frequency of the remnant BH, $f_{\rm
QNM}$. The spectra for $(Q,a) = (2,-0.5)$ and $(3, \le 0)$ in
Fig.~\ref{fig:spec1} show this feature. Note that the amplitude for
model HB-Q3M135a-5 is smaller than for model HB-Q3M135 for the frequency
range shown in Fig.~\ref{fig:spec1} because tidal disruption does not
play an important role, and Eq.~(\ref{eq:dedf}) applies throughout the
merger in both cases.

It is noteworthy that a prograde BH spin is favorable for the
gravitational-wave detection in the inspiral phase and the estimation of
$f_{\rm cut}$ in the merger phase because the prograde spin enhances the
amplitude for a given frequency in the inspiral phase and decreases the
cutoff frequency in the merger phase.  Note that the most sensitive
frequency range for ground-based detectors is $f \sim 10$ Hz--1 kHz,
which is usually lower than $f_{\rm cut}$. Thus, the features found here
are encouraging for the gravitational-wave astronomy to become an
important tool for investigating the NS radius and EOS.

\begin{figure*}
 \begin{tabular}{cc}
  \includegraphics[width=88mm,clip]{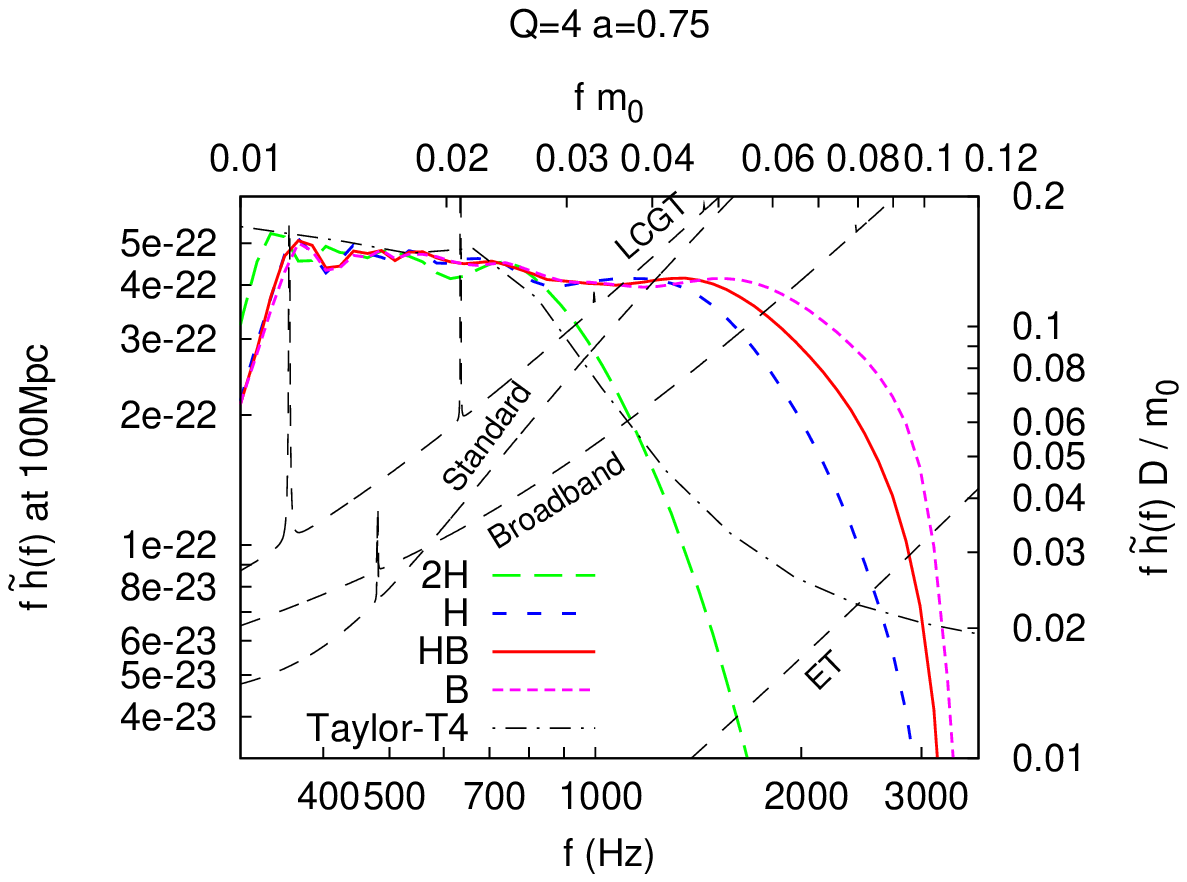} &
  \includegraphics[width=88mm,clip]{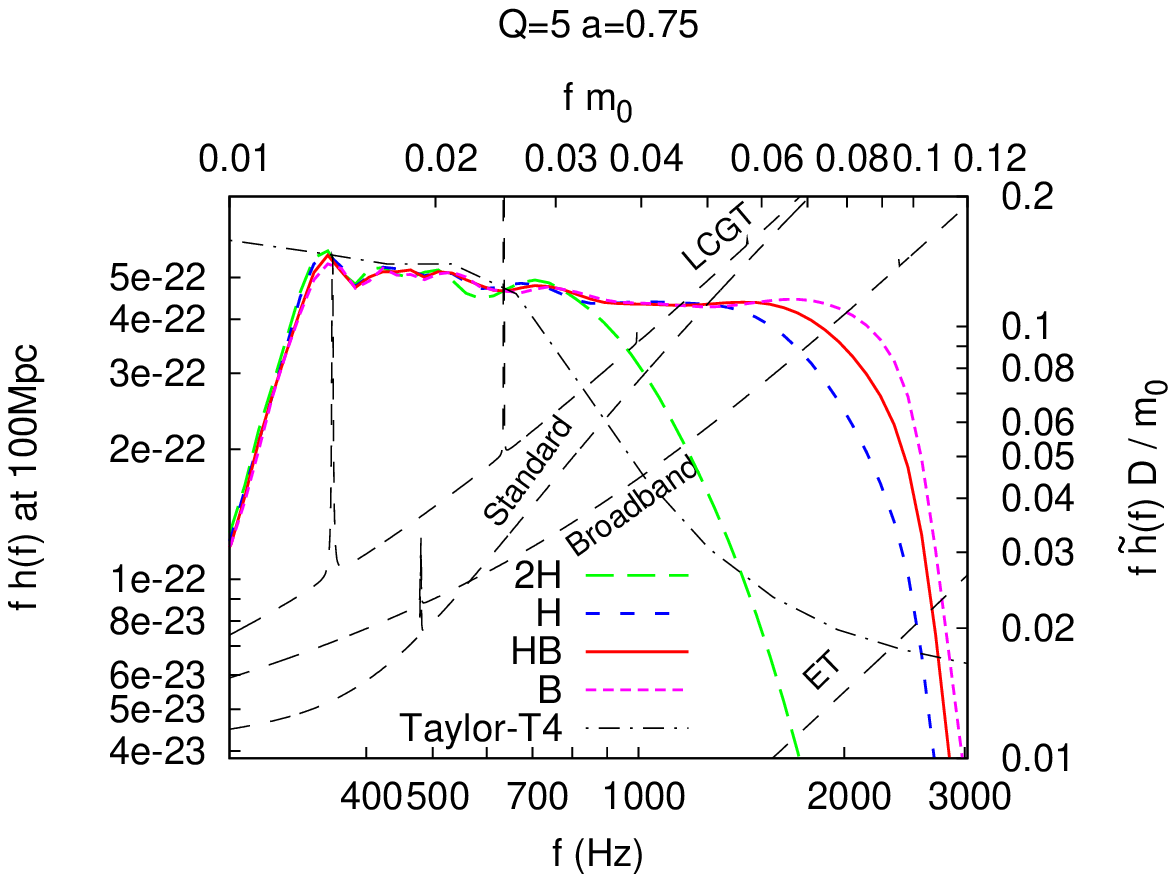}
 \end{tabular}
 \caption{The same as Fig.~\ref{fig:spec1} but with the left panel for
 $(Q,M_{\rm NS},a) = (4,1.35 M_\odot,0.75)$ and the right panel for
 $(5,1.35 M_\odot,0.75)$ with 2H, H, HB, and B EOSs. The spectrum
 derived by the Taylor-T4 formula is also included.}  \label{fig:spec3}
\end{figure*}

\begin{figure}
 \includegraphics[width=80mm,clip]{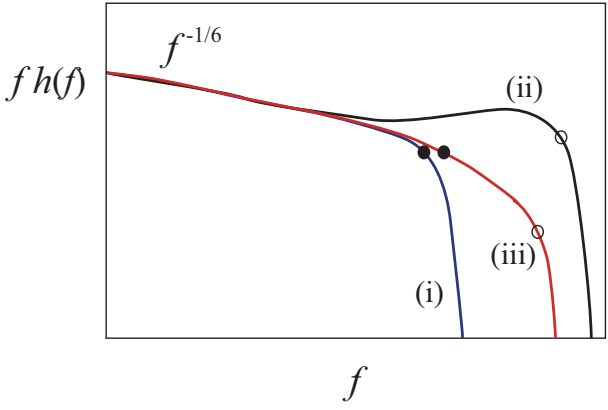} \caption{A schematic
 figure of three types of gravitational-wave spectra. Spectrum (i) is
 for the case in which tidal disruption occurs far outside the ISCO, and
 spectrum (ii) is for the case in which tidal disruption does not
 occur. Spectrum (iii) is for the case in which tidal disruption occurs
 and the QNM is also excited. The filled and open circles denote $f_{\rm
 tidal}$ and $f_{\rm QNM}$, respectively.} \label{fig:scheme2}
\end{figure}

The gravitational-wave spectra of binaries with high mass ratios show
qualitatively different behavior for a high frequency. Figure
\ref{fig:spec3} plots the gravitational-wave spectra obtained for models
with $(Q,M_{\rm NS},a) = (4,1.35 M_\odot,0.75)$ and $(5,1.35
M_\odot,0.75)$. For these binaries (except for the model with 2H EOS),
both the NS tidal disruption and the excitation of a QNM of the remnant
BH occur as is described in Sec.~\ref{subsec:res_waveform}. Hence, the
gravitational spectrum has two characteristic frequencies, i.e., $f_{\rm
tidal}$ and $f_{\rm QNM}$, simultaneously. The spectra plotted in
Fig.~\ref{fig:spec3} indeed show such features. After the NS is tidally
disrupted, the amplitude of the gravitational-wave spectrum shows a slow
damp for $f \gtrsim f_{\rm tidal} \approx 2$ kHz.  Then, the spectrum
damps steeply above the frequency of the QNM, $f \gtrsim f_{\rm QNM}
\approx 3$~kHz. A schematic figure of different spectra is depicted in
Fig.~\ref{fig:scheme2}, and the spectrum described in this paragraph
corresponds to spectrum (iii) in this figure. This suggests that the
cutoff frequency, $f_{\rm cut}$, of a high mass-ratio binary is not
determined by a unique physical process like NS tidal disruption or a
ringdown of a remnant BH, as far as both of them occur.

\begin{figure}
 \includegraphics[width=80mm,clip]{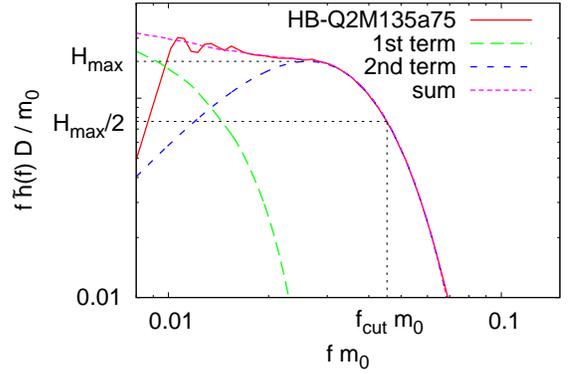} \caption{The fitting for
 model HB-Q2M135a75. The long-dashed and middle-dashed curves show the
 first and second terms of Eq.~(\ref{eq:fit}), respectively. The
 short-dashed curve is the sum of these two terms.}  \label{fig:deffcut}
\end{figure}

To estimate the cutoff frequency quantitatively, we fit the
gravitational-wave spectra by a function with seven free parameters of
the form
\begin{eqnarray}
 \frac{f \tilde{h}_{\rm fit} (f) D}{m_0} &=& \frac{f \tilde{h}_{\rm
  3PN} (f) D}{m_0} e^{-(f / f_{\rm ins})^{\sigma_{\rm ins}}}
  \nonumber \\
 &+& A e^{-(f / f_{\rm dam})^{\sigma_{\rm dam}}} [ 1 - e^{-(f / f_{\rm
  ins2})^{\sigma_{\rm ins2}}} ] , \label{eq:fit}
\end{eqnarray}
where $\tilde{h}_{\rm 3PN} (f)$ is the Fourier spectrum calculated by
the Taylor-T4 formula. The first term in Eq.~(\ref{eq:fit}) models the
inspiral spectrum, and the second term models the merger and ringdown
spectra. We determine seven free parameters $f_{\rm ins}$, $f_{\rm
ins2}$, $f_{\rm dam}$, $\sigma_{\rm ins}$, $\sigma_{\rm ins2}$,
$\sigma_{\rm dam}$, and $A$ by the condition that the following weighted
norm is minimized:
\begin{equation}
 \sum_i \left\{ [ f_i \tilde{h}_i (f_i) - f_i \tilde{h}_{\rm fit} (f_i)
	 ] f_i^{1/3} \right\}^2 .
\end{equation}
Here, {\it i} denotes the data point for the spectrum. In the previous
works \cite{skyt2009,kst2010}, we identify $f_{\rm dam}$ in
Eq.~(\ref{eq:fit}) with $f_{\rm cut}$, which is most strongly correlated
with the NS compactness for nonspinning BH-NS binaries \footnote{We
refer to $f_{\rm dam}$ as $f_{\rm cut}$ throughout in the previous work
\cite{kst2010}. In the present paper, we distinguish $f_{\rm dam}$ from
$f_{\rm cut}$ because the method for determining $f_{\rm dam}$ is
different from that for $f_{\rm cut}$}. In the present work, however, we
obtain no strong correlation between $f_{\rm dam}$ (and the other
parameters) and any parameter of physical importance, such as $a$ or
${\cal C}$. The reason may be ascribed to the inadequacy of the
functional form of Eq.~(\ref{eq:fit}), where the set of free parameters
is degenerate to some extent. In particular, such a degeneracy is severe
for a high mass-ratio binary due to two reasons. First, modeling an
inspiral spectrum by the Taylor-T4 formula is inadequate for the late
inspiral phase of a high mass-ratio binary due to the lack of
information from the Taylor-T4 formula, as is described in
Sec.~\ref{subsec:res_waveform}. Second, there is no unique, physically
motivated identification of $f_{\rm cut}$ when both the NS tidal
disruption and the QNM excitation occur. (Fortunately, these
degeneracies did not cause problems in the case of the nonspinning BH-NS
binary with a low mass ratio \cite{kst2010}.) To overcome these problems
with the fitting procedure, we redefine $f_{\rm cut}$ as the higher one
of two frequencies at which the second term in Eq.~(\ref{eq:fit}) takes
a half value of its maximum. An example of this fitting procedure is
shown in Fig.~\ref{fig:deffcut}. In this figure, $H_{\rm max}$
corresponds to the maximum value of the second term in
Eq.~(\ref{eq:fit}). We find that this definition of $f_{\rm cut}$ works
well to read off the NS compactness from the gravitational-wave
spectrum.

\begin{figure}
 \includegraphics[width=88mm,clip]{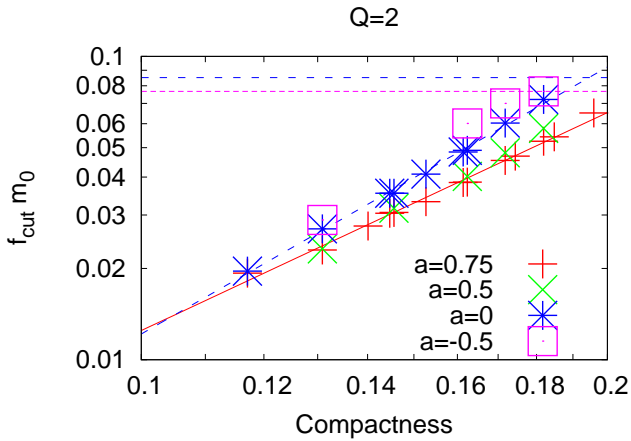} \caption{The cutoff
 frequency times the total mass $f_{\rm cut} m_0$ as a function of the
 NS compactness ${\cal C}$ for $Q=2$ binaries in logarithmic scales. The
 solid and dashed lines are obtained by linear fittings of data for
 $a=0.75$ and $a=0$, respectively. Horizontal lines denote the typical
 QNM frequencies of the remnant BHs. We note that $f_{\rm QNM} > f_{\rm
 cut}$ for $a=0.75$ as long as ${\cal C} \leq 0.2$. } \label{fig:fcut1}
\end{figure}

Figure \ref{fig:fcut1} shows $f_{\rm cut} m_0$ for spectra obtained for
all binaries with $Q=2$ as a function of the NS compactness, ${\cal C}$,
in logarithmic scales. We also plot the typical QNM frequency of the
remnant BH, $f_{\rm QNM}$, which depends primarily on $a$ of the initial
BH for a fixed value of $Q$. For each value of $a$, we find that $f_{\rm
cut} m_0$ increases monotonically as ${\cal C}$ increases, and an
approximate power law holds as
\begin{equation}
 \ln ( f_{\rm cut} m_0 ) \approx p(a) \ln {\cal C} + q(a) \; \; \; (Q=2)
  ,
\end{equation}
where $p(a) (>0)$ and $q(a)$ depend only on $a$, for any value of $a$
when $Q=2$. This monotonic relation between $f_{\rm cut} m_0$ and ${\cal
C}$ suggests us a possibility to extract the compactness, ${\cal C}$, of
a NS from the gravitational-wave observation. It is noteworthy that this
relation includes only ${\cal C}$ but neither $M_{\rm NS}$ nor $R_{\rm
NS}$ independently. It should also be noted that the simple relation
found here is a consequence of our choice for a common value of the
adiabatic index of the core EOS, $\Gamma_2 = 3$ \cite{kst2010}. The
increase of $f_{\rm cut} m_0$ with the increase of ${\cal C}$ indicates
that a more compact NS is less subject to the BH tidal effect and
disrupted at a closer orbit to the BH than a less compact NS is. The
difference in $f_{\rm cut} m_0$ due to the difference in $a$ becomes
clearer for larger values of ${\cal C}$, and conversely, $f_{\rm cut}
m_0$ depends only weakly on $a$ if the compactness is as small as
$\approx 0.12$. The weak dependence on $a$ for the small values of
${\cal C}$ is due to the fact that the effect of the BH spin at a
distant orbit, at which the NS with a large radius is disrupted, is
weak.

Figure \ref{fig:fcut1} also shows that $p(a)$ is a decreasing function
of $a$. More specifically we obtain the relations
\begin{equation}
 \ln ( f_{\rm cut} m_0 ) = ( 2.92 \pm 0.06 ) \ln {\cal C} + ( 2.32 \pm
  0.12 )
\end{equation}
for $a=0$ \footnote{The relation between $f_{\rm cut} m_0$ and ${\cal
C}$ is different from the one obtained in Ref.~\cite{kst2010} due to the
different definition of $f_{\rm cut}$.} and
\begin{equation}
 \ln ( f_{\rm cut} m_0 ) = ( 2.39 \pm 0.06 ) \ln {\cal C} + ( 1.11 \pm 0.11 )
\end{equation}
for $a=0.75$ by a linear fitting. The decreasing nature of $p(a)$ is
explained by the fact that the spin-orbit repulsive force for the
prograde BH spin, which reduces the orbital frequency at the NS tidal
disruption, works efficiently for a close orbit and hence for the NS
with a large value of ${\cal C}$. It is important that $p(a)$ is always
larger than 1.5, which is expected from the analysis of the condition
for the mass shedding,
\begin{equation}
 \Omega m_0 \propto \frac{{\cal C}^{3/2} (1+Q)^{3/2}}{\sqrt{Q}} .
\end{equation}
The large value of $p(a)$ is favorable for determining the NS
compactness from the gravitational-wave observation because the
dependence of $f_{\rm cut}$ on ${\cal C}$ becomes stronger. Note that
$f_{\rm cut}$ is always lower than the QNM frequency of the remnant BH
for a realistic range of the compactness ${\cal C} \lesssim 0.2$ for
$(Q,a) = (2, \gtrsim 0)$. If $a$ is negative, on the other hand, $f_{\rm
cut}$ for the binary with a compact NS of ${\cal C} \gtrsim 0.18$ may be
determined by the QNM frequency, $f_{\rm QNM}$, and it will be difficult
to determine the NS compactness from the cutoff frequency.

\begin{figure*}
 \begin{tabular}{cc}
  \includegraphics[width=88mm,clip]{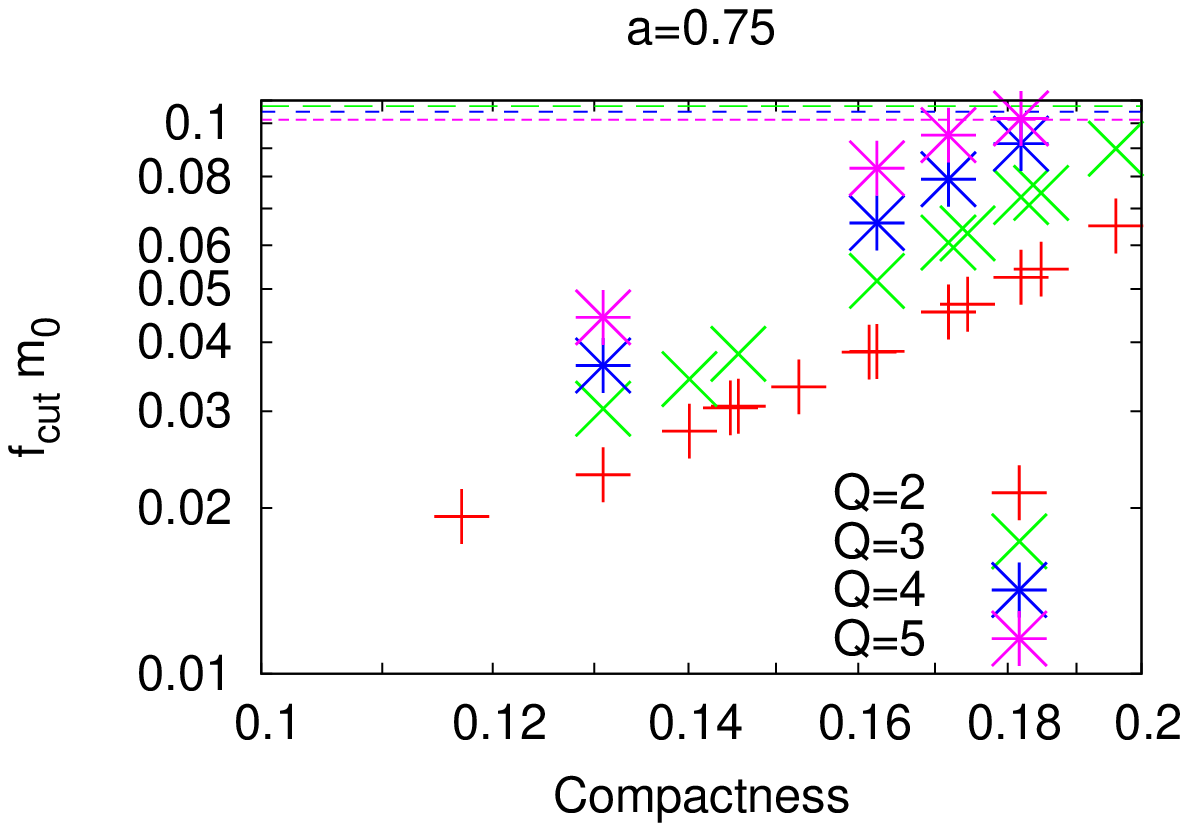} &
  \includegraphics[width=88mm,clip]{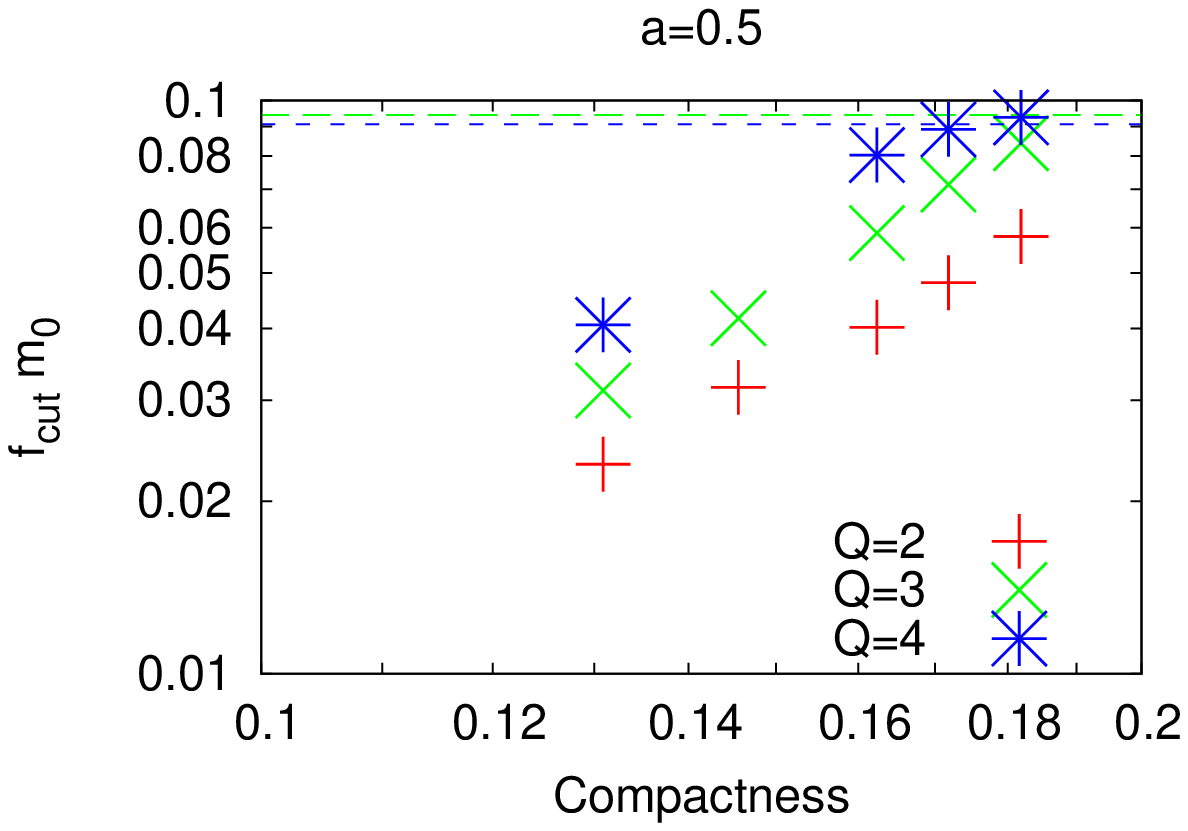} \\
 \end{tabular}
 \caption{The same as Fig.~\ref{fig:fcut1} but for $a=0.75$ (left) and
 0.5 (right). In both panels, $f_{\rm QNM} > f_{\rm cut}$ for $Q=2$.}
 \label{fig:fcut2}
\end{figure*}

Figure \ref{fig:fcut2} shows the $f_{\rm cut} m_0$--${\cal C}$ relation
of gravitational-wave spectra obtained for all binaries with $a=0.75$
and $a=0.5$. This figure, combined with Fig.~\ref{fig:fcut1}, clearly
indicates that the approximate power law of the form
\begin{equation}
 \ln ( f_{\rm cut} m_0 ) = p (Q,a) \ln {\cal C} + q (Q,a)
\end{equation}
holds for binaries of ${\cal C} \lesssim 0.2$ with $Q=5$ as far as $a
\sim 0.75$ and with $Q \le 4$ as far as $a \sim 0.5$. The striking
feature is that the cutoff frequency is lower than the QNM frequency of
the remnant BH, $f_{\rm QNM}$, for $(Q,a) = (\le 4, 0.75)$ and for
$(Q,a) = (\le 3, 0.5)$ even if a QNM is excited. Accordingly, $f_{\rm
cut}$ shows a strong correlation with ${\cal C}$. For $(Q,a) = (5,0.75)$
and (4,0.5), $f_{\rm cut}$ is lower than $f_{\rm QNM}$ as far as ${\cal
C} \lesssim 0.18$ and 0.17, respectively, and, therefore, the strong
correlation between $f_{\rm cut}$ and ${\cal C}$ is found within this
range. Although $f_{\rm cut}$ for the binary with a high mass ratio
should not be considered as $f_{\rm tidal}$ due to the QNM excitation,
monotonic relations between $f_{\rm cut} m_0$ and ${\cal C}$ gives us an
opportunity to explore the NS radius and EOS.  It should be noted that
gravitational waves from a higher mass-ratio binary are more subject to
the gravitational-wave detection due to the larger amplitude in the
inspiral phase and the lower cutoff frequency. We again note that a
massive BH of $M_{\rm BH} \gtrsim 5 M_\odot$ is an astrophysically
realistic consequence of the stellar evolution
\cite{mcclintockremillard,bbfrvvh2010}. Taking these facts into account,
we conclude that gravitational waves from the BH-NS binary are a
promising tool to investigate the NS radius and EOS in the next decade.

\subsection{Energy and angular momentum radiated by gravitational waves}
\label{subsec:res_energy}

Table \ref{table:gw} lists the total energy $\Delta E / M_0$ and angular
momentum $\Delta J / J_0$ radiated by gravitational waves. We estimate
systematic errors in the estimation of $\Delta E$ and $\Delta J$ to be
$\sim 10 \%$, which are ascribed mainly to the finite grid resolution
and partly to the finite extraction radius for $\Psi_4$. Because $\Delta
E$ and $\Delta J$ depend on the choice of $\Omega_0 m_0$, we do not
compare directly the results obtained for models with different values
of $\Omega_0 m_0$ and accordingly models with different values of $Q$
(see Table \ref{table:model}).

Contributions from all the $l = 2$--4 modes of gravitational waves are
taken into account. The $(l,|m|) = (2,2)$ mode always contributes by
$\gtrsim 85 \%$ to $\Delta E$ and $\Delta J$. $\Delta E$ and $\Delta J$
taken away by higher-mode gravitational waves are substantial for high
mass-ratio binaries. For example, the $(3,3)$ mode contributes by $\sim
2$, 5, 7.5, and 10\% for binaries with $Q=2$, 3, 4, and 5,
respectively. The $(4,4)$ mode gravitational waves contribute by $1 \sim
2 \%$ for binaries with $Q=3$--5. These values depend only weakly on $a$
and the EOS, and contributions of modes with $l \neq m$ are negligible
compared to those of $l=m$ modes.

Table \ref{table:gw} shows that $\Delta E / M_0$ and $\Delta J / J_0$
increase monotonically as the NS compactness, ${\cal C}$, increases for
binaries with fixed values of $(Q,a)$. This is the same result as that
obtained for nonspinning BH-NS binaries \cite{kst2010} and is explained
by a longer inspiral phase for a softer EOS due to the later onset of
mass shedding and the later tidal disruption. The ratio between these
two values, $( \Delta J / J_0 ) / ( \Delta E / M_0 )$, decreases as
${\cal C}$ increases. This agrees again with the result for the
nonspinning BH cases and is explained by a relation $\Delta J / \Delta E
\approx m / \Omega$ for a given angular harmonic of $m$ and by the fact
that more radiation is emitted from the orbit of a larger value of
$\Omega$ for a softer EOS. Note that these arguments are based on little
dependence of gravitational-wave luminosity in the inspiral phase on
${\cal C}$ for a fixed value of $a$; tidal correction to the luminosity
in the inspiral phase is not important.

Table \ref{table:gw} shows that $\Delta E / M_0$ does not depend
strongly on $a$, while $\Delta J / J_0$ increases as $a$ increases in
many cases for a fixed value of ${\cal C}$. Remember that $dE/df$ in the
inspiral phase increases for a large value of $a$, as is given by
Eq.~(\ref{eq:dedf}). However, the orbital frequency, $\Omega$, at the
tidal disruption decreases for a large value of $a$ due to the
spin-orbit interaction. Because of these two competing effects, the
binding energy at the tidal disruption depends only weakly on $a$, and
hence, $\Delta E / M_0$ does not change very much among different values
of $a$. The increase of $\Delta J / J_0$ for a large value of $a$ is due
to the large value of $dE/df$ in the inspiral phase, during which
$\Omega$ is relatively low, and to the approximate relation $\Delta J
\approx m \Delta E / \Omega$, which enhances the contribution of
low-frequency gravitational waves.

Finally, we comment on the linear momentum $\Delta P$ radiated by
gravitational waves and an associated kick velocity $v_{\rm kick} \equiv
\Delta P / M_0$ of the remnant BH. Because of the mass and spin
asymmetries, the remnant BH achieves the kick velocity of $\sim
100$--250 km/s when the effect of tidal disruption is weak, e.g., for
models HB-Q3M135a-5 and B-Q3M135. Although our results for $\Delta P$ do
not converge as well as those for $\Delta E$ and $\Delta J$ due to the
slow convergence of $(l,m) \neq (2,2)$ mode gravitational waves, the
values of $v_{\rm kick}$ are in reasonable agreements with the fitting
formula derived using the results of simulations for the binary BH
merger \cite{rezzolla2009,zcl2011}. By contrast, $v_{\rm kick}$ is
suppressed to $\lesssim 100$ km/s when tidal disruption occurs far
outside the ISCO. The reason for this is that the tidal disruption
suppresses significantly the gravitational radiation from the last
inspiral and merger phases, during which the linear momentum is emitted
most efficiently. This trend is consistent with the result found in our
previous work \cite{skyt2009}.

\begin{table}
 \caption{Total radiated energy $\Delta E$ and angular momentum $\Delta
 J$ carried away by gravitational waves. $\Delta E$ and $\Delta J$ are
 normalized with respect to the initial ADM mass $M_0$ and angular
 momentum $J_0$, respectively. We also show the ratio between $\Delta J$
 and $\Delta E$.}
 \begin{tabular}{cccc}
  \hline
  Model & $ \Delta E / M_0 $ (\%) & $ \Delta J / J_0 $ (\%) & $( \Delta
  J / J_0 ) / ( \Delta E / M_0 )$\\
  \hline \hline
  2H-Q2M135a75 & 0.58 & 16 & 27 \\
  1.5H-Q2M135a75 & 0.79 & 19 & 24 \\
  H-Q2M135a75 & 1.1 & 24 & 21 \\
  HB-Q2M135a75 & 1.4 & 26 & 19 \\
  B-Q2M135a75 & 1.7 & 29 & 17 \\
  \hline
  2H-Q2M135a5 & 0.60 & 17 & 26 \\
  1.5H-Q2M135a5 & 0.79 & 19 & 24 \\
  H-Q2M135a5 & 1.2 & 24 & 20 \\
  HB-Q2M135a5 & 1.4 & 26 & 19 \\
  B-Q2M135a5 & 1.7 & 28 & 17 \\
  \hline
  2H-Q2M135a-5 & 0.57 & 15 & 26 \\
  H-Q2M135a-5 & 1.1 & 19 & 16 \\
  HB-Q2M135a-5 & 1.4 & 19 & 14 \\
  B-Q2M135a-5 & 1.6 & 21 & 13 \\
  \hline
  2H-Q2M12a75 & 0.40 & 12 & 30 \\
  H-Q2M12a75 & 0.79 & 19 & 24 \\
  HB-Q2M12a75 & 0.95 & 21 & 22 \\
  B-Q2M12a75 & 1.2 & 24 & 21 \\
  \hline
  2H-Q2M145a75 & 0.73 & 19 & 25 \\
  H-Q2M145a75 & 1.5 & 27 & 19 \\
  HB-Q2M145a75 & 1.7 & 30 & 17 \\
  B-Q2M145a75 & 2.1 & 32 & 15 \\
  \hline
  2H-Q3M135a75 & 0.72 & 20 & 28 \\
  1.5H-Q3M135a75 & 0.97 & 23 & 24 \\
  H-Q3M135a75 & 1.3 & 27 & 20 \\
  HB-Q3M135a75 & 1.6 & 30 & 19 \\
  B-Q3M135a75 & 2.0 & 34 & 17 \\
  \hline
  2H-Q3M135a5 & 0.70 & 19 & 27 \\
  1.5H-Q3M135a5 & 0.94 & 22 & 23 \\
  H-Q3M135a5 & 1.4 & 26 & 19 \\
  HB-Q3M135a5 & 1.7 & 29 & 17 \\
  B-Q3M135a5 & 2.0 & 31 & 15 \\
  \hline
  HB-Q3M135a-5 & 1.3 & 19 & 14 \\
  \hline
  2H-Q3M145a75 & 0.88 & 22 & 25 \\
  H-Q3M145a75 & 1.7 & 31 & 18 \\
  HB-Q3M145a75 & 2.1 & 34 & 16 \\
  B-Q3M145a75 & 2.5 & 37 & 15 \\
  \hline
  2H-Q4M135a75 & 0.81 & 23 & 28 \\
  H-Q4M135a75 & 1.5 & 31 & 21 \\
  HB-Q4M135a75 & 1.8 & 33 & 19 \\
  B-Q4M135a75 & 2.1 & 36 & 17 \\
  \hline
  2H-Q4M135a5 & 0.72 & 19 & 27 \\
  H-Q4M135a5 & 1.5 & 27 & 19 \\
  HB-Q4M135a5 & 1.7 & 29 & 17 \\
  B-Q4M135a5 & 1.9 & 31 & 16 \\
  \hline
  2H-Q5M135a75 & 0.83 & 24 & 29 \\
  H-Q5M135a75 & 1.6 & 33 & 20 \\
  HB-Q5M135a75 & 1.9 & 35 & 19 \\
  B-Q5M135a75 & 2.1 & 36 & 17 \\
  \hline
 \end{tabular}
 \label{table:gw}
\end{table}

\section{Summary} \label{sec:summary}

We performed numerical simulations for the merger of BH-NS binaries with
various BH spins aligned or antialigned with the orbital angular
momentum, using an AMR code {\tt SACRA} with systematically chosen five
piecewise polytropic EOSs. We investigated the dependence of the merger
process, properties and structures of the remnant disk, properties of
the remnant BH, gravitational waveforms, and their spectra on the spin
of the BH and the EOS of the NS. In particular, we focused on the case
in which the BH has a prograde spin, and the tidal disruption of the NS
by a companion BH plays an important role. We adopted a number of
parameters for the mass ratio, NS mass, and BH spin. By preparing the
initial condition with a distant orbit and a small eccentricity, we
always tracked $\gtrsim 5$ quasicircular orbits in the inspiral phase
and studied the merger phase in a realistic setting. The treatment of
hydrodynamic equations and the estimation method of the disk mass are
improved in this work. In the following, we summarize the conclusions in
this paper:

\begin{enumerate}
 \item It is shown that a prograde BH spin enhances the effect of NS
       tidal disruption by the spin-orbit interaction. The mass of the
       remnant disk increases as the BH spin increases because the ISCO
       radius of the BH becomes small. A remarkable point is that the
       BH-NS binary with a high mass ratio of even $Q=5$ can form a
       sufficiently massive disk of $\gtrsim 0.1 M_\odot$ for a wide
       range of the NS compactness if the BH has a prograde spin of
       $a=0.75$. This amount of the disk mass for a high mass-ratio
       binary is hardly expected if the BH is nonspinning. This fact
       suggests that the formation of a BH-massive accretion disk system
       is a frequent outcome of the BH-NS binary merger with a prograde
       BH spin and may be encouraging for the merger scenario of a
       short-hard GRB. By contrast, the disk mass becomes very small if
       the BH has a retrograde spin.

 \item It is shown that some portion of the disrupted material can
       extend to $\gtrsim 400$ km from the BH if the massive disk is
       formed. The maximum rest-mass density in the disk is larger for
       binaries with smaller values of $Q$ because the ISCO radius is
       smaller for them. The extent of the disk could be large for a
       large value of $Q$. For such a remnant disk, the lifetime should
       be longer.

 \item The spin parameter of the remnant BH depends primarily on the
       spin parameter of the initial BH, $a$, and the mass ratio,
       $Q$. In particular, extrapolation of our results suggests that
       the merger of an extremely spinning BH and an irrotational NS
       does not form an overspinning BH.

 \item The gravitational waveform also depends strongly on the BH
       spin. The number of gravitational-wave cycles becomes larger for
       a prograde BH spin than that for a zero BH spin in the inspiral
       phase because an additional repulsive force due to the spin-orbit
       interaction reduces gravitational-wave luminosity and an
       approaching velocity of the binary. We found that the Taylor-T4
       formula does not reproduce the phase evolution in the late
       inspiral phases accurately, especially when the mass ratio is
       large.

 \item In our previous work for nonspinning BH-NS binaries, the
       waveforms are classified simply into two categories: when tidal
       disruption of the NS occurs, the waveform is composed of an
       inspiral waveform and a prompt shutdown at the tidal
       disruption. When tidal disruption does not occur, the waveform is
       composed of inspiral and QNM waveforms. However, we find that the
       NS tidal disruption and the excitation of a QNM can occur
       simultaneously for binaries with a high mass ratio and a prograde
       BH spin. This is because the disrupted material cannot become
       axisymmetric before the prompt infall due to a larger BH areal
       radius for a larger value of $Q$. As a result, the material
       accretes onto the remnant BH coherently, and, therefore, the QNM
       of the remnant BH is excited, except for the case in which the
       extremely stiff EOS is adopted.

 \item The cutoff frequency of the gravitational-wave spectrum is
       correlated with the NS compactness in a clear manner when the NS
       is disrupted, and the BH spin modifies this correlation. The
       prograde BH spin decreases the cutoff frequency for fixed values
       of ${\cal C}$ and $Q$ because the angular velocity at the tidal
       disruption becomes smaller than that for $a=0$. The cutoff
       frequency is lower for a smaller value of ${\cal C}$ for fixed
       values of $Q$ and $a$, as in the case of nonspinning BH-NS
       binaries, because the tidal effect is stronger and the disruption
       occurs at a more distant orbit. The BH spin also modifies the
       spectrum for the inspiral phase. Specifically, the spectrum
       amplitude for a given frequency in the inspiral phase becomes
       large when the BH has a prograde spin, and this is consistent
       with the PN estimation. Both the low cutoff frequency and large
       spectrum amplitude in the inspiral phase for a prograde BH spin
       are encouraging for gravitational-wave astronomy to become a tool
       to investigate the NS compactness and EOS. It is noteworthy that
       the BH-NS binary with a high mass ratio of $Q \gtrsim 5$ is a
       more promising target for ground-based gravitational-wave
       detectors if the BH has a prograde spin and the NS tidal
       disruption occurs.
\end{enumerate}

Finally, we list several issues to be explored in the future. Piecewise
polytropic EOSs with two pieces employed in this paper are not accurate
enough to model high-mass NSs with large central density of $\rho_{\rm
max} \gtrsim 10^{15} {\rm g/cm}^3$ \cite{rlof2009}. More detailed
(piecewise polytropic) EOSs are necessary to calculate gravitational
waves from a BH-relatively massive NS binary merger for which the tidal
deformation and disruption of the NS plays an important role, i.e.,
BH-NS binaries with moderately large BH spins of $a \gtrsim 0.75$.  The
implementation of detailed microphysics, such as a finite-temperature
effect and a neutrino transport process, is essential even qualitatively
to explore the evolution of the remnant BH-accretion disk system and to
discuss the jet launch such that short-hard GRBs require. Recently,
Sekiguchi developed a code to perform fully general relativistic
simulation with the finite-temperature EOS and an approximate neutrino
emission (the so-called leakage scheme) and succeed in simulating the
stellar core collapse \cite{sekiguchi2010,sekiguchishibata2010} and the
merger of binary NSs \cite{skks2011}. We plan to work on BH-NS binary
mergers along these lines.

\begin{acknowledgments}
 K. K. is deeply grateful to Eric Gourgoulhon, Nicolas Vasset, and
 Benjamin D. Lackey for valuable discussion. We also thank Tetsuro
 Yamamoto for developing {\tt SACRA}. Numerical computations of
 quasiequilibrium states were performed using the free library LORENE
 \cite{LORENE}. We thank members of the Meudon Relativity Group for
 developing LORENE. This work was supported by the Grant-in-Aid for
 Scientific Research (21340051), by the Grant-in-Aid for Scientific
 Research on Innovative Area (20105004), by the Grant-in-Aid for the
 Global COE Program ``The Next Generation of Physics, Spun from
 Universality and Emergence,'' by the HPCI strategic program of Japanese
 MEXT, by the Grant-in-Aid for Research Activity Start-up of JSPS
 (22840010), and by the Grant-in-Aid of JSPS.
\end{acknowledgments}

\appendix*

\section{Convergence of gravitational waves and the remnant disk mass}
\label{app_conv}

\begin{figure*}[tbp]
 \begin{tabular}{cc}
  \includegraphics[width=85mm,clip]{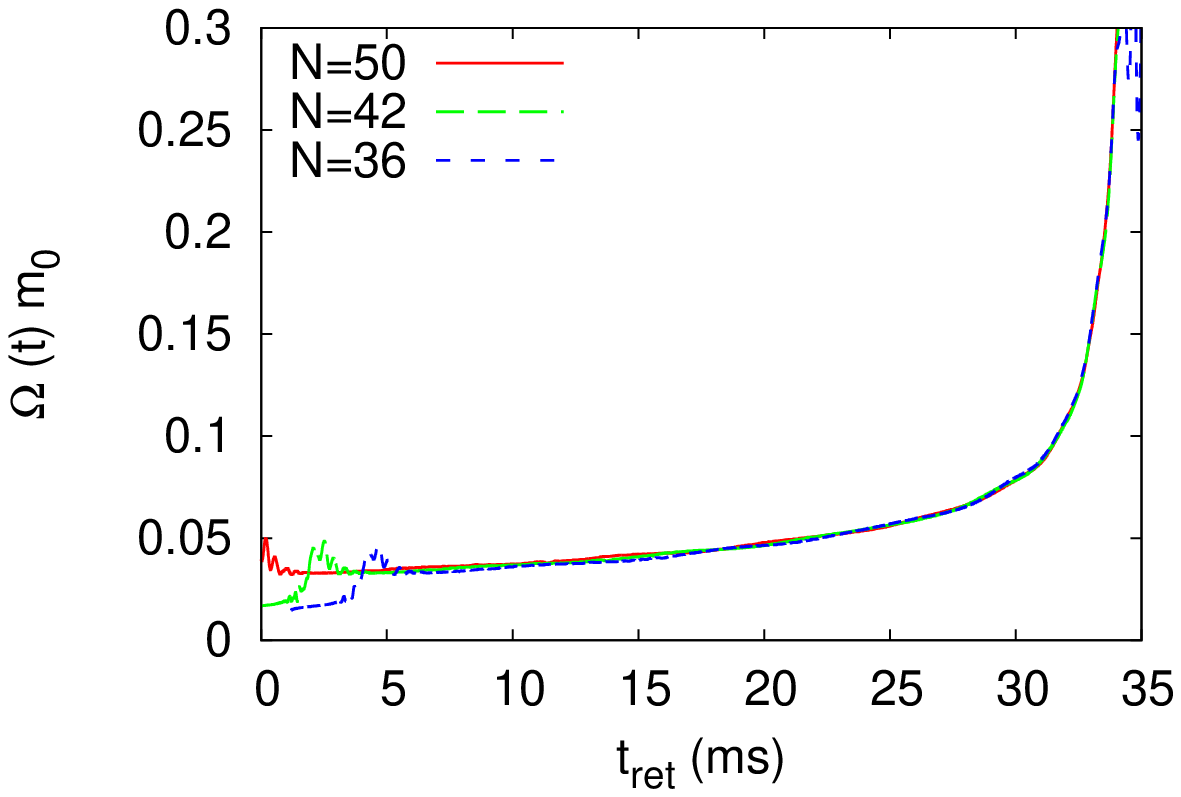} & 
  \includegraphics[width=85mm,clip]{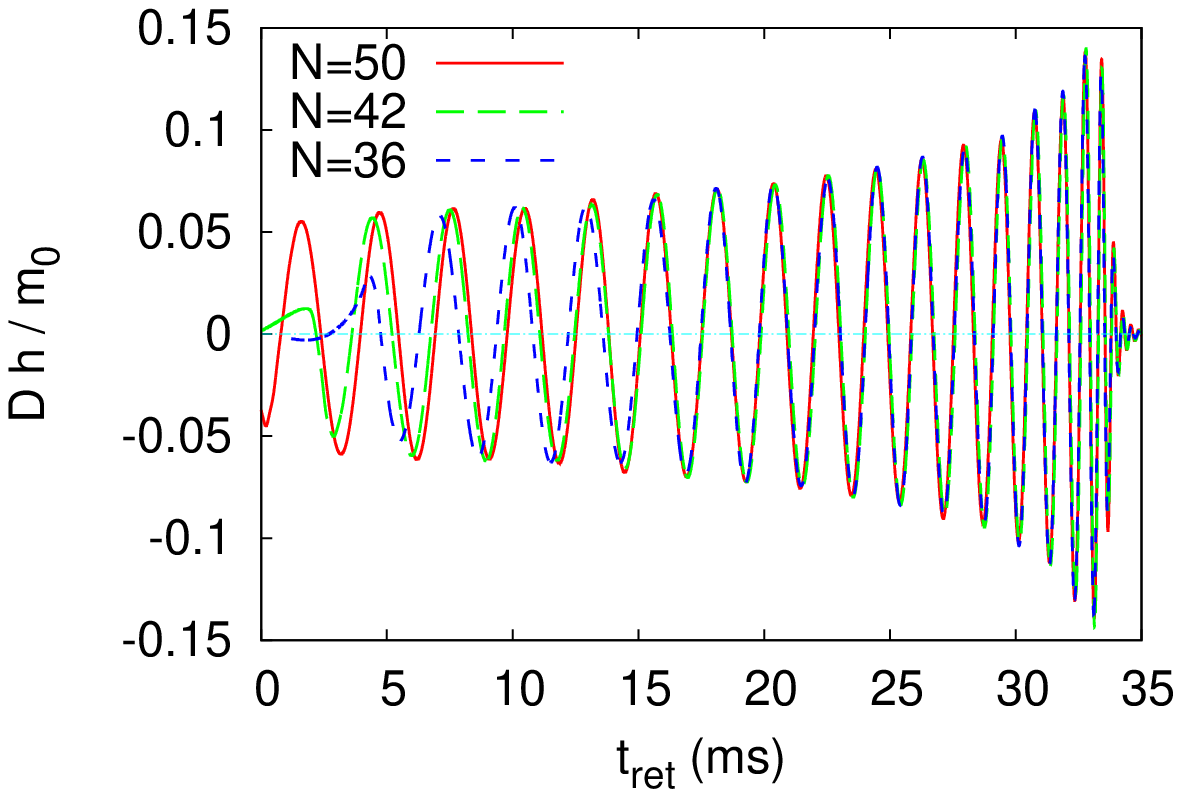} \\
  \includegraphics[width=85mm,clip]{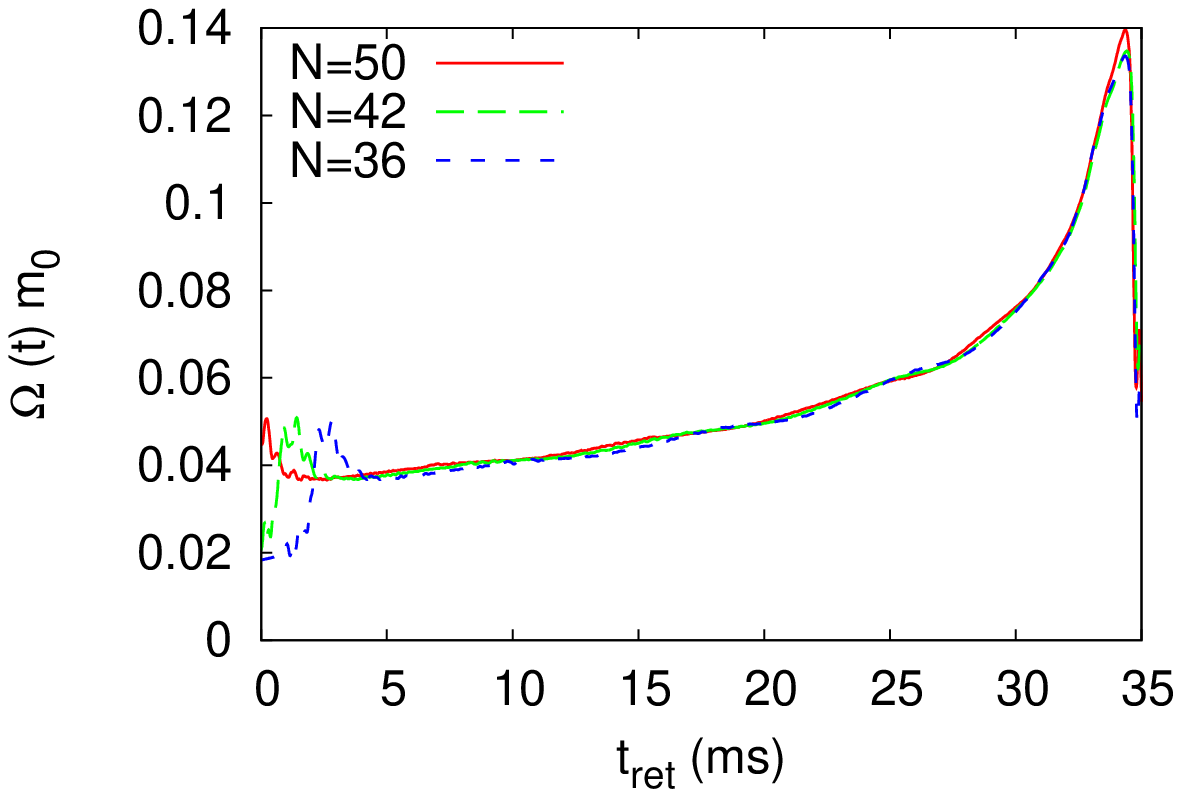} &
      \includegraphics[width=85mm,clip]{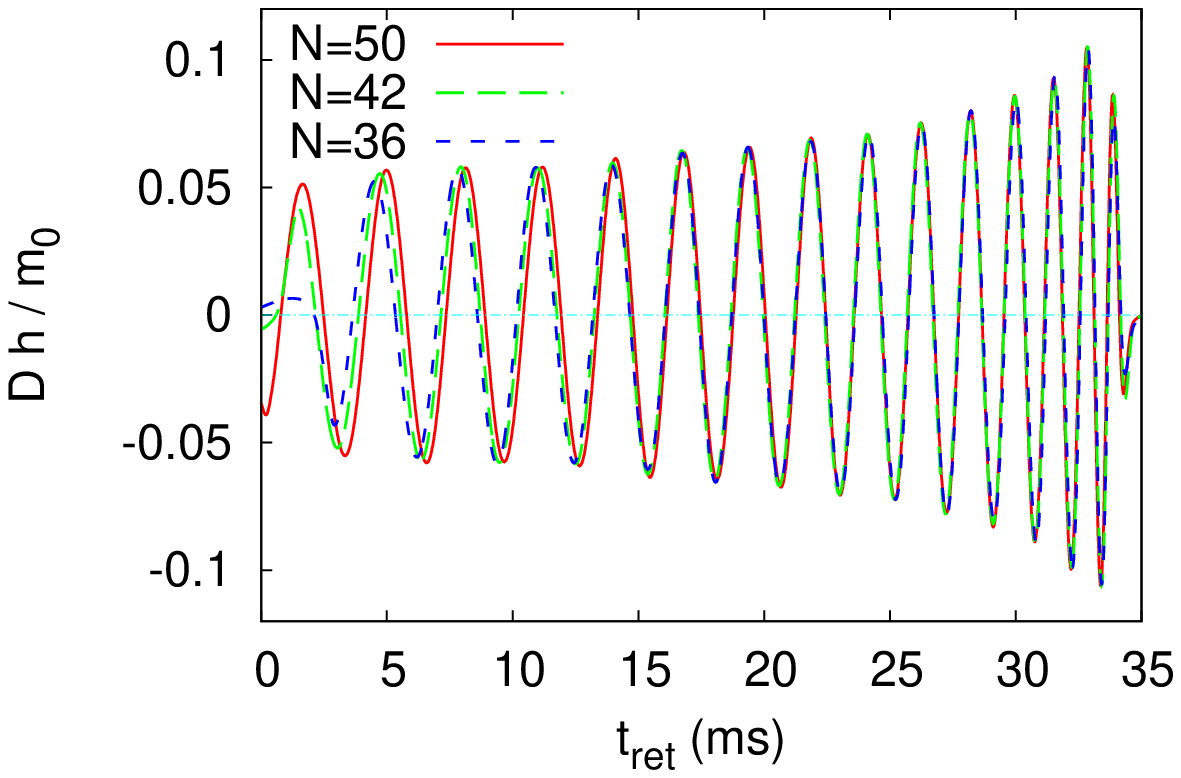}
 \end{tabular}
 \caption{Comparisons of evolution of the orbital angular velocity and
 gravitational waveforms for models HB-Q4M135a75 (top) and 2H-Q5M135a75
 (bottom) with appropriate time shifts.} \label{fig:wave_conv}
 \end{figure*}

This Appendix demonstrates that the convergence is approximately
achieved for gravitational waves and masses of the remnant disks shown
in Sec.~\ref{sec:res}. Figure \ref{fig:wave_conv} shows the evolution of
the orbital angular velocity determined by Eq.~(\ref{eq:omega}) and
gravitational waveforms obtained with different grid resolutions for
models HB-Q4M135a75 and 2H-Q5M135a75. We perform an appropriate time
shift in order to align the curves in the inspiral phase, and perform
the rotation of + and $\times$ polarization modes of gravitational
waveforms for $N=42$ and 36. We note that the time to the merger,
$t_{\rm merger}$, is systematically longer for finer grid resolutions
because numerical dissipation of the angular momentum is smaller
\cite{skyt2009,kst2010}. In both cases, the evolution of the orbital
angular velocity, $\Omega (t)$, approximately agrees up to the merger,
except for initial bursts associated with the junk radiation. The
gravitational waveforms agree very well in the final $\sim 4$ orbits of
the inspiral phase ($t \gtrsim 15$ ms), the merger phase, and the
ringdown phase if the QNM is excited. By contrast, gravitational
waveforms in the initial $\sim 2$ orbits depend strongly on the grid
resolutions, because the early inspiral phase is strongly affected by
the dissipation of the junk radiation. We conclude that the convergence
is approximately achieved for gravitational waves in the late inspiral,
merger, and ringdown phases.

\begin{figure}[tbp]
 \includegraphics[width=85mm,clip]{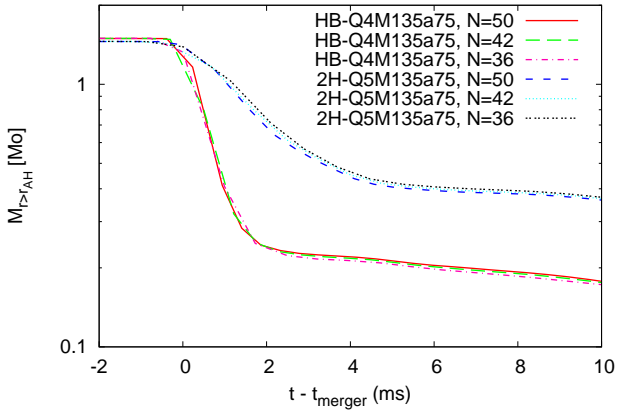}
 \caption{Comparisons of evolution of the remnant disk masses for models
 HB-Q4M135a75 and 2H-Q5M135a75.} \label{fig:disk_conv}
\end{figure}

Figure \ref{fig:disk_conv} shows the evolution of the rest mass located
outside the AH, $M_{r>r_{\rm AH}}$, with different grid resolutions for
models HB-Q4M135a75 and 2H-Q5M135a75. This figure shows that the
convergence is also approximately achieved for the mass of the remnant
disk. Quantitatively, differences in $M_{r>r_{\rm AH}}$ at $\approx 10$
ms after the merger are $\approx 2.9 \%$ and $\approx 2.6 \%$ for models
HB-Q4M135a75 and 2H-Q5M135a75, respectively. If we assume the
first-order convergence for $M_{r>r_{\rm AH}}$, the errors in the values
obtained for $N=50$ runs are $\approx 7.5 \%$ and $\approx 6.8 \%$ for
models HB-Q4M135a75 and 2H-Q5M135a75, respectively.

\begin{thebibliography}{100}%
\makeatletter
\providecommand \@ifxundefined [1]{%
 \ifx #1\undefined \expandafter \@firstoftwo
 \else \expandafter \@secondoftwo
\fi
}%
\providecommand \@ifnum [1]{%
 \ifnum #1\expandafter \@firstoftwo
 \else \expandafter \@secondoftwo
\fi
}%
\providecommand \enquote [1]{``#1''}%
\providecommand \bibnamefont  [1]{#1}%
\providecommand \bibfnamefont [1]{#1}%
\providecommand \citenamefont [1]{#1}%
\providecommand\href[0]{\@sanitize\@href}%
\providecommand\@href[1]{\endgroup\@@startlink{#1}\endgroup\@@href}%
\providecommand\@@href[1]{#1\@@endlink}%
\providecommand \@sanitize [0]{\begingroup\catcode`\&12\catcode`\#12\relax}%
\@ifxundefined \pdfoutput {\@firstoftwo}{%
 \@ifnum{\z@=\pdfoutput}{\@firstoftwo}{\@secondoftwo}%
}{%
 \providecommand\@@startlink[1]{\leavevmode}%
 \providecommand\@@endlink[0]{}%
}{%
 \providecommand\@@startlink[1]{%
  \leavevmode
  \pdfstartlink
   attr{/Border[0 0 1 ]/H/I/C[0 1 1]}%
   user{/Subtype/Link/A<</Type/Action/S/URI/URI(#1)>>}%
  \relax
 }%
 \providecommand\@@endlink[0]{\pdfendlink}%
}%
\providecommand \url  [0]{\begingroup\@sanitize \@url }%
\providecommand \@url [1]{\endgroup\@href {#1}{\urlprefix}}%
\providecommand \urlprefix [0]{URL }%
\providecommand \Eprint[0]{\href }%
\@ifxundefined \urlstyle {%
  \providecommand \doi [1]{doi:\discretionary{}{}{}#1}%
}{%
  \providecommand \doi [0]{doi:\discretionary{}{}{}\begingroup
  \urlstyle{rm}\Url }%
}%
\providecommand \doibase [0]{http://dx.doi.org/}%
\providecommand \Doi[1]{\href{\doibase#1}}%
\providecommand \bibAnnote [3]{%
  \BibitemShut{#1}%
  \begin{quotation}\noindent
    \textsc{Key:}\ #2\\\textsc{Annotation:}\ #3%
  \end{quotation}%
}%
\providecommand \bibAnnoteFile [2]{%
  \IfFileExists{#2}{\bibAnnote {#1} {#2} {\input{#2}}}{}%
}%
\providecommand \typeout [0]{\immediate \write \m@ne }%
\providecommand \selectlanguage [0]{\@gobble}%
\providecommand \bibinfo [0]{\@secondoftwo}%
\providecommand \bibfield [0]{\@secondoftwo}%
\providecommand \translation [1]{[#1]}%
\providecommand \BibitemOpen[0]{}%
\providecommand \bibitemStop [0]{}%
\providecommand \bibitemNoStop [0]{.\EOS\space}%
\providecommand \EOS [0]{\spacefactor3000\relax}%
\providecommand \BibitemShut [1]{\csname bibitem#1\endcsname}%
\bibitem{ligo2009}%
  \BibitemOpen
  \bibfield{author}{%
  \bibinfo {author} {\bibnamefont{{B. P. Abbott et al.}}},\ }%
  \bibfield{journal}{%
  \bibinfo {journal} {Rep. Prog. Phys.}\ }%
  \textbf{\bibinfo {volume} {72}},\ \bibinfo {pages} {076901} (\bibinfo {year}
  {2009})%
  \bibAnnoteFile{NoStop}{ligo2009}%
\bibitem{virgo2011}%
  \BibitemOpen
  \bibfield{author}{%
  \bibinfo {author} {\bibnamefont{{T. Accadia et al.}}},\ }%
  \bibfield{journal}{%
  \bibinfo {journal} {Class. Quantum Grav.}\ }%
  \textbf{\bibinfo {volume} {28}},\ \bibinfo {pages} {025005} (\bibinfo {year}
  {2011})%
  \bibAnnoteFile{NoStop}{virgo2011}%
\bibitem{lcgt2010}%
  \BibitemOpen
  \bibfield{author}{%
  \bibinfo {author} {\bibnamefont{{K. Kuroda and the LCGT Collaboration}}},\ }%
  \bibfield{journal}{%
  \bibinfo {journal} {Class. Quantum Grav.}\ }%
  \textbf{\bibinfo {volume} {27}},\ \bibinfo {pages} {084004} (\bibinfo {year}
  {2010})%
  \bibAnnoteFile{NoStop}{lcgt2010}%
\bibitem{lindblom1992}%
  \BibitemOpen
  \bibfield{author}{%
  \bibinfo {author} {\bibfnamefont{L.}~\bibnamefont{Lindblom}},\ }%
  \bibfield{journal}{%
  \bibinfo {journal} {Astrophys. J.}\ }%
  \textbf{\bibinfo {volume} {398}},\ \bibinfo {pages} {569} (\bibinfo {year}
  {1992})%
  \bibAnnoteFile{NoStop}{lindblom1992}%
\bibitem{vallisneri2000}%
  \BibitemOpen
  \bibfield{author}{%
  \bibinfo {author} {\bibfnamefont{M.}~\bibnamefont{Vallisneri}},\ }%
  \bibfield{journal}{%
  \bibinfo {journal} {Phys. Rev. Lett.}\ }%
  \textbf{\bibinfo {volume} {84}},\ \bibinfo {pages} {3519} (\bibinfo {year}
  {2000})%
  \bibAnnoteFile{NoStop}{vallisneri2000}%
\bibitem{rmsucf2009}%
  \BibitemOpen
  \bibfield{author}{%
  \bibinfo {author} {\bibfnamefont{J.~S.}\ \bibnamefont{Read}}, \bibinfo
  {author} {\bibfnamefont{C.}~\bibnamefont{Markakis}}, \bibinfo {author}
  {\bibfnamefont{M.}~\bibnamefont{Shibata}}, \bibinfo {author}
  {\bibfnamefont{K.}~\bibnamefont{Ury{\=u}}}, \bibinfo {author}
  {\bibfnamefont{J.~D.~E.}\ \bibnamefont{Creighton}},\ and\ \bibinfo {author}
  {\bibfnamefont{J.~L.}\ \bibnamefont{Friedman}},\ }%
  \bibfield{journal}{%
  \bibinfo {journal} {Phys. Rev. D}\ }%
  \textbf{\bibinfo {volume} {79}},\ \bibinfo {pages} {124033} (\bibinfo {year}
  {2009})%
  \bibAnnoteFile{NoStop}{rmsucf2009}%
\bibitem{fgp2010}%
  \BibitemOpen
  \bibfield{author}{%
  \bibinfo {author} {\bibfnamefont{V.}~\bibnamefont{Ferrari}}, \bibinfo
  {author} {\bibfnamefont{L.}~\bibnamefont{Gualtieri}},\ and\ \bibinfo {author}
  {\bibfnamefont{F.}~\bibnamefont{Pannarale}},\ }%
  \bibfield{journal}{%
  \bibinfo {journal} {Phys. Rev. D}\ }%
  \textbf{\bibinfo {volume} {81}},\ \bibinfo {pages} {064026} (\bibinfo {year}
  {2010})%
  \bibAnnoteFile{NoStop}{fgp2010}%
\bibitem{ptr2011}%
  \BibitemOpen
  \bibfield{author}{%
  \bibinfo {author} {\bibfnamefont{F.}~\bibnamefont{Pannarale}}, \bibinfo
  {author} {\bibfnamefont{A.}~\bibnamefont{Tonita}},\ and\ \bibinfo {author}
  {\bibfnamefont{L.}~\bibnamefont{Rezzolla}},\ }%
  \bibfield{journal}{%
  \bibinfo {journal} {Astrophys. J.}\ }%
  \textbf{\bibinfo {volume} {727}},\ \bibinfo {pages} {95} (\bibinfo {year}
  {2011})%
  \bibAnnoteFile{NoStop}{ptr2011}%
\bibitem{pror2011}%
  \BibitemOpen
  \bibfield{author}{%
  \bibinfo {author} {\bibfnamefont{F.}~\bibnamefont{Pannarale}}, \bibinfo
  {author} {\bibfnamefont{L.}~\bibnamefont{Rezzolla}}, \bibinfo {author}
  {\bibfnamefont{F.}~\bibnamefont{Ohme}},\ and\ \bibinfo {author}
  {\bibfnamefont{J.~S.}\ \bibnamefont{Read}},\ }%
  \bibinfo {journal} {arXiv:1103.3526}%
  \bibAnnoteFile{NoStop}{pror2011}%
\bibitem{dprrh2010}%
  \BibitemOpen
\bibfield{journal}{%
    }%
  \bibfield{author}{%
  \bibinfo {author} {\bibfnamefont{P.}~\bibnamefont{Demorest}}, \bibinfo
  {author} {\bibfnamefont{T.}~\bibnamefont{Pennucci}}, \bibinfo {author}
  {\bibfnamefont{S.}~\bibnamefont{Ransom}}, \bibinfo {author}
  {\bibfnamefont{M.}~\bibnamefont{Roberts}},\ and\ \bibinfo {author}
  {\bibfnamefont{J.}~\bibnamefont{Hessels}},\ }%
  \bibfield{journal}{%
  \bibinfo {journal} {Nature}\ }%
  \textbf{\bibinfo {volume} {467}},\ \bibinfo {pages} {1081} (\bibinfo {year}
  {2010})%
  \bibAnnoteFile{NoStop}{dprrh2010}%
\bibitem{slb2010}%
  \BibitemOpen
  \bibfield{author}{%
  \bibinfo {author} {\bibfnamefont{A.~W.}\ \bibnamefont{Steiner}}, \bibinfo
  {author} {\bibfnamefont{J.~M.}\ \bibnamefont{Lattimer}},\ and\ \bibinfo
  {author} {\bibfnamefont{E.~F.}\ \bibnamefont{Brown}},\ }%
  \bibfield{journal}{%
  \bibinfo {journal} {Astrophys. J.}\ }%
  \textbf{\bibinfo {volume} {722}},\ \bibinfo {pages} {33} (\bibinfo {year}
  {2010})%
  \bibAnnoteFile{NoStop}{slb2010}%
\bibitem{nakar}%
  \BibitemOpen
  \bibfield{author}{%
  \bibinfo {author} {\bibfnamefont{E.}~\bibnamefont{Nakar}},\ }%
  \bibfield{journal}{%
  \bibinfo {journal} {Phys. Rep.}\ }%
  \textbf{\bibinfo {volume} {442}},\ \bibinfo {pages} {166} (\bibinfo {year}
  {2007})%
  \bibAnnoteFile{NoStop}{nakar}%
\bibitem{leeramirezruiz}%
  \BibitemOpen
  \bibfield{author}{%
  \bibinfo {author} {\bibfnamefont{W.~H.}\ \bibnamefont{Lee}}\ and\ \bibinfo
  {author} {\bibfnamefont{E.}~\bibnamefont{Ramirez-Ruiz}},\ }%
  \bibfield{journal}{%
  \bibinfo {journal} {New J. Phys}\ }%
  \textbf{\bibinfo {volume} {9}},\ \bibinfo {pages} {17} (\bibinfo {year}
  {2007})%
  \bibAnnoteFile{NoStop}{leeramirezruiz}%
\bibitem{blandfordznajek1977}%
  \BibitemOpen
  \bibfield{author}{%
  \bibinfo {author} {\bibfnamefont{R.~D.}\ \bibnamefont{Blandford}}\ and\
  \bibinfo {author} {\bibfnamefont{R.~D.}\ \bibnamefont{Znajek}},\ }%
  \bibfield{journal}{%
  \bibinfo {journal} {Mon. Not. Roy. Astron. Soc.}\ }%
  \textbf{\bibinfo {volume} {179}},\ \bibinfo {pages} {433} (\bibinfo {year}
  {1977})%
  \bibAnnoteFile{NoStop}{blandfordznajek1977}%
\bibitem{bergeretal2005}%
  \BibitemOpen
  \bibfield{author}{%
  \bibinfo {author} {\bibnamefont{{E. Berger et al,}}},\ }%
  \bibfield{journal}{%
  \bibinfo {journal} {Nature}\ }%
  \textbf{\bibinfo {volume} {438}},\ \bibinfo {pages} {988} (\bibinfo {year}
  {2005})%
  \bibAnnoteFile{NoStop}{bergeretal2005}%
\bibitem{fbctlfgcf2010}%
  \BibitemOpen
  \bibfield{author}{%
  \bibinfo {author} {\bibfnamefont{W.}~\bibnamefont{Fong}}, \bibinfo {author}
  {\bibfnamefont{E.}~\bibnamefont{Berger}}, \bibinfo {author}
  {\bibfnamefont{R.}~\bibnamefont{Chornock}}, \bibinfo {author}
  {\bibfnamefont{N.~R.}\ \bibnamefont{Tanvir}}, \bibinfo {author}
  {\bibfnamefont{A.~J.}\ \bibnamefont{Levan}}, \bibinfo {author}
  {\bibfnamefont{J.~F.}\ \bibnamefont{Graham}}, \bibinfo {author}
  {\bibfnamefont{A.~S.}\ \bibnamefont{Fruchter}}, \bibinfo {author}
  {\bibfnamefont{A.}~\bibnamefont{Cucchiara}},\ and\ \bibinfo {author}
  {\bibfnamefont{D.~B.}\ \bibnamefont{Fox}},\ }%
  \bibfield{journal}{%
  \bibinfo {journal} {Astrophys. J.}\ }%
  \textbf{\bibinfo {volume} {730}},\ \bibinfo {pages} {26} (\bibinfo {year}
  {2011})%
  \bibAnnoteFile{NoStop}{fbctlfgcf2010}%
\bibitem{woosley1993}%
  \BibitemOpen
  \bibfield{author}{%
  \bibinfo {author} {\bibfnamefont{S.~E.}\ \bibnamefont{Woosley}},\ }%
  \bibfield{journal}{%
  \bibinfo {journal} {Astrophys. J.}\ }%
  \textbf{\bibinfo {volume} {405}},\ \bibinfo {pages} {273} (\bibinfo {year}
  {1993})%
  \bibAnnoteFile{NoStop}{woosley1993}%
\bibitem{grandclement2006}%
  \BibitemOpen
  \bibfield{author}{%
  \bibinfo {author} {\bibfnamefont{P.}~\bibnamefont{Grandcl{\'e}ment}},\ }%
  \bibfield{journal}{%
  \bibinfo {journal} {Phys. Rev. D}\ }%
  \textbf{\bibinfo {volume} {74}},\ \bibinfo {pages} {124002} (\bibinfo {year}
  {2006})%
  \bibAnnoteFile{NoStop}{grandclement2006}%
\bibitem{grandclement2006e}%
  \BibitemOpen
  \ \textbf{\bibinfo {volume} {75}},\ \bibinfo {pages} {129903(E)} (\bibinfo
  {year} {2007})%
  \bibAnnoteFile{NoStop}{grandclement2006e}%
\bibitem{tbfs2007}%
  \BibitemOpen
  \bibfield{author}{%
  \bibinfo {author} {\bibfnamefont{K.}~\bibnamefont{Taniguchi}}, \bibinfo
  {author} {\bibfnamefont{T.~W.}\ \bibnamefont{Baumgarte}}, \bibinfo {author}
  {\bibfnamefont{J.~A.}\ \bibnamefont{Faber}},\ and\ \bibinfo {author}
  {\bibfnamefont{S.~L.}\ \bibnamefont{Shapiro}},\ }%
  \bibfield{journal}{%
  \bibinfo {journal} {Phys. Rev. D}\ }%
  \textbf{\bibinfo {volume} {75}},\ \bibinfo {pages} {084005} (\bibinfo {year}
  {2007})%
  \bibAnnoteFile{NoStop}{tbfs2007}%
\bibitem{tbfs2008}%
  \BibitemOpen
  \bibfield{author}{%
  \bibinfo {author} {\bibfnamefont{K.}~\bibnamefont{Taniguchi}}, \bibinfo
  {author} {\bibfnamefont{T.~W.}\ \bibnamefont{Baumgarte}}, \bibinfo {author}
  {\bibfnamefont{J.~A.}\ \bibnamefont{Faber}},\ and\ \bibinfo {author}
  {\bibfnamefont{S.~L.}\ \bibnamefont{Shapiro}},\ }%
  \bibfield{journal}{%
  \bibinfo {journal} {Phys. Rev. D}\ }%
  \textbf{\bibinfo {volume} {77}},\ \bibinfo {pages} {044003} (\bibinfo {year}
  {2008})%
  \bibAnnoteFile{NoStop}{tbfs2008}%
\bibitem{fkpt2008}%
  \BibitemOpen
  \bibfield{author}{%
  \bibinfo {author} {\bibfnamefont{F.}~\bibnamefont{Foucart}}, \bibinfo
  {author} {\bibfnamefont{L.~E.}\ \bibnamefont{Kidder}}, \bibinfo {author}
  {\bibfnamefont{H.~P.}\ \bibnamefont{Pfeiffer}},\ and\ \bibinfo {author}
  {\bibfnamefont{S.~A.}\ \bibnamefont{Teukolsky}},\ }%
  \bibfield{journal}{%
  \bibinfo {journal} {Phys. Rev. D}\ }%
  \textbf{\bibinfo {volume} {77}},\ \bibinfo {pages} {124051} (\bibinfo {year}
  {2008})%
  \bibAnnoteFile{NoStop}{fkpt2008}%
\bibitem{kst2009}%
  \BibitemOpen
  \bibfield{author}{%
  \bibinfo {author} {\bibfnamefont{K.}~\bibnamefont{Kyutoku}}, \bibinfo
  {author} {\bibfnamefont{M.}~\bibnamefont{Shibata}},\ and\ \bibinfo {author}
  {\bibfnamefont{K.}~\bibnamefont{Taniguchi}},\ }%
  \bibfield{journal}{%
  \bibinfo {journal} {Phys. Rev. D}\ }%
  \textbf{\bibinfo {volume} {79}},\ \bibinfo {pages} {124018} (\bibinfo {year}
  {2009})%
  \bibAnnoteFile{NoStop}{kst2009}%
\bibitem{shibatauryu2006}%
  \BibitemOpen
  \bibfield{author}{%
  \bibinfo {author} {\bibfnamefont{M.}~\bibnamefont{Shibata}}\ and\ \bibinfo
  {author} {\bibfnamefont{K.}~\bibnamefont{Ury{\=u}}},\ }%
  \bibfield{journal}{%
  \bibinfo {journal} {Phys. Rev. D}\ }%
  \textbf{\bibinfo {volume} {74}},\ \bibinfo {pages} {121503(R)} (\bibinfo
  {year} {2006})%
  \bibAnnoteFile{NoStop}{shibatauryu2006}%
\bibitem{shibatauryu2007}%
  \BibitemOpen
  \bibfield{author}{%
  \bibinfo {author} {\bibfnamefont{M.}~\bibnamefont{Shibata}}\ and\ \bibinfo
  {author} {\bibfnamefont{K.}~\bibnamefont{Ury{\=u}}},\ }%
  \bibfield{journal}{%
  \bibinfo {journal} {Class. Quant. Grav.}\ }%
  \textbf{\bibinfo {volume} {24}},\ \bibinfo {pages} {S125} (\bibinfo {year}
  {2007})%
  \bibAnnoteFile{NoStop}{shibatauryu2007}%
\bibitem{shibatataniguchi2008}%
  \BibitemOpen
  \bibfield{author}{%
  \bibinfo {author} {\bibfnamefont{M.}~\bibnamefont{Shibata}}\ and\ \bibinfo
  {author} {\bibfnamefont{K.}~\bibnamefont{Taniguchi}},\ }%
  \bibfield{journal}{%
  \bibinfo {journal} {Phys. Rev. D}\ }%
  \textbf{\bibinfo {volume} {77}},\ \bibinfo {pages} {084015} (\bibinfo {year}
  {2008})%
  \bibAnnoteFile{NoStop}{shibatataniguchi2008}%
\bibitem{eflstb2008}%
  \BibitemOpen
  \bibfield{author}{%
  \bibinfo {author} {\bibfnamefont{Z.~B.}\ \bibnamefont{Etienne}}, \bibinfo
  {author} {\bibfnamefont{J.~A.}\ \bibnamefont{Faber}}, \bibinfo {author}
  {\bibfnamefont{Y.~T.}\ \bibnamefont{Liu}}, \bibinfo {author}
  {\bibfnamefont{S.~L.}\ \bibnamefont{Shapiro}}, \bibinfo {author}
  {\bibfnamefont{K.}~\bibnamefont{Taniguchi}},\ and\ \bibinfo {author}
  {\bibfnamefont{T.~W.}\ \bibnamefont{Baumgarte}},\ }%
  \bibfield{journal}{%
  \bibinfo {journal} {Phys. Rev. D}\ }%
  \textbf{\bibinfo {volume} {77}},\ \bibinfo {pages} {084002} (\bibinfo {year}
  {2008})%
  \bibAnnoteFile{NoStop}{eflstb2008}%
\bibitem{dfkpst2008}%
  \BibitemOpen
  \bibfield{author}{%
  \bibinfo {author} {\bibfnamefont{M.~D.}\ \bibnamefont{Duez}}, \bibinfo
  {author} {\bibfnamefont{F.}~\bibnamefont{Foucart}}, \bibinfo {author}
  {\bibfnamefont{L.~E.}\ \bibnamefont{Kidder}}, \bibinfo {author}
  {\bibfnamefont{H.~P.}\ \bibnamefont{Pfeiffer}}, \bibinfo {author}
  {\bibfnamefont{M.~A.}\ \bibnamefont{Scheel}},\ and\ \bibinfo {author}
  {\bibfnamefont{S.~A.}\ \bibnamefont{Teukolsky}},\ }%
  \bibfield{journal}{%
  \bibinfo {journal} {Phys. Rev. D}\ }%
  \textbf{\bibinfo {volume} {78}},\ \bibinfo {pages} {104015} (\bibinfo {year}
  {2008})%
  \bibAnnoteFile{NoStop}{dfkpst2008}%
\bibitem{elsb2009}%
  \BibitemOpen
  \bibfield{author}{%
  \bibinfo {author} {\bibfnamefont{Z.~B.}\ \bibnamefont{Etienne}}, \bibinfo
  {author} {\bibfnamefont{Y.~T.}\ \bibnamefont{Liu}}, \bibinfo {author}
  {\bibfnamefont{S.~L.}\ \bibnamefont{Shapiro}},\ and\ \bibinfo {author}
  {\bibfnamefont{T.~W.}\ \bibnamefont{Baumgarte}},\ }%
  \bibfield{journal}{%
  \bibinfo {journal} {Phys. Rev. D}\ }%
  \textbf{\bibinfo {volume} {79}},\ \bibinfo {pages} {044024} (\bibinfo {year}
  {2009})%
  \bibAnnoteFile{NoStop}{elsb2009}%
\bibitem{skyt2009}%
  \BibitemOpen
  \bibfield{author}{%
  \bibinfo {author} {\bibfnamefont{M.}~\bibnamefont{Shibata}}, \bibinfo
  {author} {\bibfnamefont{K.}~\bibnamefont{Kyutoku}}, \bibinfo {author}
  {\bibfnamefont{T.}~\bibnamefont{Yamamoto}},\ and\ \bibinfo {author}
  {\bibfnamefont{K.}~\bibnamefont{Taniguchi}},\ }%
  \bibfield{journal}{%
  \bibinfo {journal} {Phys. Rev. D}\ }%
  \textbf{\bibinfo {volume} {79}},\ \bibinfo {pages} {044030} (\bibinfo {year}
  {2009})%
  \bibAnnoteFile{NoStop}{skyt2009}%
\bibitem{dfkot2010}%
  \BibitemOpen
  \bibfield{author}{%
  \bibinfo {author} {\bibfnamefont{M.~D.}\ \bibnamefont{Duez}}, \bibinfo
  {author} {\bibfnamefont{F.}~\bibnamefont{Foucart}}, \bibinfo {author}
  {\bibfnamefont{L.~E.}\ \bibnamefont{Kidder}}, \bibinfo {author}
  {\bibfnamefont{C.~D.}\ \bibnamefont{Ott}},\ and\ \bibinfo {author}
  {\bibfnamefont{S.~A.}\ \bibnamefont{Teukolsky}},\ }%
  \bibfield{journal}{%
  \bibinfo {journal} {Class. Quant. Grav.}\ }%
  \textbf{\bibinfo {volume} {27}},\ \bibinfo {pages} {114106} (\bibinfo {year}
  {2010})%
  \bibAnnoteFile{NoStop}{dfkot2010}%
\bibitem{kst2010}%
  \BibitemOpen
  \bibfield{author}{%
  \bibinfo {author} {\bibfnamefont{K.}~\bibnamefont{Kyutoku}}, \bibinfo
  {author} {\bibfnamefont{M.}~\bibnamefont{Shibata}},\ and\ \bibinfo {author}
  {\bibfnamefont{K.}~\bibnamefont{Taniguchi}},\ }%
  \bibfield{journal}{%
  \bibinfo {journal} {Phys. Rev. D}\ }%
  \textbf{\bibinfo {volume} {82}},\ \bibinfo {pages} {044049} (\bibinfo {year}
  {2010})%
  \bibAnnoteFile{NoStop}{kst2010}%
\bibitem{cabllmn2010}%
  \BibitemOpen
  \bibfield{author}{%
  \bibinfo {author} {\bibfnamefont{S.}~\bibnamefont{Chawla}}, \bibinfo {author}
  {\bibfnamefont{M.}~\bibnamefont{Anderson}}, \bibinfo {author}
  {\bibfnamefont{M.}~\bibnamefont{Besselman}}, \bibinfo {author}
  {\bibfnamefont{L.}~\bibnamefont{Lehner}}, \bibinfo {author}
  {\bibfnamefont{S.~L.}\ \bibnamefont{Liebling}}, \bibinfo {author}
  {\bibfnamefont{P.~M.}\ \bibnamefont{Motl}},\ and\ \bibinfo {author}
  {\bibfnamefont{D.}~\bibnamefont{Neilsen}},\ }%
  \bibfield{journal}{%
  \bibinfo {journal} {Phys. Rev. Lett.}\ }%
  \textbf{\bibinfo {volume} {105}},\ \bibinfo {pages} {111101} (\bibinfo {year}
  {2010})%
  \bibAnnoteFile{NoStop}{cabllmn2010}%
\bibitem{LORENE}%
  \BibitemOpen
  \bibfield{author}{%
  \bibinfo {author} {\bibnamefont{{LORENE website}}},\ }%
  \url{http://www.lorene.obspm.fr/.}%
  \bibAnnoteFile{Stop}{LORENE}%
\bibitem{cook}%
  \BibitemOpen
  \bibfield{author}{%
  \bibinfo {author} {\bibfnamefont{G.~B.}\ \bibnamefont{Cook}},\ }%
  \bibfield{journal}{%
  \bibinfo {journal} {Living Rev. Relativity}\ }%
  \textbf{\bibinfo {volume} {3}},\ \bibinfo {pages} {5} (\bibinfo {year}
  {2000})%
  \bibAnnoteFile{NoStop}{cook}%
\bibitem{bildstencutler1992}%
  \BibitemOpen
  \bibfield{author}{%
  \bibinfo {author} {\bibfnamefont{L.}~\bibnamefont{Bildsten}}\ and\ \bibinfo
  {author} {\bibfnamefont{C.}~\bibnamefont{Cutler}},\ }%
  \bibfield{journal}{%
  \bibinfo {journal} {Astrophys. J.}\ }%
  \textbf{\bibinfo {volume} {400}},\ \bibinfo {pages} {175} (\bibinfo {year}
  {1992})%
  \bibAnnoteFile{NoStop}{bildstencutler1992}%
\bibitem{kochanek1992}%
  \BibitemOpen
  \bibfield{author}{%
  \bibinfo {author} {\bibfnamefont{C.~S.}\ \bibnamefont{Kochanek}},\ }%
  \bibfield{journal}{%
  \bibinfo {journal} {Astrophys. J.}\ }%
  \textbf{\bibinfo {volume} {398}},\ \bibinfo {pages} {234} (\bibinfo {year}
  {1992})%
  \bibAnnoteFile{NoStop}{kochanek1992}%
\bibitem{york1999}%
  \BibitemOpen
  \bibfield{author}{%
  \bibinfo {author} {\bibfnamefont{J.~W.}\ \bibnamefont{York}},\ }%
  \bibfield{journal}{%
  \bibinfo {journal} {Phys. Rev. Lett.}\ }%
  \textbf{\bibinfo {volume} {82}},\ \bibinfo {pages} {1350} (\bibinfo {year}
  {1999})%
  \bibAnnoteFile{NoStop}{york1999}%
\bibitem{pfeifferyork2003}%
  \BibitemOpen
  \bibfield{author}{%
  \bibinfo {author} {\bibfnamefont{H.~P.}\ \bibnamefont{Pfeiffer}}\ and\
  \bibinfo {author} {\bibfnamefont{J.~W.}\ \bibnamefont{York}},\ }%
  \bibfield{journal}{%
  \bibinfo {journal} {Phys. Rev. D}\ }%
  \textbf{\bibinfo {volume} {67}},\ \bibinfo {pages} {044022} (\bibinfo {year}
  {2003})%
  \bibAnnoteFile{NoStop}{pfeifferyork2003}%
\bibitem{brandtbrugmann1997}%
  \BibitemOpen
  \bibfield{author}{%
  \bibinfo {author} {\bibfnamefont{S.}~\bibnamefont{Brandt}}\ and\ \bibinfo
  {author} {\bibfnamefont{B.}~\bibnamefont{Br{\"u}gmann}},\ }%
  \bibfield{journal}{%
  \bibinfo {journal} {Phys. Rev. Lett.}\ }%
  \textbf{\bibinfo {volume} {78}},\ \bibinfo {pages} {3606} (\bibinfo {year}
  {1997})%
  \bibAnnoteFile{NoStop}{brandtbrugmann1997}%
\bibitem{clmz2006}%
  \BibitemOpen
  \bibfield{author}{%
  \bibinfo {author} {\bibfnamefont{M.}~\bibnamefont{Campanelli}}, \bibinfo
  {author} {\bibfnamefont{C.~O.}\ \bibnamefont{Lousto}}, \bibinfo {author}
  {\bibfnamefont{P.}~\bibnamefont{Marronetti}},\ and\ \bibinfo {author}
  {\bibfnamefont{Y.}~\bibnamefont{Zlochower}},\ }%
  \bibfield{journal}{%
  \bibinfo {journal} {Phys. Rev. Lett.}\ }%
  \textbf{\bibinfo {volume} {96}},\ \bibinfo {pages} {111101} (\bibinfo {year}
  {2006})%
  \bibAnnoteFile{NoStop}{clmz2006}%
\bibitem{bcckm2006}%
  \BibitemOpen
  \bibfield{author}{%
  \bibinfo {author} {\bibfnamefont{J.~G.}\ \bibnamefont{Baker}}, \bibinfo
  {author} {\bibfnamefont{J.}~\bibnamefont{Centrella}}, \bibinfo {author}
  {\bibfnamefont{D.-I.}\ \bibnamefont{Choi}}, \bibinfo {author}
  {\bibfnamefont{M.}~\bibnamefont{Koppitz}},\ and\ \bibinfo {author}
  {\bibfnamefont{J.}~\bibnamefont{van Meter}},\ }%
  \bibfield{journal}{%
  \bibinfo {journal} {Phys. Rev. Lett.}\ }%
  \textbf{\bibinfo {volume} {96}},\ \bibinfo {pages} {111102} (\bibinfo {year}
  {2006})%
  \bibAnnoteFile{NoStop}{bcckm2006}%
\bibitem{beig1978}%
  \BibitemOpen
  \bibfield{author}{%
  \bibinfo {author} {\bibfnamefont{R.}~\bibnamefont{Beig}},\ }%
  \bibfield{journal}{%
  \bibinfo {journal} {Phys. Lett. A}\ }%
  \textbf{\bibinfo {volume} {69}},\ \bibinfo {pages} {153} (\bibinfo {year}
  {1978})%
  \bibAnnoteFile{NoStop}{beig1978}%
\bibitem{ashtekarashtekar1979}%
  \BibitemOpen
  \bibfield{author}{%
  \bibinfo {author} {\bibfnamefont{A.}~\bibnamefont{Ashtekar}}\ and\ \bibinfo
  {author} {\bibfnamefont{A.}~\bibnamefont{Magnon-Ashtekar}},\ }%
  \bibfield{journal}{%
  \bibinfo {journal} {J. Math. Phys.}\ }%
  \textbf{\bibinfo {volume} {20}},\ \bibinfo {pages} {793} (\bibinfo {year}
  {1979})%
  \bibAnnoteFile{NoStop}{ashtekarashtekar1979}%
\bibitem{bowenyork1980}%
  \BibitemOpen
  \bibfield{author}{%
  \bibinfo {author} {\bibfnamefont{J.~M.}\ \bibnamefont{Bowen}}\ and\ \bibinfo
  {author} {\bibfnamefont{J.~W.}\ \bibnamefont{York}},\ }%
  \bibfield{journal}{%
  \bibinfo {journal} {Phys. Rev. D}\ }%
  \textbf{\bibinfo {volume} {21}},\ \bibinfo {pages} {2047} (\bibinfo {year}
  {1980})%
  \bibAnnoteFile{NoStop}{bowenyork1980}%
\bibitem{ashtekarkrishnan}%
  \BibitemOpen
  \bibfield{author}{%
  \bibinfo {author} {\bibfnamefont{A.}~\bibnamefont{Ashtekar}}\ and\ \bibinfo
  {author} {\bibfnamefont{B.}~\bibnamefont{Krishnan}},\ }%
  \bibfield{journal}{%
  \bibinfo {journal} {Living Rev. Relativity}\ }%
  \textbf{\bibinfo {volume} {7}},\ \bibinfo {pages} {10} (\bibinfo {year}
  {2004})%
  \bibAnnoteFile{NoStop}{ashtekarkrishnan}%
\bibitem{gourgoulhonjaramillo}%
  \BibitemOpen
  \bibfield{author}{%
  \bibinfo {author} {\bibfnamefont{E.}~\bibnamefont{Gourgoulhon}}\ and\
  \bibinfo {author} {\bibfnamefont{J.~L.}\ \bibnamefont{Jaramillo}},\ }%
  \bibfield{journal}{%
  \bibinfo {journal} {Phys. Rep.}\ }%
  \textbf{\bibinfo {volume} {423}},\ \bibinfo {pages} {159} (\bibinfo {year}
  {2006})%
  \bibAnnoteFile{NoStop}{gourgoulhonjaramillo}%
\bibitem{linnovak2007}%
  \BibitemOpen
  \bibfield{author}{%
  \bibinfo {author} {\bibfnamefont{L.-M.}\ \bibnamefont{Lin}}\ and\ \bibinfo
  {author} {\bibfnamefont{J.}~\bibnamefont{Novak}},\ }%
  \bibfield{journal}{%
  \bibinfo {journal} {Class. Quant. Grav.}\ }%
  \textbf{\bibinfo {volume} {24}},\ \bibinfo {pages} {2665} (\bibinfo {year}
  {2007})%
  \bibAnnoteFile{NoStop}{linnovak2007}%
\bibitem{cookwhiting2007}%
  \BibitemOpen
  \bibfield{author}{%
  \bibinfo {author} {\bibfnamefont{G.~B.}\ \bibnamefont{Cook}}\ and\ \bibinfo
  {author} {\bibfnamefont{B.~F.}\ \bibnamefont{Whiting}},\ }%
  \bibfield{journal}{%
  \bibinfo {journal} {Phys. Rev. D}\ }%
  \textbf{\bibinfo {volume} {76}},\ \bibinfo {pages} {041501} (\bibinfo {year}
  {2007})%
  \bibAnnoteFile{NoStop}{cookwhiting2007}%
\bibitem{lopc2008}%
  \BibitemOpen
  \bibfield{author}{%
  \bibinfo {author} {\bibfnamefont{G.}~\bibnamefont{Lovelace}}, \bibinfo
  {author} {\bibfnamefont{R.}~\bibnamefont{Owen}}, \bibinfo {author}
  {\bibfnamefont{H.~P.}\ \bibnamefont{Pfeiffer}},\ and\ \bibinfo {author}
  {\bibfnamefont{T.}~\bibnamefont{Chu}},\ }%
  \bibfield{journal}{%
  \bibinfo {journal} {Phys. Rev. D}\ }%
  \textbf{\bibinfo {volume} {78}},\ \bibinfo {pages} {084017} (\bibinfo {year}
  {2008})%
  \bibAnnoteFile{NoStop}{lopc2008}%
\bibitem{dlz2008}%
  \BibitemOpen
  \bibfield{author}{%
  \bibinfo {author} {\bibfnamefont{S.}~\bibnamefont{Dain}}, \bibinfo {author}
  {\bibfnamefont{C.~O.}\ \bibnamefont{Lousto}},\ and\ \bibinfo {author}
  {\bibfnamefont{Y.}~\bibnamefont{Zlochower}},\ }%
  \bibfield{journal}{%
  \bibinfo {journal} {Phys. Rev. D}\ }%
  \textbf{\bibinfo {volume} {78}},\ \bibinfo {pages} {024039} (\bibinfo {year}
  {2008})%
  \bibAnnoteFile{NoStop}{dlz2008}%
\bibitem{bgm1997}%
  \BibitemOpen
  \bibfield{author}{%
  \bibinfo {author} {\bibfnamefont{S.}~\bibnamefont{Bonazzola}}, \bibinfo
  {author} {\bibfnamefont{E.}~\bibnamefont{Gourgoulhon}},\ and\ \bibinfo
  {author} {\bibfnamefont{J.-A.}\ \bibnamefont{Marck}},\ }%
  \bibfield{journal}{%
  \bibinfo {journal} {Phys. Rev. D}\ }%
  \textbf{\bibinfo {volume} {56}},\ \bibinfo {pages} {7740} (\bibinfo {year}
  {1997})%
  \bibAnnoteFile{NoStop}{bgm1997}%
\bibitem{asada1998}%
  \BibitemOpen
  \bibfield{author}{%
  \bibinfo {author} {\bibfnamefont{H.}~\bibnamefont{Asada}},\ }%
  \bibfield{journal}{%
  \bibinfo {journal} {Phys. Rev. D}\ }%
  \textbf{\bibinfo {volume} {57}},\ \bibinfo {pages} {7292} (\bibinfo {year}
  {1998})%
  \bibAnnoteFile{NoStop}{asada1998}%
\bibitem{shibata1998}%
  \BibitemOpen
  \bibfield{author}{%
  \bibinfo {author} {\bibfnamefont{M.}~\bibnamefont{Shibata}},\ }%
  \bibfield{journal}{%
  \bibinfo {journal} {Phys. Rev. D}\ }%
  \textbf{\bibinfo {volume} {58}},\ \bibinfo {pages} {024012} (\bibinfo {year}
  {1998})%
  \bibAnnoteFile{NoStop}{shibata1998}%
\bibitem{teukolsky1998}%
  \BibitemOpen
  \bibfield{author}{%
  \bibinfo {author} {\bibfnamefont{S.~A.}\ \bibnamefont{Teukolsky}},\ }%
  \bibfield{journal}{%
  \bibinfo {journal} {Astrophys. J.}\ }%
  \textbf{\bibinfo {volume} {504}},\ \bibinfo {pages} {442} (\bibinfo {year}
  {1998})%
  \bibAnnoteFile{NoStop}{teukolsky1998}%
\bibitem{fbb2006}%
  \BibitemOpen
  \bibfield{author}{%
  \bibinfo {author} {\bibfnamefont{G.}~\bibnamefont{Faye}}, \bibinfo {author}
  {\bibfnamefont{L.}~\bibnamefont{Blanchet}},\ and\ \bibinfo {author}
  {\bibfnamefont{A.}~\bibnamefont{Buonanno}},\ }%
  \bibfield{journal}{%
  \bibinfo {journal} {Phys. Rev. D}\ }%
  \textbf{\bibinfo {volume} {74}},\ \bibinfo {pages} {104033} (\bibinfo {year}
  {2006})%
  \bibAnnoteFile{NoStop}{fbb2006}%
\bibitem{bbf2006}%
  \BibitemOpen
  \bibfield{author}{%
  \bibinfo {author} {\bibfnamefont{L.}~\bibnamefont{Blanchet}}, \bibinfo
  {author} {\bibfnamefont{A.}~\bibnamefont{Buonanno}},\ and\ \bibinfo {author}
  {\bibfnamefont{G.}~\bibnamefont{Faye}},\ }%
  \bibfield{journal}{%
  \bibinfo {journal} {Phys. Rev. D}\ }%
  \textbf{\bibinfo {volume} {74}},\ \bibinfo {pages} {104034} (\bibinfo {year}
  {2006})%
  \bibAnnoteFile{NoStop}{bbf2006}%
\bibitem{blanchet2002}%
  \BibitemOpen
  \bibfield{author}{%
  \bibinfo {author} {\bibfnamefont{L.}~\bibnamefont{Blanchet}},\ }%
  \bibfield{journal}{%
  \bibinfo {journal} {Phys. Rev. D}\ }%
  \textbf{\bibinfo {volume} {65}},\ \bibinfo {pages} {124009} (\bibinfo {year}
  {2002})%
  \bibAnnoteFile{NoStop}{blanchet2002}%
\bibitem{lattimerprakash}%
  \BibitemOpen
  \bibfield{author}{%
  \bibinfo {author} {\bibfnamefont{J.~M.}\ \bibnamefont{Lattimer}}\ and\
  \bibinfo {author} {\bibfnamefont{M.}~\bibnamefont{Prakash}},\ }%
  \bibfield{journal}{%
  \bibinfo {journal} {Science}\ }%
  \textbf{\bibinfo {volume} {304}},\ \bibinfo {pages} {536} (\bibinfo {year}
  {2004})%
  \bibAnnoteFile{NoStop}{lattimerprakash}%
\bibitem{rlof2009}%
  \BibitemOpen
  \bibfield{author}{%
  \bibinfo {author} {\bibfnamefont{J.~S.}\ \bibnamefont{Read}}, \bibinfo
  {author} {\bibfnamefont{B.~D.}\ \bibnamefont{Lackey}}, \bibinfo {author}
  {\bibfnamefont{B.~J.}\ \bibnamefont{Owen}},\ and\ \bibinfo {author}
  {\bibfnamefont{J.~L.}\ \bibnamefont{Friedman}},\ }%
  \bibfield{journal}{%
  \bibinfo {journal} {Phys. Rev. D}\ }%
  \textbf{\bibinfo {volume} {79}},\ \bibinfo {pages} {124032} (\bibinfo {year}
  {2009})%
  \bibAnnoteFile{NoStop}{rlof2009}%
\bibitem{stairs2004}%
  \BibitemOpen
  \bibfield{author}{%
  \bibinfo {author} {\bibfnamefont{I.~H.}\ \bibnamefont{Stairs}},\ }%
  \bibfield{journal}{%
  \bibinfo {journal} {Science}\ }%
  \textbf{\bibinfo {volume} {304}},\ \bibinfo {pages} {547} (\bibinfo {year}
  {2004})%
  \bibAnnoteFile{NoStop}{stairs2004}%
\bibitem{lattimerprakash2001}%
  \BibitemOpen
  \bibfield{author}{%
  \bibinfo {author} {\bibfnamefont{J.~M.}\ \bibnamefont{Lattimer}}\ and\
  \bibinfo {author} {\bibfnamefont{M.}~\bibnamefont{Prakash}},\ }%
  \bibfield{journal}{%
  \bibinfo {journal} {Astrophys. J.}\ }%
  \textbf{\bibinfo {volume} {550}},\ \bibinfo {pages} {426} (\bibinfo {year}
  {2001})%
  \bibAnnoteFile{NoStop}{lattimerprakash2001}%
\bibitem{Note1}%
  \BibitemOpen
  \bibinfo {note} {In this paper, we determine the stiffness simply as the
  magnitude of the pressure for the nuclear-density region.}%
  \bibAnnoteFile{Stop}{Note1}%
\bibitem{lksbf}%
  \BibitemOpen
  \bibfield{author}{%
  \bibinfo {author} {\bibfnamefont{B.~D.}\ \bibnamefont{Lackey}}, \bibinfo
  {author} {\bibfnamefont{K.}~\bibnamefont{Kyutoku}}, \bibinfo {author}
  {\bibfnamefont{M.}~\bibnamefont{Shibata}}, \bibinfo {author}
  {\bibfnamefont{P.~R.}\ \bibnamefont{Brady}},\ and\ \bibinfo {author}
  {\bibfnamefont{J.~L.}\ \bibnamefont{Friedman}},\ }%
  \bibinfo {howpublished} {to be published}%
  \bibAnnoteFile{NoStop}{lksbf}%
\bibitem{shibata1996}%
  \BibitemOpen
  \bibfield{author}{%
  \bibinfo {author} {\bibfnamefont{M.}~\bibnamefont{Shibata}},\ }%
  \bibfield{journal}{%
  \bibinfo {journal} {Prog. Theor. Phys.}\ }%
  \textbf{\bibinfo {volume} {96}},\ \bibinfo {pages} {917} (\bibinfo {year}
  {1996})%
  \bibAnnoteFile{NoStop}{shibata1996}%
\bibitem{wigginslai2000}%
  \BibitemOpen
  \bibfield{author}{%
  \bibinfo {author} {\bibfnamefont{P.}~\bibnamefont{Wiggins}}\ and\ \bibinfo
  {author} {\bibfnamefont{D.}~\bibnamefont{Lai}},\ }%
  \bibfield{journal}{%
  \bibinfo {journal} {Astrophys. J.}\ }%
  \textbf{\bibinfo {volume} {532}},\ \bibinfo {pages} {530} (\bibinfo {year}
  {2000})%
  \bibAnnoteFile{NoStop}{wigginslai2000}%
\bibitem{Note2}%
  \BibitemOpen
  \bibinfo {note} {In this paper, ``the ISCO radius'' always represents ``the
  ISCO radius in the Boyer-Lindquist coordinates,'' which is physical in the
  sense that it gives the proper circumferential length for the equatorial
  circular orbit. It should be noted that the coordinate radius of the ISCO in
  our simulation is different from the Boyer-Lindquist one.}%
  \bibAnnoteFile{Stop}{Note2}%
\bibitem{bpt1972}%
  \BibitemOpen
  \bibfield{author}{%
  \bibinfo {author} {\bibfnamefont{J.~M.}\ \bibnamefont{Bardeen}}, \bibinfo
  {author} {\bibfnamefont{W.~H.}\ \bibnamefont{Press}},\ and\ \bibinfo {author}
  {\bibfnamefont{S.~A.}\ \bibnamefont{Teukolsky}},\ }%
  \bibfield{journal}{%
  \bibinfo {journal} {Astrophys. J.}\ }%
  \textbf{\bibinfo {volume} {178}},\ \bibinfo {pages} {347} (\bibinfo {year}
  {1972})%
  \bibAnnoteFile{NoStop}{bpt1972}%
\bibitem{blanchet}%
  \BibitemOpen
  \bibfield{author}{%
  \bibinfo {author} {\bibfnamefont{L.}~\bibnamefont{Blanchet}},\ }%
  \bibfield{journal}{%
  \bibinfo {journal} {Living Rev. Relativity}\ }%
  \textbf{\bibinfo {volume} {9}},\ \bibinfo {pages} {4} (\bibinfo {year}
  {2006})%
  \bibAnnoteFile{NoStop}{blanchet}%
\bibitem{yst2008}%
  \BibitemOpen
  \bibfield{author}{%
  \bibinfo {author} {\bibfnamefont{T.}~\bibnamefont{Yamamoto}}, \bibinfo
  {author} {\bibfnamefont{M.}~\bibnamefont{Shibata}},\ and\ \bibinfo {author}
  {\bibfnamefont{K.}~\bibnamefont{Taniguchi}},\ }%
  \bibfield{journal}{%
  \bibinfo {journal} {Phys. Rev. D}\ }%
  \textbf{\bibinfo {volume} {78}},\ \bibinfo {pages} {064054} (\bibinfo {year}
  {2008})%
  \bibAnnoteFile{NoStop}{yst2008}%
\bibitem{shibatanakamura1995}%
  \BibitemOpen
  \bibfield{author}{%
  \bibinfo {author} {\bibfnamefont{M.}~\bibnamefont{Shibata}}\ and\ \bibinfo
  {author} {\bibfnamefont{T.}~\bibnamefont{Nakamura}},\ }%
  \bibfield{journal}{%
  \bibinfo {journal} {Phys. Rev. D}\ }%
  \textbf{\bibinfo {volume} {52}},\ \bibinfo {pages} {5428} (\bibinfo {year}
  {1995})%
  \bibAnnoteFile{NoStop}{shibatanakamura1995}%
\bibitem{baumgarteshapiro1998}%
  \BibitemOpen
  \bibfield{author}{%
  \bibinfo {author} {\bibfnamefont{T.~W.}\ \bibnamefont{Baumgarte}}\ and\
  \bibinfo {author} {\bibfnamefont{S.~L.}\ \bibnamefont{Shapiro}},\ }%
  \bibfield{journal}{%
  \bibinfo {journal} {Phys. Rev. D}\ }%
  \textbf{\bibinfo {volume} {59}},\ \bibinfo {pages} {024007} (\bibinfo {year}
  {1998})%
  \bibAnnoteFile{NoStop}{baumgarteshapiro1998}%
\bibitem{bghhst2008}%
  \BibitemOpen
  \bibfield{author}{%
  \bibinfo {author} {\bibfnamefont{B.}~\bibnamefont{Br{\"u}gmann}}, \bibinfo
  {author} {\bibfnamefont{J.~A.}\ \bibnamefont{Gonz{\'a}lez}}, \bibinfo
  {author} {\bibfnamefont{M.}~\bibnamefont{Hannam}}, \bibinfo {author}
  {\bibfnamefont{S.}~\bibnamefont{Husa}}, \bibinfo {author}
  {\bibfnamefont{U.}~\bibnamefont{Sperhake}},\ and\ \bibinfo {author}
  {\bibfnamefont{W.}~\bibnamefont{Tichy}},\ }%
  \bibfield{journal}{%
  \bibinfo {journal} {Phys. Rev. D}\ }%
  \textbf{\bibinfo {volume} {77}},\ \bibinfo {pages} {024027} (\bibinfo {year}
  {2008})%
  \bibAnnoteFile{NoStop}{bghhst2008}%
\bibitem{kurganovtadmor2000}%
  \BibitemOpen
  \bibfield{author}{%
  \bibinfo {author} {\bibfnamefont{A.}~\bibnamefont{Kurganov}}\ and\ \bibinfo
  {author} {\bibfnamefont{E.}~\bibnamefont{Tadmor}},\ }%
  \bibfield{journal}{%
  \bibinfo {journal} {J. Comput. Phys.}\ }%
  \textbf{\bibinfo {volume} {160}},\ \bibinfo {pages} {241} (\bibinfo {year}
  {2000})%
  \bibAnnoteFile{NoStop}{kurganovtadmor2000}%
\bibitem{reisswigpollney2010}%
  \BibitemOpen
  \bibfield{author}{%
  \bibinfo {author} {\bibfnamefont{C.}~\bibnamefont{Reisswig}}\ and\ \bibinfo
  {author} {\bibfnamefont{D.}~\bibnamefont{Pollney}},\ }%
  \bibinfo {journal} {arXiv:1006.1632}%
  \bibAnnoteFile{NoStop}{reisswigpollney2010}%
\bibitem{bbkmpsct2007}%
  \BibitemOpen
\bibfield{journal}{%
    }%
  \bibfield{author}{%
  \bibinfo {author} {\bibfnamefont{M.}~\bibnamefont{Boyle}}, \bibinfo {author}
  {\bibfnamefont{D.~A.}\ \bibnamefont{Brown}}, \bibinfo {author}
  {\bibfnamefont{L.~E.}\ \bibnamefont{Kidder}}, \bibinfo {author}
  {\bibfnamefont{A.~H.}\ \bibnamefont{Mrou{\'e}}}, \bibinfo {author}
  {\bibfnamefont{H.~P.}\ \bibnamefont{Pfeiffer}}, \bibinfo {author}
  {\bibfnamefont{M.~A.}\ \bibnamefont{Scheel}}, \bibinfo {author}
  {\bibfnamefont{G.~B.}\ \bibnamefont{Cook}},\ and\ \bibinfo {author}
  {\bibfnamefont{S.~A.}\ \bibnamefont{Teukolsky}},\ }%
  \bibfield{journal}{%
  \bibinfo {journal} {Phys. Rev. D}\ }%
  \textbf{\bibinfo {volume} {76}},\ \bibinfo {pages} {124038} (\bibinfo {year}
  {2007})%
  \bibAnnoteFile{NoStop}{bbkmpsct2007}%
\bibitem{santamariaetal2010}%
  \BibitemOpen
  \bibfield{author}{%
  \bibinfo {author} {\bibfnamefont{L.}~\bibnamefont{Santamaria}}, \bibinfo
  {author} {\bibfnamefont{F.}~\bibnamefont{Ohme}}, \bibinfo {author}
  {\bibfnamefont{P.}~\bibnamefont{Ajith}}, \bibinfo {author}
  {\bibfnamefont{B.}~\bibnamefont{Br{\"u}gmann}}, \bibinfo {author}
  {\bibfnamefont{N.}~\bibnamefont{Dorband}}, \bibinfo {author}
  {\bibfnamefont{M.}~\bibnamefont{Hannam}}, \bibinfo {author}
  {\bibfnamefont{S.}~\bibnamefont{Husa}}, \bibinfo {author}
  {\bibfnamefont{P.}~\bibnamefont{M{\"o}sta}}, \bibinfo {author}
  {\bibfnamefont{D.}~\bibnamefont{Pollney}}, \bibinfo {author}
  {\bibfnamefont{C.}~\bibnamefont{Reisswig}}, \bibinfo {author}
  {\bibfnamefont{E.~L.}\ \bibnamefont{Robinson}}, \bibinfo {author}
  {\bibfnamefont{J.}~\bibnamefont{Seiler}},\ and\ \bibinfo {author}
  {\bibfnamefont{B.}~\bibnamefont{Krishnan}},\ }%
  \bibfield{journal}{%
  \bibinfo {journal} {Phys. Rev. D}\ }%
  \textbf{\bibinfo {volume} {82}},\ \bibinfo {pages} {064016} (\bibinfo {year}
  {2010})%
  \bibAnnoteFile{NoStop}{santamariaetal2010}%
\bibitem{kidder2008}%
  \BibitemOpen
  \bibfield{author}{%
  \bibinfo {author} {\bibfnamefont{L.~E.}\ \bibnamefont{Kidder}},\ }%
  \bibfield{journal}{%
  \bibinfo {journal} {Phys. Rev. D}\ }%
  \textbf{\bibinfo {volume} {77}},\ \bibinfo {pages} {044016} (\bibinfo {year}
  {2008})%
  \bibAnnoteFile{NoStop}{kidder2008}%
\bibitem{damournagar}%
  \BibitemOpen
  \bibfield{author}{%
  \bibinfo {author} {\bibfnamefont{T.}~\bibnamefont{Damour}}\ and\ \bibinfo
  {author} {\bibfnamefont{A.}~\bibnamefont{Nagar}},\ }%
  \bibinfo {journal} {arXiv:0906.1769}%
  \bibAnnoteFile{NoStop}{damournagar}%
\bibitem{kidder1995}%
  \BibitemOpen
\bibfield{journal}{%
    }%
  \bibfield{author}{%
  \bibinfo {author} {\bibfnamefont{L.~E.}\ \bibnamefont{Kidder}},\ }%
  \bibfield{journal}{%
  \bibinfo {journal} {Phys. Rev. D}\ }%
  \textbf{\bibinfo {volume} {52}},\ \bibinfo {pages} {821} (\bibinfo {year}
  {1995})%
  \bibAnnoteFile{NoStop}{kidder1995}%
\bibitem{fishbone1973}%
  \BibitemOpen
  \bibfield{author}{%
  \bibinfo {author} {\bibfnamefont{L.~G.}\ \bibnamefont{Fishbone}},\ }%
  \bibfield{journal}{%
  \bibinfo {journal} {Astrophys. J.}\ }%
  \textbf{\bibinfo {volume} {185}},\ \bibinfo {pages} {43} (\bibinfo {year}
  {1973})%
  \bibAnnoteFile{NoStop}{fishbone1973}%
\bibitem{marck1983}%
  \BibitemOpen
  \bibfield{author}{%
  \bibinfo {author} {\bibfnamefont{J.-A.}\ \bibnamefont{Marck}},\ }%
  \bibfield{journal}{%
  \bibinfo {journal} {Proc. Roy. Soc. London}\ }%
  \textbf{\bibinfo {volume} {385}},\ \bibinfo {pages} {431} (\bibinfo {year}
  {1983})%
  \bibAnnoteFile{NoStop}{marck1983}%
\bibitem{ism2005}%
  \BibitemOpen
  \bibfield{author}{%
  \bibinfo {author} {\bibfnamefont{M.}~\bibnamefont{Ishii}}, \bibinfo {author}
  {\bibfnamefont{M.}~\bibnamefont{Shibata}},\ and\ \bibinfo {author}
  {\bibfnamefont{Y.}~\bibnamefont{Mino}},\ }%
  \bibfield{journal}{%
  \bibinfo {journal} {Phys. Rev. D}\ }%
  \textbf{\bibinfo {volume} {71}},\ \bibinfo {pages} {044017} (\bibinfo {year}
  {2005})%
  \bibAnnoteFile{NoStop}{ism2005}%
\bibitem{mcclintockremillard}%
  \BibitemOpen
  \bibfield{author}{%
  \bibinfo {author} {\bibfnamefont{J.~E.}\ \bibnamefont{McClintock}}\ and\
  \bibinfo {author} {\bibfnamefont{R.~A.}\ \bibnamefont{Remillard}},\ }%
  in\ \emph{\bibinfo {booktitle} {Compact Stellar X-ray Sources}},\ \bibinfo
  {editor} {edited by\ \bibinfo {editor} {\bibfnamefont{W.~H.~G.}\
  \bibnamefont{Lewin}}\ and\ \bibinfo {editor}
  {\bibfnamefont{M.}~\bibnamefont{van~der Klis}}}\ (\bibinfo {publisher}
  {Cambridge University Press},\ \bibinfo {year} {2006})%
  \bibAnnoteFile{NoStop}{mcclintockremillard}%
\bibitem{bbfrvvh2010}%
  \BibitemOpen
  \bibfield{author}{%
  \bibinfo {author} {\bibfnamefont{K.}~\bibnamefont{Belczynski}}, \bibinfo
  {author} {\bibfnamefont{T.}~\bibnamefont{Bulik}}, \bibinfo {author}
  {\bibfnamefont{C.~L.}\ \bibnamefont{Fryer}}, \bibinfo {author}
  {\bibfnamefont{A.}~\bibnamefont{Ruiter}}, \bibinfo {author}
  {\bibfnamefont{F.}~\bibnamefont{Valsecchi}}, \bibinfo {author}
  {\bibfnamefont{J.~S.}\ \bibnamefont{Vink}},\ and\ \bibinfo {author}
  {\bibfnamefont{J.~R.}\ \bibnamefont{Hurley}},\ }%
  \bibfield{journal}{%
  \bibinfo {journal} {Astrophys. J.}\ }%
  \textbf{\bibinfo {volume} {714}},\ \bibinfo {pages} {1217} (\bibinfo {year}
  {2010})%
  \bibAnnoteFile{NoStop}{bbfrvvh2010}%
\bibitem{fdkt2011}%
  \BibitemOpen
  \bibfield{author}{%
  \bibinfo {author} {\bibfnamefont{F.}~\bibnamefont{Foucart}}, \bibinfo
  {author} {\bibfnamefont{M.~D.}\ \bibnamefont{Duez}}, \bibinfo {author}
  {\bibfnamefont{L.~E.}\ \bibnamefont{Kidder}},\ and\ \bibinfo {author}
  {\bibfnamefont{S.~A.}\ \bibnamefont{Teukolsky}},\ }%
  \bibfield{journal}{%
  \bibinfo {journal} {Phys. Rev. D}\ }%
  \textbf{\bibinfo {volume} {83}},\ \bibinfo {pages} {024005} (\bibinfo {year}
  {2011})%
  \bibAnnoteFile{NoStop}{fdkt2011}%
\bibitem{penrose}%
  \BibitemOpen
  \bibfield{author}{%
  \bibinfo {author} {\bibfnamefont{R.}~\bibnamefont{Penrose}},\ }%
  in\ \emph{\bibinfo {booktitle} {general relativity: an Einstein centenary
  survey}},\ \bibinfo {editor} {edited by\ \bibinfo {editor}
  {\bibfnamefont{S.~W.}\ \bibnamefont{Hawking}}\ and\ \bibinfo {editor}
  {\bibfnamefont{W.}~\bibnamefont{Israel}}}\ (\bibinfo {year} {1979})%
  \bibAnnoteFile{NoStop}{penrose}%
\bibitem{bcs2009}%
  \BibitemOpen
  \bibfield{author}{%
  \bibinfo {author} {\bibfnamefont{E.}~\bibnamefont{Berti}}, \bibinfo {author}
  {\bibfnamefont{V.}~\bibnamefont{Cardoso}},\ and\ \bibinfo {author}
  {\bibfnamefont{A.~O.}\ \bibnamefont{Starinets}},\ }%
  \bibfield{journal}{%
  \bibinfo {journal} {Class. Quantum Grav.}\ }%
  \textbf{\bibinfo {volume} {26}},\ \bibinfo {pages} {163001} (\bibinfo {year}
  {2009})%
  \bibAnnoteFile{NoStop}{bcs2009}%
\bibitem{ulfgs2006}%
  \BibitemOpen
  \bibfield{author}{%
  \bibinfo {author} {\bibfnamefont{K.}~\bibnamefont{Ury{\=u}}}, \bibinfo
  {author} {\bibfnamefont{F.}~\bibnamefont{Limousin}}, \bibinfo {author}
  {\bibfnamefont{J.~L.}\ \bibnamefont{Friedman}}, \bibinfo {author}
  {\bibfnamefont{E.}~\bibnamefont{Gourgoulhon}},\ and\ \bibinfo {author}
  {\bibfnamefont{M.}~\bibnamefont{Shibata}},\ }%
  \bibfield{journal}{%
  \bibinfo {journal} {Phys. Rev. Lett.}\ }%
  \textbf{\bibinfo {volume} {97}},\ \bibinfo {pages} {171101} (\bibinfo {year}
  {2006})%
  \bibAnnoteFile{NoStop}{ulfgs2006}%
\bibitem{ulfgs2009}%
  \BibitemOpen
  \bibfield{author}{%
  \bibinfo {author} {\bibfnamefont{K.}~\bibnamefont{Ury{\=u}}}, \bibinfo
  {author} {\bibfnamefont{F.}~\bibnamefont{Limousin}}, \bibinfo {author}
  {\bibfnamefont{J.~L.}\ \bibnamefont{Friedman}}, \bibinfo {author}
  {\bibfnamefont{E.}~\bibnamefont{Gourgoulhon}},\ and\ \bibinfo {author}
  {\bibfnamefont{M.}~\bibnamefont{Shibata}},\ }%
  \bibfield{journal}{%
  \bibinfo {journal} {Phys. Rev. D}\ }%
  \textbf{\bibinfo {volume} {80}},\ \bibinfo {pages} {124004} (\bibinfo {year}
  {2009})%
  \bibAnnoteFile{NoStop}{ulfgs2009}%
\bibitem{saijonakamura2000}%
  \BibitemOpen
  \bibfield{author}{%
  \bibinfo {author} {\bibfnamefont{M.}~\bibnamefont{Saijo}}\ and\ \bibinfo
  {author} {\bibfnamefont{T.}~\bibnamefont{Nakamura}},\ }%
  \bibfield{journal}{%
  \bibinfo {journal} {Phys. Rev. Lett.}\ }%
  \textbf{\bibinfo {volume} {85}},\ \bibinfo {pages} {2665} (\bibinfo {year}
  {2000})%
  \bibAnnoteFile{NoStop}{saijonakamura2000}%
\bibitem{saijonakamura2001}%
  \BibitemOpen
  \bibfield{author}{%
  \bibinfo {author} {\bibfnamefont{M.}~\bibnamefont{Saijo}}\ and\ \bibinfo
  {author} {\bibfnamefont{T.}~\bibnamefont{Nakamura}},\ }%
  \bibfield{journal}{%
  \bibinfo {journal} {Phys. Rev. D}\ }%
  \textbf{\bibinfo {volume} {63}},\ \bibinfo {pages} {064004} (\bibinfo {year}
  {2001})%
  \bibAnnoteFile{NoStop}{saijonakamura2001}%
\bibitem{et2010}%
  \BibitemOpen
  \bibfield{author}{%
  \bibinfo {author} {\bibnamefont{{M. Punturo et al.}}},\ }%
  \bibfield{journal}{%
  \bibinfo {journal} {Class. Quantum Grav.}\ }%
  \textbf{\bibinfo {volume} {27}},\ \bibinfo {pages} {194002} (\bibinfo {year}
  {2010})%
  \bibAnnoteFile{NoStop}{et2010}%
\bibitem{bdgnr2010}%
  \BibitemOpen
  \bibfield{author}{%
  \bibinfo {author} {\bibfnamefont{L.}~\bibnamefont{Baiotti}}, \bibinfo
  {author} {\bibfnamefont{T.}~\bibnamefont{Damour}}, \bibinfo {author}
  {\bibfnamefont{B.}~\bibnamefont{Giacomazzo}}, \bibinfo {author}
  {\bibfnamefont{A.}~\bibnamefont{Nagar}},\ and\ \bibinfo {author}
  {\bibfnamefont{L.}~\bibnamefont{Rezzolla}},\ }%
  \bibfield{journal}{%
  \bibinfo {journal} {Phys. Rev. Lett.}\ }%
  \textbf{\bibinfo {volume} {105}},\ \bibinfo {pages} {261101} (\bibinfo {year}
  {2010})%
  \bibAnnoteFile{NoStop}{bdgnr2010}%
\bibitem{bdgnr2011}%
  \BibitemOpen
  \bibfield{author}{%
  \bibinfo {author} {\bibfnamefont{L.}~\bibnamefont{Baiotti}}, \bibinfo
  {author} {\bibfnamefont{T.}~\bibnamefont{Damour}}, \bibinfo {author}
  {\bibfnamefont{B.}~\bibnamefont{Giacomazzo}}, \bibinfo {author}
  {\bibfnamefont{A.}~\bibnamefont{Nagar}},\ and\ \bibinfo {author}
  {\bibfnamefont{L.}~\bibnamefont{Rezzolla}},\ }%
  \bibfield{journal}{%
  \bibinfo {journal} {Phys. Rev. D}\ }%
  \textbf{\bibinfo {volume} {84}},\ \bibinfo {pages} {024017} (\bibinfo {year}
  {2011})%
  \bibAnnoteFile{NoStop}{bdgnr2011}%
\bibitem{Note3}%
  \BibitemOpen
  \bibinfo {note} {We refer to $f_{\protect \rm dam}$ as $f_{\protect \rm cut}$
  throughout in the previous work \cite {kst2010}. In the present paper, we
  distinguish $f_{\protect \rm dam}$ from $f_{\protect \rm cut}$ because the
  method for determining $f_{\protect \rm dam}$ is different from that for
  $f_{\protect \rm cut}$}%
  \bibAnnoteFile{NoStop}{Note3}%
\bibitem{Note4}%
  \BibitemOpen
  \bibinfo {note} {The relation between $f_{\protect \rm cut} m_0$ and
  ${\protect \cal C}$ is different from the one obtained in Ref.~\cite
  {kst2010} due to the different definition of $f_{\protect \rm cut}$.}%
  \bibAnnoteFile{Stop}{Note4}%
\bibitem{rezzolla2009}%
  \BibitemOpen
  \bibfield{author}{%
  \bibinfo {author} {\bibfnamefont{L.}~\bibnamefont{Rezzolla}},\ }%
  \bibfield{journal}{%
  \bibinfo {journal} {Class. Quantum Grav.}\ }%
  \textbf{\bibinfo {volume} {26}},\ \bibinfo {pages} {094023} (\bibinfo {year}
  {2009})%
  \bibAnnoteFile{NoStop}{rezzolla2009}%
\bibitem{zcl2011}%
  \BibitemOpen
  \bibfield{author}{%
  \bibinfo {author} {\bibfnamefont{Y.}~\bibnamefont{Zlochower}}, \bibinfo
  {author} {\bibfnamefont{M.}~\bibnamefont{Campanelli}},\ and\ \bibinfo
  {author} {\bibfnamefont{C.~O.}\ \bibnamefont{Lousto}},\ }%
  \bibfield{journal}{%
  \bibinfo {journal} {Class. Quantum Grav.}\ }%
  \textbf{\bibinfo {volume} {28}},\ \bibinfo {pages} {114015} (\bibinfo {year}
  {2011})%
  \bibAnnoteFile{NoStop}{zcl2011}%
\bibitem{sekiguchi2010}%
  \BibitemOpen
  \bibfield{author}{%
  \bibinfo {author} {\bibfnamefont{Y.}~\bibnamefont{Sekiguchi}},\ }%
  \bibfield{journal}{%
  \bibinfo {journal} {Prog. Theor. Phys.}\ }%
  \textbf{\bibinfo {volume} {124}},\ \bibinfo {pages} {331} (\bibinfo {year}
  {2010})%
  \bibAnnoteFile{NoStop}{sekiguchi2010}%
\bibitem{sekiguchishibata2010}%
  \BibitemOpen
  \bibfield{author}{%
  \bibinfo {author} {\bibfnamefont{Y.}~\bibnamefont{Sekiguchi}}\ and\ \bibinfo
  {author} {\bibfnamefont{M.}~\bibnamefont{Shibata}},\ }%
  \bibfield{journal}{%
  \bibinfo {journal} {Astrophys. J.}\ }%
  \textbf{\bibinfo {volume} {737}},\ \bibinfo {pages} {6} (\bibinfo {year}
  {2011})%
  \bibAnnoteFile{NoStop}{sekiguchishibata2010}%
\bibitem{skks2011}%
  \BibitemOpen
  \bibfield{author}{%
  \bibinfo {author} {\bibfnamefont{Y.}~\bibnamefont{Sekiguchi}}, \bibinfo
  {author} {\bibfnamefont{K.}~\bibnamefont{Kiuchi}}, \bibinfo {author}
  {\bibfnamefont{K.}~\bibnamefont{Kyutoku}},\ and\ \bibinfo {author}
  {\bibfnamefont{M.}~\bibnamefont{Shibata}},\ }%
  \bibfield{journal}{%
  \bibinfo {journal} {Phys. Rev. Lett.}\ }%
  \textbf{\bibinfo {volume} {107}},\ \bibinfo {pages} {051102} (\bibinfo {year}
  {2011})%
  \bibAnnoteFile{NoStop}{skks2011}%
\end{thebibliography}
%

\end{document}